\documentclass[aps,prc,amsfonts,preprintnumbers,superscriptaddress,showpacs,nofootinbib]{revtex4}

\usepackage{graphics}
\usepackage{amssymb}
\usepackage{psfrag}
\usepackage{epsfig}
\usepackage{epsf}
\usepackage{float}
\usepackage{color}

\newcommand{\fet}[1]{\mbox{\boldmath $#1$}}

\newcommand{\beq}{\begin{equation}}
\newcommand{\eeq}{\end{equation}}
\newcommand{\beqa}{\begin{eqnarray}}
\newcommand{\eeqa}{\end{eqnarray}}
\newcommand{\nn}{\nonumber \\ }
\newcommand{\nnrl}{\right. \nonumber \\ &&  \left. }
\newcommand{\nnrlrl}{\right. \right. \nonumber \\ && \left. \left.}
\newcommand{\nnrlrlrl}{\right. \right. \right. \nonumber \\ && 
\left. \left. \left.}

\begin{document}

\title{Chiral three-nucleon force at N$^4$LO II: 
Intermediate-range contributions}

\author{H.~Krebs}
\email[]{Email: hermann.krebs@rub.de}
\affiliation{Institut f\"ur Theoretische Physik II, Ruhr-Universit\"at Bochum,
  D-44780 Bochum, Germany}
\author{A.~Gasparyan}
\email[]{Email: ashotg@tp2.rub.de}
\affiliation{Institut f\"ur Theoretische Physik II, Ruhr-Universit\"at Bochum,
  D-44780 Bochum, Germany}
\affiliation{FSBI SSC RF ITEP, Bolshaya Cheremushkinskaya 25, 117218 Moscow, Russia}
\author{E.~Epelbaum}
\email[]{Email: evgeny.epelbaum@rub.de}
\affiliation{Institut f\"ur Theoretische Physik II, Ruhr-Universit\"at Bochum,
  D-44780 Bochum, Germany}
\date{\today}

\begin{abstract}
We derive the subleading contributions to the
two-pion-one-pion exchange and ring three-nucleon force topologies
emerging at next-to-next-to-next-to-next-to-leading order in
chiral effective field theory.  The resulting 
expressions do not involve any unknown parameters. To study 
convergence of the chiral expansion  
we work out the most general operator structure of a local
isospin-invariant three-nucleon force. Using the resulting 
operator basis with 22 independent structures, we compare the strength of the corresponding potentials 
in configuration space for individual topologies at various orders in the chiral expansion. 
As expected, the subleading contributions from the two-pion-one-pion-exchange and ring diagrams 
are large which can be understood in terms of intermediate excitation of the $\Delta$(1232)
isobar. 
\end{abstract}


\maketitle

\vspace{-0.2cm}

%%%%%%%%%%%%%%%%%%%%%%%%%%%%%%%%%%%%%%%%%%%%%%%%%%%%%%%%%%%%%%%%%%%%%%%%%%%%%%%%%
\section{Introduction}
\def\theequation{\arabic{section}.\arabic{equation}}
\label{sec:intro}

Three-nucleon forces (3NF) are presently subject to intense research, see 
Refs.~\cite{Stephan:2010zz,Ciepal:2012zz,Viviani:2010mf,Navratil:2007we,Gazit:2008ma,Roth:2011ar,Hebeler:2010jx,Hagen:2012sh,Holt:2012fr,Tews:2012fj}
for a selection of recent few- and
many-body calculations along these lines and
Refs.~\cite{KalantarNayestanaki:2011wz,Hammer:2012id}  for review articles.
%On the one hand, this interest is stimulated by the
%substantial progress achieved in recent years in many-body theory. 
On the one hand, rapidly increasing
computational resources coupled with sophisticated  few- and
many-body methods allow nowadays for reliable and accurate nuclear
structure calculations for light and even medium-mass nuclei. One can,
therefore, relate the properties of the nuclear
Hamiltonian to observables in a reliable way and without invoking any uncontrollable
approximations. On the other hand, considerable progress 
has also been reached towards quantitative description of  
nuclear forces using the framework of chiral effective field theory
(EFT), see recent review articles
\cite{Epelbaum:2008ga,Machleidt:2011zz,Epelbaum:2012vx,Hammer:2012id} and references
therein. In particular, nucleon-nucleon (NN) potentials at
next-to-next-to-next-to-leading order (N$^3$LO) in the chiral expansion were
developed \cite{Entem:2003ft,Epelbaum:2004fk}, which allow for an
accurate description of NN scattering data up to laboratory energies of the order
of $E_{\rm lab} \sim 200$ MeV. For heavier systems, 
the accuracy of theoretical predictions is currently limited by the
3NFs for which
only the dominant contributions at next-to-next-to-leading
order (N$^2$LO) in the chiral expansion of the nuclear Hamiltonian
have so far been  employed in few- and many-body  calculations. 

\begin{figure}[t]
\vskip 1 true cm
%  \begin{center} 
%    \epsfxsize=4.3cm
%    \epsffile{fig2.eps}
\includegraphics[width=0.9\textwidth,keepaspectratio,angle=0,clip]{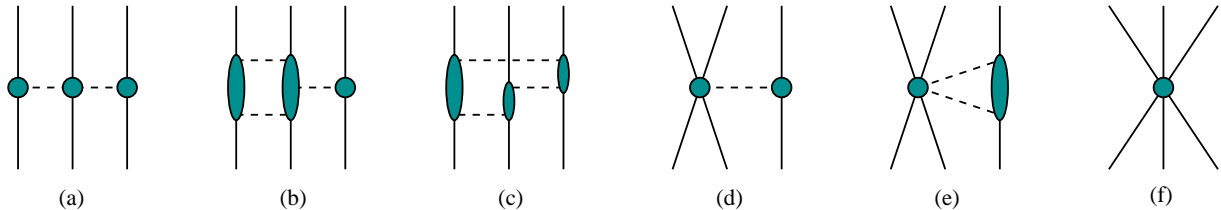}
    \caption{
          Various topologies contributing to the 3NF 
up-to-and-including N$^4$LO:  two-pion ($2\pi$) exchange (a), two-pion-one-pion ($2\pi$-$1\pi$)
         exchange (b), ring (c), one-pion-exchange-contact (d),
         two-pion-exchange-contact (e) and purely contact (f) diagrams.
Solid and dashed lines
         represent nucleons and pions, respectively. 
Shaded blobs represent the corresponding amplitudes. 
%         Long-range topologies in the three-nucleon force up to
%         N$^4$LO: two-pion ($2\pi$) exchange (a), two-pion-one-pion ($2\pi$-$1\pi$)
%         exchange (b) and ring (c) diagrams. Solid and dashed lines
%         represent nucleons and pions, respectively. Shaded blobs are
%         the corresponding amplitudes. 
\label{fig0} 
 }
%  \end{center}
\end{figure}

The chiral expansion of the 3NF at one-loop level, i.e.~up to
 next-to-next-to-next-to-next-to-leading order (N$^4$LO), 
can be described in terms of six topologies depicted in
Fig.~\ref{fig0}. The first nonvanishing contributions to the 3NF
emerge at N$^2$LO from tree-level diagrams corresponding to the
$2\pi$-exchange, one-pion-exchange-contact and purely contact graphs
(a), (d) and (f), respectively \cite{vanKolck:1994yi,Epelbaum:2002vt}. The shorter-range terms emerging from 
diagrams (d) and (f) depend on one unknown low-energy constant (LEC) each which can be
determined from suitable few-nucleon observables, see
e.g.~\cite{Navratil:2007we,Gazit:2008ma,Epelbaum:2002vt,Epelbaum:2009zsa}. 
The long-range contribution (a) is, on the other
hand, parameter-free since the LECs $c_{1}$,   $c_{3}$ and  $c_{4}$
accompanying the subleading $\pi \pi NN$ vertices can be 
extracted from pion-nucleon scattering, see
\cite{Bernard:1996gq,Fettes:1998ud,Fettes:2000xg,Buettiker:1999ap} for 
heavy-baryon results, Refs.~\cite{Chen:2012nx,Alarcon:2012kn} for some more recent calculations using
manifestly covariant formulations of chiral perturbation theory as
well as Refs.~\cite{Rentmeester:2003mf}  for an attempt to determine
these LECs from proton-proton and neutron-proton partial wave
analyses. The resulting 3NF at N$^2$LO has been intensively explored in three-
and four-nucleon scattering calculations, see
\cite{KalantarNayestanaki:2011wz} and references therein.  One
finds a good description of low-energy nucleon-deuteron  scattering
observables except for the well-known, long-standing puzzles such as
the vector analyzing power in elastic nucleon-deuteron scattering (the so-called
$A_y$-puzzle) and  the cross section in the space-star
breakup configuration, see Ref.~\cite{KalantarNayestanaki:2011wz} for more
details. Promising results for low-energy four-nucleon
scattering observables based on the chiral 3NF, especially in
connection with the $A_y$-puzzle, are reported in
Ref.~\cite{Viviani:2010mf}.  While nucleon-deuteron scattering data at higher energies are
also well described, the theoretical uncertainty increases rapidly 
reflecting similar pattern in the two-nucleon sector at this order in the
chiral expansion. Promising results based on chiral nuclear forces were also obtained in
nuclear structure calculations showing, in particular, sensitivity to
the individual terms of the N$^2$LO  3NF, see \cite{KalantarNayestanaki:2011wz}
and references therein. All these findings clearly underline the need
to include corrections to the 3NF beyond the leading terms at N$^2$LO.  

The first corrections to the 3NF emerge at N$^3$LO from all possible one-loop
diagrams of type (a)-(e) involving solely the lowest-order vertices
from the effective chiral Lagrangian. The resulting parameter-free
expressions for the long-range and intermediate-range contributions of
types (a),  and (b), (c), respectively, 
can be found in Ref.~\cite{Bernard:2007sp}, see also
\cite{Ishikawa:2007zz} where the leading one-loop corrections to the  longest-range two-pion exchange terms
of type (a) are calculated within the infrared-regularized version of
chiral perturbation theory.  N$^3$LO contributions to shorter-range
diagrams of types (d) and (e) as well as the leading
relativistic corrections are given in Ref.~\cite{Bernard:2011zr}, see
also \cite{Robilotta:2006xq} for a related work. Notice
that these shorter-range terms are driven by the leading four-nucleon
contact operators which also contribute to nucleon-nucleon S-wave
scattering. Thus, they do not involve any unknown LECs. Finally, there are no
corrections to the purely short-range topology of type (f) at this
order.  An interesting feature of the N$^3$LO 3NF corrections is their
rather rich isospin-spin-momentum structure emerging primarily from
the ring topology (c) in Fig.~1. This is in  contrast with the quite
restricted operator structure of the N$^2$LO 3NF. The impact of these 
novel 3NF terms on nucleon-deuteron scattering and nuclear structure
observables is unknown which makes the complete N$^3$LO calculations
even more urgent, especially in connection with the already mentioned 
unsolved puzzles. Numerical implementation of the new terms in the
3NF at N$^3$LO requires their partial wave
decomposition  which is a nontrivial
task. In Ref.~\cite{Golak:2009ri}, a novel method to perform partial-wave
decomposition of any type of the 3NF by carrying out five-dimensional
angular integrations numerically was introduced.  This approach is
quite general in the sense that it can be applied to any type of 3NF
but requires substantial computational resources. The partial wave
decomposition of the N$^3$LO  3NF using this new technique is in
progress, see Ref.~\cite{Skibinski:2011vi} for some first (but still incomplete) results. 

Meanwhile, one may ask whether the derived expressions for the  3NF at
subleading order in the chiral expansion are already converged or, at
least, provide a reasonable approximation to the converged
result. This applies especially to new operator structures emerging
from the genuine loop topologies (b) and (c), whose chiral expansion 
starts at N$^3$LO rather than N$^2$LO. At this order, the resulting
contributions still miss physics associated with intermediate
$\Delta$(1232) excitations. In the standard chiral EFT formulation
based on pions and nucleons as the only explicit degrees of freedom,
all effects of the $\Delta$ (and heavier resonances as well as heavy mesons) are hidden
in the (renormalized) values of certain LECs starting from the
subleading effective Lagrangian. The major part of 
the $\Delta$ contributions to the nuclear forces is known to be well
represented in terms of resonance saturation of
the LECs $c_{2,3,4}$ accompanying the subleading $\pi \pi NN$ vertices
\cite{Bernard:1996gq,Ordonez:1993tn,Kaiser:1998wa,Krebs:2007rh} 
(see, however, the last two references for examples of the
$\Delta$-contributions that go beyond the saturation of
$c_{2,3,4}$). The values of these LECs are known to be largely driven
by the $\Delta$ and appear to be large in magnitude. 
As a consequence, one observes a rather unnatural convergence
pattern in the chiral expansion of the two-pion exchange
nucleon-nucleon potential $V_{NN}^{2\pi}$ with by far the strongest 
contribution emerging from the formally subleading triangle diagram
proportional to $c_3$ \cite{Kaiser:1997mw}. The (formally) leading contribution to
$V_{NN}^{2\pi}$ does not provide a good approximation to the potential so
that one needs to go to higher orders in the chiral expansion and/or 
include the $\Delta$-isobar as an explicit degree of
freedom. One expects similar convergence pattern for 
the chiral expansion of the $2\pi$-$1\pi$
exchange and ring 3NF topologies, see also the discussion in
Ref.~\cite{Machleidt:2010kb}. For the ring topology, this expectation
is in line with the phenomenological study of
Ref.~\cite{Pieper:2001ap}. All this suggests that one should not
truncate the chiral expansion of the 3NF at N$^3$LO but rather  
go to (at least) N$^4$LO in the standard $\Delta$-less EFT approach
and/or include the $\Delta$-isobar as an explicit degree of freedom. 
In the latter case, first contributions of the $\Delta$ to the
$2\pi$-$1\pi$ exchange and ring 3NF topologies would appear already at N$^3$LO.  
It should be understood that the strategies outlined above are, to some extent, complementary 
to each other. This is because N$^4$LO 3NF corrections in the $\Delta$-less 
theory only take into account (some) effects due to single $\Delta$-excitation but not due to  
double and triple $\Delta$-excitations which appear first at N$^5$LO and
N$^6$LO, respectively. While these effects are included at
N$^3$LO in the $\Delta$-full approach, 
N$^4$LO contributions not related to $\Delta$-excitations are
certainly not. We further emphasize that in both cases a number of
unknown LECs will appear. It remains to be 
seen which strategy will turn out to be most efficient in practical
terms. 

In our recent work \cite{Krebs:2012yv} we already made a first step in this
direction and worked out N$^4$LO corrections to the longest-range
$2\pi$-exchange topology in the delta-less approach. Apart from
relativistic corrections (which in our power counting scheme appear at
N$^3$LO but turn out to vanish at N$^4$LO), the general form of
the $2\pi$-exchange 3NF can be parametrized in terms of two scalar
functions ${\cal A} (q_2)$ and ${\cal B} (q_2)$ which depend on
the momentum transfer $q_2 \equiv | \vec q_2|$ of, say,  the second
nucleon. In spite of this simple structure, this topology turns out to
be  most
challenging to calculate. The pion-nucleon scattering amplitude enters here at the
subleading one-loop order so that the N$^4$LO correction depends not
only on the pion decay constant $F_\pi$ and the pion-nucleon
coupling $g_{\pi NN}$ but also on 13 independent (linear combinations
of the) LECs from higher-order
effective Lagrangians: $c_{1,2,3,4}$ from $\mathcal{L}_{\pi
  N}^{(2)}$, $\bar d_1 + \bar d_2$, $\bar d_3$, $\bar d_5$,
$\bar d_{14} - \bar d_{15}$ from $\mathcal{L}_{\pi N}^{(3)}$  and
$\bar e_{14,15,16,17,18}$ from $\mathcal{L}_{\pi N}^{(4)}$. The
explicit form of the heavy-baryon pion-nucleon effective
Lagrangians $\mathcal{L}_{\pi N}^{(n)}$ of chiral dimension $n$ needed
in the derivation can be found in  \cite{Krebs:2012yv} while the
complete pion-nucleon Lagrangian   $\mathcal{L}_{\pi N}^{(4)}$ is
constructed in Ref.~\cite{Fettes:2000gb}.
In order to determine these LECs  we re-analyzed 
pion-nucleon scattering at subleading one-loop order employing exactly
the same power counting scheme as in the derivation of the nuclear
forces. We used the available partial wave analyses of the
pion-nucleon scattering data to determine all relevant LECs. 
With all LECs being fixed from pion-nucleon scattering as discussed
above, we found a good (reasonable) convergence of the chiral
expansion for the functions  ${\cal A} (q_2)$ (${\cal B} (q_2)$). 
This is to be expected given that effects of the $\Delta$-isobar are,
to a large extent, accounted for already in the leading contribution 
to ${\cal A} (q_2)$ and ${\cal B} (q_2)$ at N$^2$LO through resonance
saturation of the LECs $c_{3,4}$. As pointed out above, this situation
is different for the $2\pi$-$1\pi$ exchange and ring 3NF topologies,
whose leading contributions at N$^3$LO completely miss effects of the
$\Delta$-isobar which lets one expect large N$^4$LO corrections. 

In the present work we calculate the intermediate-range
contributions to the 3NF at N$^4$LO, namely the ones corresponding to diagrams (b) and
(c) in Fig.~\ref{fig0}, and analyze in detail convergence of the chiral
expansion for long-range tail of the 3NF by comparing the  
coordinate-space potentials associated with individual 
isospin-spin-position structures. In order to carry out such a comparison
in a meaningful way, we worked out the most general structure of a
local isospin-invariant 3NF both in momentum and configuration spaces
and defined the minimal sets of linearly independent operators. 
 Our paper is organized as
follows. In section \ref{sec:TPE} we carry out Fourier transformation
of the momentum-space expressions for the $2\pi$-exchange 3NF of
Ref.~\cite{Krebs:2012yv}.  Sections \ref{sec:TPEOPE} and
\ref{sec:Ring} are devoted to the calculation of the N$^4$LO
corrections to the $2\pi$-$1\pi$-exchange and ring topologies,
respectively. For $2\pi$-$1\pi$-exchange contributions we provide
results both in momentum and coordinate spaces. For the ring topology
we give compact expressions in coordinate space while the rather
lengthy expressions in
momentum space are delegated to appendix
\ref{app1}. The most general operator structure of a local 3NF is
worked out in section \ref{sec:structure} where we also define the
basis of 22 isospin-spin-momentum operators. We use corresponding
coordinate-space version of this basis when discussing numerical results for various
potentials in section \ref{sec:numerics} in connection with 
convergence of the chiral expansion. The findings of our work are briefly
summarized in section \ref{sec:summary}.

%%%%%%%%%%%%%%%%%%%%%%%%%%%%%%%%%%%%%%%%%%%%%%%%%%%%%%%%%%%%%%%%%%%%%%%%%%%%%%%%%
\section{Two-pion-exchange 3NF in configuration space }
\def\theequation{\arabic{section}.\arabic{equation}}
\label{sec:TPE}

The $2\pi$-exchange topology (a) generates the longest-range contribution
to the 3NF. In the isospin and static limits, i.e.~the limit of
infinitely heavy nucleons, its general structure in momentum space 
has the following form (modulo terms of a shorter range corresponding 
to other topologies):
\beq
\label{2pi_general}
V_{2 \pi} (\vec q_1 , \, \vec q_3 ) = \frac{\vec \sigma_1 \cdot \vec q_1\,  \vec \sigma_3 \cdot
  \vec q_3}{[q_1^2 + M_\pi^2 ] \, [q_3^2 + M_\pi^2 ]}  \Big( \fet
  \tau_1 \cdot \fet \tau_3  \, {\cal A}(q_2) + \fet \tau_1 \times  \fet
\tau_3 \cdot \fet \tau_2  \,   \vec q_1 \times  \vec q_3   \cdot \vec
\sigma_2 \, {\cal B}(q_2) \Big)  \,,
\eeq
where $M_\pi$ stays for the pion mass, $\vec \sigma_i$  denote the Pauli
spin matrices for the nucleon $i$ and $\vec q_{i} = \vec p_i \,
' - \vec p_i$,  with $\vec p_i \, '$ and $\vec p_i$ being the final and initial momenta of the nucleon $i$. 
Here and in what follows, we use the notation: $q_i \equiv | \vec q_i
|$.  Notice that the momentum transfers are not independent and
related to each other via the condition $\vec q_1 + \vec q_2 +\vec q_3 =0$. 
The quantities ${\cal A} (q_2)$ and 
${\cal B} (q_2)$ in Eq.~(\ref{2pi_general}) are scalar
functions of the momentum transfer $q_2$ of the second nucleon whose explicit form is
determined by means of the chiral expansion, i.e.~the expansion in powers
of the soft scale $Q \sim M_\pi$.  
Unless stated otherwise, the expressions for the 3NF results are always
given for a particular choice of the nucleon labels. The complete result 
can then be found by taking into account all possible permutations of the
nucleons
\beq
\label{def_perm}
V_{\rm 3N}^{\rm full} = V_{\rm 3N} + \mbox{5 permutations}\,.
\eeq
The explicit expressions for the functions ${\cal A} (q_2)$ and 
${\cal B} (q_2)$ at first three nonvanishing orders in the chiral
expansion, i.e.~N$^2$LO [$Q^3$],  N$^3$LO [$Q^4$] and N$^4$LO
[$Q^5$] \footnote{Notice that the overall ``chiral dimension'' is a
  matter of convention. In the context of nuclear chiral EFT, one
  usually uses the convention in which the leading-order (LO) one-pion exchange
  nucleon-nucleon potential is assigned the chiral dimension $Q^0$.}
are given in Ref.~\cite{Krebs:2012yv}.  The functions ${\cal A} (q_2)$ and 
${\cal B} (q_2)$ resulting at different orders in the chiral expansion
are plotted versus the values of $q_2$ in Fig.~5 of that work. While
in the case of the $2\pi$-exchange topology it is possible to address 
the convergence of the chiral expansion in momentum space thanks to the
particularly simple parametrization in Eq.~(\ref{2pi_general}), this is
generally not possible for the more complicated cases of the
$2\pi$-$1\pi$ exchange and ring diagrams. This is because there is, in
general, no
easy way to separate the truly long-range components, which are
unambiguously predicted in terms of the chiral expansion, from
scheme-dependent short-range contributions. Such a
separation is naturally achieved by looking at the corresponding
coordinate-space potentials at sufficiently large distances. It is, 
therefore, advantageous and, in fact, also quite natural to switch to
coordinate space in order to study the convergence of the chiral
expansion for nuclear forces.    

We define the coordinate space
representation of a static 3NF by means of the Fourier-transform 
\beqa
\label{temp5}
\tilde V_{3N} (\vec r_{12}, \, \vec r_{32} \, ) &=& \int \frac{d^3 q_1}{(2
  \pi)^3} \,  \frac{d^3 q_3}{(2 \pi)^3} \, e^{i \vec
  q_1 \cdot \vec r_{12}} \; e^{i \vec   q_3 \cdot \vec r_{32}} \;V_{3N}(\vec q_{1}, \, \vec q_{3} ). 
\eeqa
For the two-pion-exchange contribution, we obtain from Eq.~(\ref{2pi_general})
\beq
\tilde V_{\rm 2\pi} (\vec r_{12}, \, \vec r_{32} \,
)=-\vec\sigma_1\cdot\vec\nabla_{12}\;\vec\sigma_3\cdot\vec\nabla_{32}
\left(\fet\tau_1\cdot\fet\tau_3\;\tilde{\cal A}(\vec r_{12}, \vec r_{32})
-\fet \tau_1\times\fet\tau_3\cdot\fet\tau_2\; \vec
\nabla_{12}\times\vec\nabla_{32}\cdot\vec\sigma_2\; \tilde{\cal B}(\vec r_{12}, \vec r_{32})\right),
\eeq
where $\vec r_{ij}=\vec r_i-\vec r_j$ denotes the distance between the
nucleons $i$ and $j$. The differential operators $\vec \nabla_{ij}$ are
defined in terms of dimensionless variables $\vec x_{ij}=\vec r_{ij}
M_\pi$; the functions
$\tilde {\cal A}$ and $\tilde {\cal B}$ are given by
\beqa
\tilde{\cal A}(\vec r_{12}, \vec r_{32})& =&\int \frac{d^3
  q_1}{(2\pi)^3}\frac{d^3 q_3}{(2\pi)^3} \, e^{i \vec
  q_1 \cdot \vec r_{12}} \; e^{i \vec   q_3 \cdot \vec r_{32}}
\;\frac{1}{q_1^2+M_\pi^2}\frac{1}{q_3^2+M_\pi^2}\;{\cal A}(q_2),\nn
\tilde{\cal B}(\vec r_{12}, \vec r_{32})& =&\int \frac{d^3
  q_1}{(2\pi)^3}\frac{d^3 q_3}{(2\pi)^3} \, e^{i \vec
  q_1 \cdot \vec r_{12}} \; e^{i \vec   q_3 \cdot \vec r_{32}}
\;\frac{1}{q_1^2+M_\pi^2}\frac{1}{q_3^2+M_\pi^2}\;{\cal B}(q_2).
\eeqa
The N$^2$LO expressions for ${\cal A}$ and ${\cal B}$ corresponding to
${\cal A}^{(3)} (q_2)$ and ${\cal B}^{(3)}(q_2)$ from
Eq.~(3.5) of Ref.~\cite{Krebs:2012yv}
are given by
\beqa
\tilde{\cal A}^{(3)}(\vec r_{12}, \vec r_{32})&=&\frac{g_A^2 M_\pi^6}{128\pi^2 F_\pi^4} \Big(2 c_3 - 4 c_1
  -  c_3 
(\vec\nabla_{12}+\vec\nabla_{32})^2
\Big) U_1(x_{12})U_1(x_{32}),\nn
\tilde{\cal B}^{(3)}(\vec r_{12}, \vec r_{32})&=&\frac{g_A^2 M_\pi^6 c_4}{128\pi^2 F_\pi^4} U_1(x_{12})U_1(x_{32}),
\eeqa
where $g_A$ denotes the nucleon axial vector coupling and the Yukawa function $U_1$ is defined as 
\beq
U_1(x)=\frac{4\pi}{M_\pi}\int \frac{d^3 q}{(2\pi)^3}\frac{e^{i\vec q\cdot
  \vec x/M_\pi} }{q^2+M_\pi^2}=\frac{e^{-x}}{x}.
\eeq
Here and in what follows, the superscripts of ${\cal A}$,  ${\cal B}$, $\tilde {\cal A}$,  $\tilde {\cal B}$
as well as other functions parametrizing 3NF matrix elements
refer to the chiral dimension, i.e. to the associated power of the soft
scale $Q$.  

The first corrections to $\tilde {\cal A}$ and $\tilde {\cal B}$
emerge from Fourier-transforming the expressions ${\cal
  A}^{(4)} (q_2)$ and ${\cal B}^{(4)}(q_2)$ given in 
Eq.~(3.4) of Ref.~\cite{Krebs:2012yv}. We obtain 
\beqa
\tilde{\cal A}^{(4)}(\vec r_{12}, \vec r_{32})&=&\frac{g_A^4M_\pi^7}{4096 \pi^3  F_\pi^6} \Big\{\big[\left(4 g_A^2+1\right)-2 \left(g_A^2+1\right)  
(\vec\nabla_{12}+\vec\nabla_{32})^2\big] U_1(x_{12})U_1(x_{32})\nn
&+&
\frac{1}{4\pi}\left(2 -5 (\vec\nabla_{12}+\vec\nabla_{32})^2+2 (\vec\nabla_{12}+\vec\nabla_{32})^4
\right)\int d^3x \;U_1(|\vec x_{12}+\vec x|) \;W_1(x)\; U_1(|\vec
x_{32}+\vec x|) \Big\},\nn
\tilde{\cal B}^{(4)}(\vec r_{12}, \vec r_{32})&=& -\frac{g_A^4M_\pi^7}{4096 \pi^3  F_\pi^6} \Big\{(2 g_A^2 
+1)  U_1(x_{12})U_1(x_{32}) \nn
&+& \frac{1}{4\pi}\left(4 - (\vec\nabla_{12}+\vec\nabla_{32})^2\right) \int d^3x \;U_1(|\vec x_{12}+\vec x|) \;W_1(x)\; U_1(|\vec
x_{32}+\vec x|) \Big\}.
\label{AB4}
\eeqa
The profile function $W_1$ is given in terms of the
Fourier-transform of the loop function $A(q)$ appearing in  $ {\cal A}^{(4)}
(q_2)$ and ${\cal B}^{(4)}(q_2)$:  
\beq
W_1(x) = \frac{4 \pi }{M_\pi^2} \int \frac{d^3 q}{(2 \pi )^3} \, e^{i
    \vec q \cdot \vec x/M_\pi}\, A (q)  = \frac{e^{-2 x}}{2 x^2}\,, 
\quad \quad 
\mbox{with}
\quad 
A(q) = \frac{1}{2 q} \arctan \frac{q}{2 M_\pi}\,.
\eeq

To give the coordinate space expressions for N$^4$LO contributions we
need to Fourier-transform another loop function, namely
\beq
L(q)  =  \frac{\sqrt{q^2 + 4 M_\pi^2}}{q} \log \frac{\sqrt{q^2 + 4
    M_\pi^2} + q}{2 M_\pi} \,,
\eeq
which enters the expressions for $ {\cal A}^{(5)}
(q_2)$ and ${\cal B}^{(5)}(q_2)$ in Eq.~(3.14) of Ref.~\cite{Krebs:2012yv}.
This can be most easily  achieved by using the spectral
representation of $L$  given by
\beq
\label{spectral_f_L}
L(q)=1+\int_{2M_\pi}^\infty d\mu\frac{q^2}{\mu^2+q^2}\frac{1}{\mu^2}\sqrt{\mu^2-4M_\pi^2}.
\eeq
The Fourier-transform of the square-integrable part of $L$ is given by
\beq
V_1(x)=\frac{4\pi}{M_\pi}\int \frac{d^3 q}{(2 \pi )^3} \, e^{i
    \vec q \cdot \vec x/M_\pi}\, \int_{2M_\pi}^\infty
  d\mu\frac{1}{\mu^2+q^2}\frac{1}{\mu^2}\sqrt{\mu^2-4M_\pi^2}=\frac{1}{x}\int_{2}^\infty
  d\mu \frac{e^{-x\,\mu}}{\mu^2}\sqrt{\mu^2-4}.
\label{v1_definition}
\eeq
With these preparations, we obtain the following result for the
Fourier-transform of $ {\cal A}^{(5)}
(q_2)$ and ${\cal B}^{(5)}(q_2)$:
\beqa
\tilde{\cal A}^{(5)}(\vec r_{12}, \vec r_{32})&=&\frac{g_A M_\pi^8}{73728 \pi^4 F_\pi^6}\Big[ -(\vec\nabla_{12}+\vec\nabla_{32})^2 \big(F_\pi^2 \left(2304 \pi^2 g_A (4 
\bar{e}_{14}+2 \bar{e}_{19}-\bar{e}_{22}-\bar{e}_{36})-2304 \pi^2 
\bar{d}_{18} c_3\right)\nn\
&+&g_A (144 c_1-53 c_2-90 c_3)\big)+ 
\left. F_\pi^2 \left(4608 \pi^2 \bar{d}_{18} (2 c_1-c_3)+4608 \pi^2 
g_A (2 \bar{e}_{14}+2 \bar{e}_{19}-\bar{e}_{36}-4 \bar{e}_{38})
\right)\right.\nn\
&+&\left. g_A \left(72 \left(64 \pi^2 \bar{l}_{3}+1\right) c_1-24 c_2-36 
c_3\right.\right)\nn
&+&(\vec\nabla_{12}+\vec\nabla_{32})^4 \left(2304 \pi^2 \bar{e}_{14} F_\pi^2 g_A-2 
g_A (5 c_2+18 c_3)\right)\Big]U_1(x_{12})U_1(x_{32})\nn
&-&\frac{g_A^2 M_\pi^8}{12288 \pi^4 F_\pi^6}
 \left(1-2 (\vec\nabla_{12}+\vec\nabla_{32})^2\right) \Big(4 (6 c_1-c_2-3 
c_3)-(\vec\nabla_{12}+\vec\nabla_{32})^2 (-c_2-6 c_3)\Big)
U_1(x_{12})U_1(x_{32})\nn
&+&\frac{g_A^2 M_\pi^8}{49152 \pi^5 F_\pi^6}
\left(1-2 (\vec\nabla_{12}+\vec\nabla_{32})^2\right) \Big(4 (6 c_1-c_2-3 
c_3)\nn
&-&(\vec\nabla_{12}+\vec\nabla_{32})^2 (-c_2-6
c_3)\Big)(\vec\nabla_{12}+\vec\nabla_{32})^2
 \int d^3x \;U_1(|\vec x_{12}+\vec x|) \;V_1(x)\; U_1(|\vec
x_{32}+\vec x|),\nn
\tilde{\cal B}^{(5)}(\vec r_{12}, \vec r_{32})&=&-\frac{g_A
  M_\pi^8}{36864 \pi^4 F_\pi^6} \Big[ \left. F_\pi^2 \left(1152 \pi^2
\bar{d}_{18} c_4-1152 \pi^2 g_A (2 \bar{e}_{17}+2 
\bar{e}_{21}-\bar{e}_{37})\right)+108 g_A^3 c_4+24 g_A c_4
\right. \nn\
&-&(\vec\nabla_{12}+\vec\nabla_{32})^2 \left(5 g_A c_4-1152 \pi^2 \bar{e}_{17} F_\pi^2 g_A
\right)\Big] U_1(x_{12})U_1(x_{32})\nn
&+& \frac{g_A^2 c_4 M_\pi^8}{6144 \pi^4 
F_\pi^6} \left(4 -(\vec\nabla_{12}+\vec\nabla_{32})^2\right)
U_1(x_{12})U_1(x_{32})\nn
&-& \frac{g_A^2 c_4 M_\pi^8}{24576 \pi^5 
F_\pi^6} \left(4 -(\vec\nabla_{12}+\vec\nabla_{32})^2\right) (\vec\nabla_{12}+\vec\nabla_{32})^2 \int d^3x \;U_1(|\vec x_{12}+\vec x|) \;V_1(x)\; U_1(|\vec
x_{32}+\vec x|).
\label{AB5}
\eeqa
It remains to emphasize that while the momentum space representation
of the functions ${\cal A}$ and ${\cal B}$ depends on just one
variable $q_2$, the 
%corresponding 
coordinate-space functions $\tilde
{\cal A}$ and  $\tilde {\cal B}$ depend on three scalar arguments. We
will discuss the convergence of the chiral expansion for the
coordinate space potentials in section \ref{sec:numerics}.

%%%%%%%%%%%%%%%%%%%%%%%%%%%%%%%%%%%%%%%%%%%%%%%%%%%%%%%%%%%%%%%%%%%%%%%%%%%%%%%%%
\section{Two-pion-one-pion exchange 3NF at N$^4$LO}
\def\theequation{\arabic{section}.\arabic{equation}}
\label{sec:TPEOPE}

We now turn to the $2\pi$-$1\pi$ exchange topology. In contrast to the
longest-range $2\pi$ exchange  topology discussed in the previous
section, its chiral expansion starts at N$^3$LO. At
this order, one has to evaluate all one-loop diagrams made out of the 
lowest-order pion-nucleon vertices. This was achieved in
Ref.~\cite{Bernard:2007sp}, see Eqs.~(2.16)-(2.23) of that work.  
As pointed out in Ref.~\cite{Krebs:2012yv}, the decomposition of
momentum-space 3NF expressions according to the type of the topology 
is not unique as e.g.~some parts of the $2\pi$ exchange
contributions can be reshuffled into $2\pi$-$1\pi$ exchange  and
shorter-range terms by canceling pion propagators with the
corresponding expressions in the numerator. In
Ref.~\cite{Krebs:2012yv} we introduced  a ``minimal''
parametrization of the $2\pi$ exchange 3NF which corresponds to 
Eq.~(\ref{2pi_general}) and which is adopted here and in what follows. 
 
The structure of two-pion-one-pion exchange contributions up to
N$^4$LO 
in the chiral expansion has the form 
\beqa
\label{two_pion_one_pion_general}
V_{\rm 2\pi \mbox{-}1\pi} &=& 
 \frac{\vec \sigma_3 \cdot \vec q_3}{q_3^2 + M_\pi^2} \Big[ \fet \tau_1
  \cdot \fet \tau_3  \; \left[ \vec \sigma_2 \cdot \vec q_1 \; \vec q_1 \cdot
    \vec q_3 \; F_1 (q_1)  + \vec \sigma_2 \cdot \vec q_1 \; F_2 (q_1) + 
  \vec \sigma_2 \cdot \vec q_3 \;  F_3 (q_1)  \right] + \fet \tau_2
  \cdot \fet \tau_3 \; [  \vec \sigma_1 \cdot \vec q_1 \;  \vec q_1 \cdot
    \vec q_3 \; F_4 (q_1) \nn
&& {} +  \vec \sigma_1 \cdot \vec q_3 \; F_5  (q_1) +
\vec \sigma_2 \cdot \vec q_1 \; \vec q_1 \cdot \vec q_3 \;  F_6 (q_1) +  
\vec \sigma_2 \cdot \vec q_1 \; F_7(q_1)
+  \vec \sigma_2 \cdot \vec q_3 \;  \vec q_1 \cdot \vec q_3 \; F_{8} (q_1)
+\vec \sigma_2 \cdot \vec q_3 \; F_{9}(q_1)]  \nn\
&& {} + \fet \tau_1
  \times \fet \tau_2 \cdot \fet \tau_3 \left[ \vec \sigma_1 \times \vec \sigma_2
  \cdot \vec q_1 \;  (\vec q_1 \cdot
    \vec q_3 \; F_{10} (q_1)+ F_{11}(q_1))+\vec q_1 \times \vec q_3
    \cdot \vec \sigma_1 \; \vec q_1 \cdot \vec \sigma_2 \; F_{12}(q_1)\right]
\Big]\,, 
\eeqa
where $F_{1\ldots 12} (q_1)$ are scalar functions to be calculated.
Notice that we use a slightly different notation compared to
our early paper \cite{Bernard:2007sp}, which is now also
valid at N$^4$LO. 
First non-vanishing contributions to the structure functions $F_i$
are generated at N$^3$LO by diagrams shown in Fig.~3 of that work.
Adjusting  the expressions obtained in Ref.~\cite{Bernard:2007sp} to
our new notation and taking into account terms induced by reshuffling
the $2\pi$ exchange contributions as explained above we obtain the
following results for the functions $F_i (q_1)$: 
\beqa
\label{TPEOPEQ4MOM}
F_1^{(4)}(q_1)&=&\frac{g_A^4 }{256 \pi  F_\pi^6 q_1^2} \Big[A(q_1) \left(\left(8 g_A^2-4\right) M_\pi^2+\left(g_A^2+1
\right) q_1^2\right)-\frac{M_\pi }{4 M_\pi^2+q_1^2}
\left(\left(8 g_A^2-4\right) M_\pi^2+\left(3 g_A^2-1\right)
  q_1^2\right)\Big], \nn
F_2^{(4)}(q_1)&=&\frac{g_A^4 }{128 \pi  F_\pi^6} A(q_1) \left(2
  M_\pi^2+q_1^2\right),\nn
F_3^{(4)}(q_1)&=&-\frac{g_A^4 }{256 \pi  F_\pi^6} A(q_1) \left(\left(8 g_A^2-4\right) M_\pi^2+\left(3 \
g_A^2-1\right) q_1^2\right), \nn
F_4^{(4)}(q_1)&=&-\frac{F_5^{(4)}(q_1)}{q_1^2}=-\frac{g_A^6 }{128 \pi
  F_\pi^6}A(q_1),\nn
F_6^{(4)}(q_1)&=&F_8^{(4)}(q_1) = F_9^{(4)}(q_1) = F_{10}^{(4)}(q_1) =
F_{12}^{(4)}(q_1) = 0, \nn 
F_7^{(4)}(q_1)&=&\frac{g_A^4 }{128 \pi  F_\pi^6} A(q_1) \left(2
  M_\pi^2+q_1^2\right), \nn
F_{11}^{(4)}(q_1)&=&-\frac{g_A^4 }{512 \pi  F_\pi^6} A(q_1) \left(4 M_\pi^2+q_1^2\right).
\eeqa
Notice that we give here only non-polynomial parts as  the polynomial
ones simply lead to shifts of the low-energy constants $D$ and $E$
 from N$^2$LO three-body force. 

\begin{figure}[tb]
\vskip 1 true cm
%  \begin{center} 
%    \epsfxsize=4.3cm
%    \epsffile{fig2.eps}
\includegraphics[width=11.0cm,keepaspectratio,angle=0,clip]{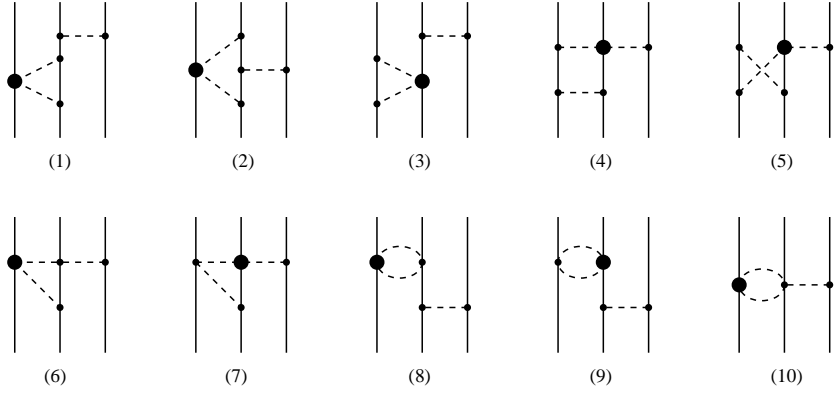}
    \caption{
         Two-pion-one-pion exchange 3N diagrams at N$^4$LO. Solid dots
         and filled circles denote the leading and subleading
         pion-nucleon vertices, respectively. Graphs resulting from
         the interchange of the nucleon lines and/or applying time
         reversal operation are not shown. For remaining 
 notation see Fig.~\ref{fig0}.
\label{fig3} 
 }
%  \end{center}
\end{figure}

First corrections to these results emerge at N$^4$LO from diagrams
shown in Fig.~\ref{fig3},
which involve a single insertion of $c_i$-vertices from the subleading
pion-nucleon Lagrangian. Evaluating the irreducible contributions of
these diagrams following the lines of Refs.~\cite{Bernard:2007sp,Bernard:2011zr}
and keeping only terms non-polynomial in $q_1$ we obtain the following expressions: 
\beqa
\label{TPEOPEQ5MOM}
F_1^{(5)}(q_1)&=&-\frac{g_A^2 c_4 }{96 \pi^2 F_\pi^6 q_1^2 \left(4
    M_\pi^2+q_1^2
\right)} L(q_1) \left(8 \left(4 g_A^2-1\right) M_\pi^4+2 
\left(5 g_A^2+1\right) M_\pi^2 q_1^2-\left(g_A^2-1\right) q_1^4
\right)\nn\
&-&\frac{\left(1-4 g_A^2\right) g_A^2 c_4 M_\pi^2}{48 \pi^2 
F_\pi^6 q_1^2},\nn
F_2^{(5)}(q_1)&=& F_8^{(5)}(q_1) = F_{11}^{(5)}(q_1)  =    0,\nn
F_3^{(5)}(q_1)&=&-\frac{g_A^2 c_4 }{48 \pi^2 F_\pi^6 \left(4 M_\pi^2+q_1^2\right)} L(q_1) \left(4 \left(4 g_A^2-1\right) 
M_\pi^4+\left(17 g_A^2-5\right) M_\pi^2 q_1^2+\left(4 g_A^2-1\right) 
q_1^4\right),\nn\
F_4^{(5)}(q_1)&=&-\frac{F_5^{(5)}(q_1)}{q_1^2}=-\frac{g_A^4 c_4 }{16
  \pi^2 F_\pi^6}L(q_1), \nn
F_6^{(5)}(q_1)&=&\frac{g_A^4 M_\pi^2 (6
  c_1+c_2-3 c_3)}{96 \pi^2 F_\pi^6 q_1^2}\nn\
&+&\frac{g_A^4 L(q_1)}{192 \pi^2 F_\pi^6 q_1^2 \left(4 M_\pi^2+q_1^2\right)} \left(-48 c_1 M_\pi^4+c_2 \left(-8 M_\pi^4+2 
M_\pi^2 q_1^2+q_1^4\right)+12 c_3 M_\pi^2 \left(2 M_\pi^2+q_1^2
\right)\right),\nn\
F_7^{(5)}(q_1)&=&-\frac{g_A^2 }{192 \pi^2 F_\pi^6} L(q_1) \left(24 c_1 M_\pi^2-c_2 \left(4
      M_\pi^2+q_1^2
\right)-6 c_3 \left(2 M_\pi^2+q_1^2\right)\right), \nn\
F_9^{(5)}(q_1)&=&-\frac{g_A^4 L(q_1) }{128 \pi^2 
F_\pi^6 \left(4 M_\pi^2+q_1^2\right)} \Big[-32 c_1 M_\pi^2 \left(3
  M_\pi^2+q_1^2
\right)+c_2 \left(16 M_\pi^4+16 M_\pi^2 q_1^2+3 q_1^4\right)\nn\
&+&c_3 
\left(80 M_\pi^4+68 M_\pi^2 q_1^2+13 q_1^4\right)\Big],\nn
F_{10}^{(5)}(q_1)&=&F_{12}^{(5)}(q_1)\; = \; \frac{g_A^4 c_4 L(q_1)}{64 \pi^2 F_\pi^6}.
\eeqa

It is straightforward to transform these results into coordinate
space. The general structure corresponding to the momentum-space
expression in Eq.~(\ref{two_pion_one_pion_general}) has the form:
\beqa
\tilde V_{\rm 2\pi \mbox{-}1\pi}(\vec r_{12}, \vec r_{32}) &=& 
\vec \sigma_3 \cdot \vec \nabla_{32} \left[ \fet \tau_1
  \cdot \fet \tau_3  \; \left( \vec \sigma_2 \cdot \vec \nabla_{12} \; \vec \nabla_{12} \cdot
    \vec \nabla_{32} \; \tilde F_1 (x_{12})  - \vec \sigma_2 \cdot \vec
    \nabla_{12} \; \tilde F_2 (x_{12}) - 
  \vec \sigma_2 \cdot \vec \nabla_{32} \;  \tilde F_3 (x_{12})
\right)\right. \nn 
&+& \left.  \fet \tau_2
  \cdot \fet \tau_3 \; \left (  \vec \sigma_1 \cdot \vec \nabla_{12} \;  \vec \nabla_{12} \cdot
    \vec \nabla_{32} \; \tilde F_4 (x_{12})  -  \vec \sigma_1 \cdot \vec \nabla_{32} \; \tilde F_5  (x_{12}) +
\vec \sigma_2 \cdot \vec \nabla_{12} \; \vec \nabla_{12} \cdot \vec \nabla_{32} \;
\tilde F_6 (x_{12})\right. \right. \nn 
&-&  \left. \left.
\vec \sigma_2 \cdot \vec \nabla_{12} \; \tilde F_7(x_{12})
+  \vec \sigma_2 \cdot \vec \nabla_{32} \;  \vec \nabla_{12} \cdot \vec
\nabla_{32} \; \tilde F_{8} (x_{12})
-\vec \sigma_2 \cdot \vec \nabla_{32} \; \tilde F_{9}(x_{12})\right)
\right. \nn\
&+& \left. \fet \tau_1
  \times \fet \tau_2 \cdot \fet \tau_3 \left( \vec \sigma_1 \times \vec \sigma_2
  \cdot \vec \nabla_{12} \;  (\vec \nabla_{12} \cdot
    \vec \nabla_{32} \; \tilde F_{10} (x_{12})- \tilde
    F_{11}(x_{12}))\right. \right. \nn
&+&\left. \left. \vec \nabla_{12} \times \vec \nabla_{32} \cdot \vec \sigma_1 \;
\vec \nabla_{12} \cdot \vec \sigma_2 \tilde F_{12}(x_{12})\right)
\right] U_1(x_{32})\,.
\eeqa
In order to calculate the functions $F_i$ it is convenient to employ
the spectral representations of the function $L(q)$, see Eq.~(\ref{spectral_f_L}), and $A(q)$ given
by
\beq
\label{spectral_f_A}
A(q)=\frac{1}{2}\int_{2M_\pi}^\infty d\mu \frac{1}{\mu^2+q^2}.
\eeq
Following Ref.~\cite{Bernard:2007sp}, we define the profile
function $W_3(x)$ via  
\beq
W_3(x) =\frac{4 \pi }{M_\pi^2} \int \frac{d^3 q}{(2 \pi )^3} \; e^{i
    \vec q \cdot \vec x/M_\pi}\, \left[ \frac{M_\pi}{q^2} - \frac{4
      M_\pi^2}{q^2} A (q)  \right]  = 2 Ei (-2 x) + \frac{e^{-2 x}}{x}\,,
\eeq
where 
\beq
Ei (x) \equiv - \int_{-x}^\infty \frac{e^{-t} \, dt}{t}\,.
\eeq
Fourier transform of terms in Eq.~(\ref{TPEOPEQ5MOM}) involving
the function $L(q)$ can be expressed using the profile functions
$V_1(x)$ from Eq.~(\ref{v1_definition}) and  $V_2
(x)$ which is defined according to  
\beq
V_2(x)=\frac{1}{x}\int_{2}^\infty d\mu\frac{e^{-x
    \,\mu}}{\mu^2\sqrt{\mu^2-4}}\,.
\eeq
With these definitions,  the N$^3$LO contributions to the $\tilde
F_i$-functions are given by
\beqa
\tilde F_1^{(4)}(x_{12})&=&-\frac{g_A^4M_\pi^7 }{4096 \pi^3 F_\pi^6}
\left(2 g_A^2 U_1(2 
x_{12})-\left(g_A^2+1\right) W_1(x_{12})+\left(2 g_A^2-1\right) 
W_3(x_{12})\right), \nn
\tilde F_2^{(4)}(x_{12})&=&-\frac{
g_A^4 M_\pi^7 
}{2048 \pi^3 F_\pi^6}
\left(\nabla_{12}^2-2\right)  
W_1(x_{12}), \nn
\tilde F_3^{(4)}(x_{12})&=&\frac{g_A^4 M_\pi^7 
}{4096 \pi^3 F_\pi^6}
\left(-\nabla_{12}^2+\left(3
      \nabla_{12}^2-8\right) g_A^2+4\right) 
  W_1(x_{12}),\nn
\tilde F_4^{(4)}(x_{12})&=&-\frac{g_A^6 M_\pi^7 
}{2048
  \pi^3 F_\pi^6}
 W_1(x_{12}),\nn
\tilde F_5^{(4)}(x_{12})&=&-\frac{ g_A^6 M_\pi^7 
}{2048 \
\pi^3 F_\pi^6}
\nabla_{12}^2 W_1(x_{12}), \nn 
\tilde F_6^{(4)}(x_{12})&=&\tilde F_8^{(4)}(x_{12})=\tilde
F_9^{(4)}(x_{12})=\tilde F_{10}^{(4)}(x_{12})=\tilde F_{12}^{(4)}(x_{12})=0, \nn
\tilde F_7^{(4)}(x_{12})&=&-\frac{ g_A^4 M_\pi^7
}{2048 \pi^3 F_\pi^6}
\left(\nabla_{12}^2-2\right)  \
W_1(x_{12}),\nn
\tilde F_{11}^{(4)}(x_{12})&=&\frac{g_A^4 M_\pi^7 
}{8192 \pi^3 F_\pi^6}
\left(\nabla_{12}^2-4\right)   \
W_1(x_{12}),
\eeqa
while for the N$^4$LO contributions we obtain the following results: 
\beqa
\tilde F_1^{(5)}(x_{12})&=&\frac{g_A^2 c_4 M_\pi^8 }{3072 
\pi^4 F_\pi^6}\left[\left(\nabla_{12}^2-4\right) \left(-
\nabla_{12}^2+\left(\nabla_{12}^2+10\right) g_A^2+2\right) U_1(2 
x_{12})\right.  \nn
&-&\left.  2 \left(-\nabla_{12}^4+2 \nabla_{12}^2+\left(\nabla_{12}^4+10 
\nabla_{12}^2-32\right) g_A^2+8\right) V_2(x_{12})\right],\nn
\tilde F_2^{(5)}(x_{12})&=& \tilde  F_8^{(5)}(x_{12})=   \tilde
F_{11}^{(5)}(x_{12})=                          0,\nn
\tilde F_3^{(5)}(x_{12})&=&\frac{g_A^2 c_4 M_\pi^8 }{1536 \pi^4 F_\pi^6}\left(-\nabla_{12}^4+5 \nabla_{12}^2+\left(4 
\nabla_{12}^4-17 \nabla_{12}^2+16\right) g_A^2-4\right) \nn
&\times&\left(\left(
\nabla_{12}^2-4\right) U_1(2 x_{12})-2 \nabla_{12}^2 V_2(x_{12})
\right),\nn
\tilde F_4^{(5)}(x_{12})&=&\frac{ g_A^4 c_4 M_\pi^8  
}{256 
\pi^4 F_\pi^6}
\nabla_{12}^2 V_1(x_{12}),\nn
\tilde F_5^{(5)}(x_{12})&=&\frac{ g_A^4 c_4 M_\pi^8  
}{256 
\pi^4 F_\pi^6}
\nabla_{12}^4 V_1(x_{12}),\nn
\tilde F_6^{(5)}(x_{12})&=&\frac{g_A^4 M_\pi^8 }{6144 
\pi^4 F_\pi^6}\left[2 \left(48 c_1+\left(-
\nabla_{12}^4+2 \nabla_{12}^2+8\right) c_2+12 \left(\nabla_{12}^2-2
\right) c_3\right) V_2(x_{12})\right. \nn
&+&\left. \left(\nabla_{12}^2-4\right) 
\left(\left(\nabla_{12}^2-2\right) c_2-12 c_3\right) U_1(2 x_{12})\right],\nn
\tilde F_7^{(5)}(x_{12})&=&\frac{ g_A^2 M_\pi^8  
}{3072 \pi^4 F_\pi^6}
\nabla_{12}^2\left(24 
c_1+\left(\nabla_{12}^2-4\right) c_2+6 \left(\nabla_{12}^2-2\right) 
c_3\right) V_1(x_{12}),\nn
\tilde F_9^{(5)}(x_{12})&=&\frac{g_A^4 M_\pi^8 }{4096 \pi^4 F_\pi^6}\left(32 \left(\nabla_{12}^2-3\right) c_1+\left(3 
\nabla_{12}^4-16 \nabla_{12}^2+16\right) c_2+\left(13 
\nabla_{12}^4-68 \nabla_{12}^2+80\right) c_3\right)\nn
&\times& \left(\left(
\nabla_{12}^2-4\right) U_1(2 x_{12})-2 \nabla_{12}^2 V_2(x_{12})
\right),\nn 
\tilde F_{10}^{(5)}(x_{12})&=&\tilde F_{12}^{(5)}(x_{12})\; =\; -\frac{ g_A^4 c_4 M_\pi^8
 }{1024 \pi^4 F_\pi^6}
\nabla_{12}^2 V_1(x_{12}).
\eeqa

%%%%%%%%%%%%%%%%%%%%%%%%%%%%%%%%%%%%%%%%%%%%%%%%%%%%%%%%%%%%%%%%%%%%%%%%%%%%%%%%%
\section{Ring diagrams at N$^4$LO}
\def\theequation{\arabic{section}.\arabic{equation}}
\label{sec:Ring}

Finally, we consider the ring topology. The leading contributions
emerge at N$^3$LO from diagrams shown in Fig.~4 of
Ref.~\cite{Bernard:2007sp}. As explained in that paper, only diagrams proportional to
$g_A^6$ and $g_A^4$ generate nonvanishing 3NFs:   
\beqa
V_{\rm ring}^{(4)}&=&V_{\rm ring}^{(4), g_A^6}+V_{\rm
  ring}^{(4), g_A^4}. 
\eeqa
Evaluating the corresponding loop integrals in momentum space we
obtained  
complicated expressions involving three-point function which are given
explicitly in appendix of Ref.~\cite{Bernard:2007sp}.\footnote{We
  emphasize that several symmetry factors are missing in Eq.~(A1) of
  that work. The corrected equation has the form:  
\beqa
\label{ringR}
V_{\rm ring}&=&
\vec{\sigma}_1\cdot\vec{\sigma}_2 \; {\fet \tau}_2\cdot{\fet\tau}_3  \; R_1+
\vec{\sigma}_1\cdot\vec{q}_1\vec{\sigma}_2\cdot\vec{q}_1 \; {\fet
  \tau}_2\cdot{\fet\tau}_3  \; R_2+
\vec{\sigma}_1\cdot\vec{q}_1\vec{\sigma}_2\cdot\vec{q}_3 \; {\fet
  \tau}_2\cdot{\fet\tau}_3  \; R_3+
\vec{\sigma}_1\cdot\vec{q}_3\vec{\sigma}_2\cdot\vec{q}_1 \; {\fet
  \tau}_2\cdot{\fet\tau}_3  \; R_4\nonumber\\
&+&\vec{\sigma}_1\cdot\vec{q}_3\vec{\sigma}_2\cdot\vec{q}_3 \; {\fet
  \tau}_2\cdot{\fet\tau}_3  \; R_5+ \frac{1}{2} {\fet\tau}_1\cdot{\fet\tau}_3 \;  R_6
+\vec{\sigma}_1\cdot\vec{q}_1\vec{\sigma}_3\cdot\vec{q}_1  \; R_7
+ \frac{1}{2}\vec{\sigma}_1\cdot\vec{q}_1\vec{\sigma}_3\cdot\vec{q}_3  \; R_8
+ \frac{1}{2}\vec{\sigma}_1\cdot\vec{q}_3\vec{\sigma}_3\cdot\vec{q}_1  \; R_9\nn
&+& \frac{1}{2}\vec{\sigma}_1\cdot\vec{\sigma}_3  \; R_{10}
+ \frac{1}{2}\vec{q}_1\cdot \vec{q}_3\times\vec{\sigma}_2 \; 
{\fet\tau}_1\cdot{\fet\tau}_2\times{\fet\tau}_3 \;  R_{11}.
\eeqa}
The results
in coordinate space are much more compact and have
the form:  
\beqa
\label{ring1}
V_{\rm ring}^{(4), g_A^6} (\vec r_{12}, \, \vec r_{32}\, )  &=& \left( \frac{g_A}{2 F_\pi} \right)^6 \int \, 
\frac{d^3 l_1}{(2 \pi )^3} \, \frac{d^3 l_2}{(2 \pi )^3} \, \frac{d^3 l_3}{(2 \pi )^3} \; 
e^{i \vec l_1 \cdot \vec r_{23}} \; e^{i \vec l_2 \cdot \vec r_{31}} \; 
e^{i \vec l_3 \cdot \vec r_{12}} \; \frac{v}{[l_1^2 + M_\pi^2] \, [l_2^2 +
  M_\pi^2]^2 \, 
[l_3^2 + M_\pi^2]}
\nn
&=&  - \frac{g_A^6 \, M_\pi^7}{4096\,  \pi^3 \, F_\pi^6} \Big[
-4 \fet \tau_1 \cdot \fet \tau_2 \; \vec \nabla_{23} \times \vec \nabla_{12}
\cdot \vec \sigma_2 \;  \vec \nabla_{23} \times \vec \nabla_{31}
\cdot \vec \sigma_3 \; \vec \nabla_{31} \cdot \vec \nabla_{12} \nn 
&& {} - \; 2 \fet \tau_1 \cdot \fet \tau_3 \;   \vec \nabla_{23} \cdot \vec \nabla_{31}
\; \vec \nabla_{23} \cdot \vec \nabla_{12} \;  \vec \nabla_{31} \cdot \vec
\nabla_{12}  
\;  + \; \fet \tau_1 \times \fet \tau_2 \cdot \fet
\tau_3 \; \vec \nabla_{23} \times \vec \nabla_{12}
\cdot \vec \sigma_2 \; \vec \nabla_{23} \cdot \vec \nabla_{31} \; 
\vec \nabla_{31} \cdot \vec \nabla_{12}  \nn 
&& {}  + \; 3 \vec \nabla_{31} \times \vec \nabla_{12}
\cdot \vec \sigma_1 \;  \vec \nabla_{23} \times \vec \nabla_{31}
\cdot \vec \sigma_3 \; \vec \nabla_{23} \cdot \vec \nabla_{12} \Big] \;
U_1 (x_{23}) \; U_2 (x_{31}) \; U_1 (x_{12}) \,,\nn
V_{\rm ring}^{(4), g_A^4} (\vec r_{12}, \, \vec r_{32}\, )  &=& \frac{g_A^4 \, M_\pi^7}{2048\,  \pi^3 \, F_\pi^6} \Big[
2  \fet \tau_1 \cdot \fet \tau_2 \, 
\big( \vec \nabla_{23} \cdot \vec \nabla_{31} \; \vec \nabla_{31} \cdot \vec \nabla_{12}
-  \vec \nabla_{31} \times \vec \nabla_{12}  \cdot \vec \sigma_1 \; 
 \vec \nabla_{23} \times \vec \nabla_{31}  \cdot \vec \sigma_3 \big) \nn
&& {} +  \fet \tau_1 \times \fet \tau_2 \cdot \fet \tau_3 
\; \vec \nabla_{31} \times \vec \nabla_{12} \cdot \vec \sigma_1 \; \vec
\nabla_{23} \cdot \vec \nabla_{31} \Big] 
U_1 (x_{23}) \; U_1 (x_{31}) \; U_1 (x_{12}), 
\eeqa
where the derivatives should be evaluated as if the variables $\vec
x_{12}$, $\vec x_{23}$
and $\vec x_{31}$ were independent\footnote{Clearly, the relative distances $\vec
r_{12}$, $\vec r_{23}$
and $\vec r_{31}$ are related via $\vec
r_{12} + \vec r_{23} +\vec r_{31} = 0$.} 
and the numerator $v$ in the first line is given by  
\beqa
v &=& - 8 \fet \tau_1 \cdot \fet \tau_2 \;  \vec l_1 \times \vec l_3  \cdot
\vec \sigma_2\;  \vec l_1 \times \vec l_2  \cdot \vec \sigma_3 \; \vec l_2
\cdot \vec l_3 \; - \;  4 \fet \tau_1 \cdot \fet \tau_3 \; 
\vec l_1 \cdot \vec l_2 \; \vec l_1 \cdot \vec l_3 \; \vec l_2 \cdot \vec l_3
\;  + \; 2  \fet \tau_1 \times \fet \tau_2  \cdot \fet \tau_3 
\;  \vec l_1 \times \vec l_3  \cdot \vec \sigma_2 \;
\vec l_1 \cdot \vec l_2 \; \vec l_2 \cdot \vec l_3 \nn
&& {}  +  
6  \vec l_2 \times \vec l_3  \cdot
\vec \sigma_1\;  \vec l_1 \times \vec l_2  \cdot \vec \sigma_3 \; \vec l_1
\cdot \vec l_3 \,.
\eeqa

\begin{figure}[tb]
\vskip 1 true cm
%  \begin{center} 
%    \epsfxsize=4.3cm
%    \epsffile{fig2.eps}
\includegraphics[width=8.5cm,keepaspectratio,angle=0,clip]{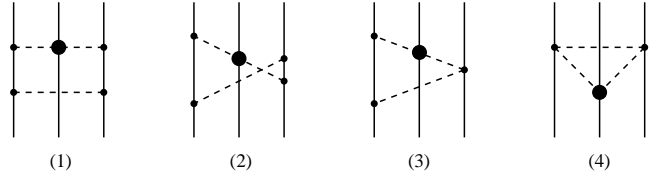}
    \caption{
         Ring diagrams at N$^4$LO. For notation
         see Figs.~\ref{fig0}, \ref{fig3}.
\label{fig4} 
 }
%  \end{center}
\end{figure}

At N$^4$LO, one only needs to evaluate the contributions of the four
diagrams shown in Fig.~\ref{fig4},
\beqa
V_{\rm ring}^{(5)}&=&V_{\rm ring}^{(5), g_A^4}+V_{\rm
  ring}^{(5), g_A^2}+V_{\rm ring}^{(5), g_A^0}\,.
\eeqa 
We were again able to obtain fairly
compact expressions in coordinate space, which, however, involve now 
a single scalar integral over the mass of the exchanged particles:  
\beqa
V_{\rm ring}^{(5), g_A^4}&=&-\frac{g_A^4 M_\pi^8 }{1024 \pi^4 F_\pi^6} \int_{-\infty}^\infty ds
\Big[2 \vec \nabla_{12}\cdot \vec \nabla_{23} \left(\vec \nabla_{23}\cdot 
\vec \nabla_{31} \left(12 c_1 \vec \sigma_2\cdot \vec \sigma_3-4 c_2 s^2 
\vec \sigma_2\cdot \vec \sigma_3 \fet \tau_2\cdot \fet \tau_3+6 c_2 
s^2\right. \right. \nn
&+&\left. \left. \vec \nabla_{12}\cdot \vec \nabla_{31} (\vec \sigma_1\cdot \vec \sigma_3 
(-3 c_3+c_4 \fet \tau_1\cdot \fet \tau_2+c_4 \fet \tau_2\cdot \fet 
\tau_3)+2 (\vec \sigma_2\cdot \vec \sigma_3 (-3 c_3+c_4 \fet 
\tau_1\cdot \fet \tau_2+c_4 \fet \tau_1\cdot \fet
\tau_3)\right. \right. \nn
&+&\left. \left. c_3 (2 \fet 
\tau_1\cdot \fet \tau_2+\fet \tau_1\cdot \fet \tau_3)))+2 
\vec \nabla_{12}\cdot \vec \sigma_1 \vec \nabla_{31}\cdot \vec \sigma_2 (3 
c_3-c_4 \fet \tau_1\cdot \fet \tau_3-c_4 \fet \tau_2\cdot \fet 
\tau_3)-4 c_3 s^2 \vec \sigma_2\cdot \vec \sigma_3 \fet \tau_2\cdot 
\fet \tau_3\right. \right. \nn
&+&\left. \left. 6 c_3 s^2+c_4 \vec \nabla_{12}\cdot \vec \nabla_{31}\times \vec 
\sigma_1 \fet \tau_1\cdot \fet \tau_2\times \fet \tau_3\right)-2 
\left(2 \vec \nabla_{12}\cdot \vec \sigma_1 \vec \nabla_{31}\cdot \vec \sigma_2 
\left(3 c_1-s^2 (c_2+c_3) \fet \tau_1\cdot \fet
  \tau_2\right)\right.\right.  \nn
&+&\left. \left. \left(
\vec \nabla_{31}\cdot \vec \sigma_1 \vec \nabla_{31}\cdot \vec 
\sigma_3-\left(s^2+1\right) \vec \sigma_1\cdot \vec \sigma_3\right) 
\left(s^2 (c_2+c_3) \fet \tau_1\cdot \fet \tau_3-3 c_1\right)+
\vec \nabla_{12}\cdot \vec \nabla_{31} (4 c_1 \fet \tau_1\cdot \fet
\tau_2\right. \right. \nn
&+&\left. \left. 
\vec \nabla_{23}\cdot \vec \sigma_1 \vec \nabla_{31}\cdot \vec \sigma_3 (-3 
c_3+c_4 \fet \tau_1\cdot \fet \tau_2+c_4 \fet \tau_2\cdot \fet 
\tau_3))\right)-\left(s^2+1\right) \vec \nabla_{12}\cdot \vec \sigma_3 
\vec \nabla_{23}\cdot \vec \sigma_1 (3 c_3-c_4 \fet \tau_1\cdot \fet 
\tau_2\right. \nn
&-&\left. c_4 \fet \tau_2\cdot \fet \tau_3)\right)+\vec \nabla_{12}\cdot 
\vec \nabla_{31} \left(8 \left(\vec \nabla_{12}\cdot \vec \sigma_2 
\vec \nabla_{23}\cdot \vec \sigma_1 \left(s^2 (c_2+c_3) \fet \tau_1\cdot 
\fet \tau_2-3 c_1\right)\right.\right.  \nn
&+&\left. \left. \vec \nabla_{23}\cdot \vec \sigma_1 
\vec \nabla_{31}\cdot \vec \sigma_3 \left(s^2 (c_2+c_3) \fet \tau_1\cdot 
\fet \tau_3-3 c_1\right)+\left(s^2+1\right) \vec \sigma_2\cdot \vec 
\sigma_3 \left(s^2 (c_2+c_3) \fet \tau_2\cdot \fet \tau_3-3 c_1
\right)\right)\right. \nn
&-&\left. \vec \nabla_{23}\cdot \vec \nabla_{31} \left(4 \vec \sigma_1\cdot 
\vec \sigma_3 \left(s^2 (c_2+c_3) \fet \tau_1\cdot \fet \tau_3-3 c_1
\right)+8 c_1 \fet \tau_1\cdot \fet \tau_3-6 c_2 s^2+4 \vec \nabla_{12}\cdot 
\vec \sigma_2 \vec \nabla_{23}\cdot \vec \sigma_1\right. \right.  \nn
&\times&\left. \left. (-3 c_3+c_4 \fet 
\tau_1\cdot \fet \tau_3+c_4 \fet \tau_2\cdot \fet \tau_3)-6 c_3 s^2+c_4 
\vec \nabla_{12}\cdot \vec \nabla_{23}\times \vec \sigma_2 \fet \tau_1\cdot 
\fet \tau_2\times \fet \tau_3\right)\right)\nn
&+&4 \left(\left(s^2+1
\right) \vec \nabla_{23}\cdot \vec \sigma_1 \left(\vec \nabla_{31}\cdot \vec 
\sigma_2 \left(6 c_1-2 s^2 (c_2+c_3) \fet \tau_1\cdot \fet \tau_2+
\vec \nabla_{23}\cdot \vec \nabla_{31} (-3 c_3+c_4 \fet \tau_1\cdot \fet 
\tau_3+c_4 \fet \tau_2\cdot \fet \tau_3)\right)\right. \right. \nn
&+&\left. \left. \vec \nabla_{12}\cdot \vec 
\sigma_3 \left(3 c_1-s^2 (c_2+c_3) \fet \tau_1\cdot \fet \tau_3
\right)\right)+\vec \nabla_{12}\cdot \vec \sigma_1 \vec \nabla_{12}\cdot \vec 
\sigma_2 \vec \nabla_{23}\cdot \vec \nabla_{31} \left(6 c_1-2 s^2 (c_2+c_3) 
\fet \tau_1\cdot \fet \tau_2\right. \right. \nn
&+&\left. \left. \vec \nabla_{23}\cdot \vec \nabla_{31} (-3 c_3+c_4 
\fet \tau_1\cdot \fet \tau_3+c_4 \fet \tau_2\cdot \fet \tau_3)\right)
\right)-2 (\vec \nabla_{12}\cdot \vec \nabla_{23})^2 \left(\vec \nabla_{31}\cdot 
\vec \sigma_1 \vec \nabla_{31}\cdot \vec \sigma_3-\left(s^2+1\right) \vec 
\sigma_1\cdot \vec \sigma_3\right)\nn
&\times& (3 c_3-c_4 \fet \tau_1\cdot \fet 
\tau_2-c_4 \fet \tau_2\cdot \fet \tau_3)+4 \left(s^2+1\right) (
\vec \nabla_{12}\cdot \vec \nabla_{31})^2 \vec \sigma_2\cdot \vec \sigma_3 (3 
c_3-c_4 \fet \tau_1\cdot \fet \tau_2-c_4 \fet \tau_1\cdot \fet 
\tau_3) \Big]\nn
&\times& U_1^s(x_{12}) U_1^s(x_{23}) U_2^s(x_{31}), 
\eeqa
\beqa
V_{\rm ring}^{(5), g_A^2}&=&\frac{g_A^2 M_\pi^8 }{1024 \pi^4 F_\pi^6}\int_{-\infty}^\infty ds \Big[8 
c_1 \vec \nabla_{12}\cdot \vec \nabla_{23} \fet \tau_2\cdot \fet 
\tau_3+8 c_1 \vec \nabla_{23}\cdot \vec \nabla_{31} \fet \tau_2\cdot 
\fet \tau_3+4 c_2 s^2 \vec \nabla_{12}\cdot \vec \nabla_{23} \fet 
\tau_2\cdot \fet \tau_3\nn
&+&4 c_2 s^2 \vec \nabla_{23}\cdot \vec 
\nabla_{31} \fet \tau_2\cdot \fet \tau_3+\vec \nabla_{12}\cdot \vec 
\nabla_{31} \left(-4 c_3 \vec \nabla_{12}\cdot \vec \nabla_{23} \fet 
\tau_2\cdot \fet \tau_3-4 c_3 \vec \nabla_{23}\cdot \vec \nabla_{31} 
\fet \tau_2\cdot \fet \tau_3\right. \nn
&+&\left. c_4 \vec \nabla_{12}\cdot \vec 
\nabla_{23}\times \vec \sigma_2 \fet \tau_1\cdot \fet \tau_2\times \fet 
\tau_3+4 c_4 \vec \nabla_{23}\cdot \vec \sigma_1 \vec 
\nabla_{31}\cdot \vec \sigma_3 (\fet \tau_1\cdot \fet \tau_2+\fet 
\tau_2\cdot \fet \tau_3)\right. \nn
&+&\left. 2 c_4 \left(s^2+1\right) \vec \sigma_2\cdot 
\vec \sigma_3 (\fet \tau_1\cdot \fet \tau_2+\fet \tau_1\cdot \fet 
\tau_3)\right)+4 c_3 s^2 \vec \nabla_{12}\cdot \vec \nabla_{23} \fet 
\tau_2\cdot \fet \tau_3+4 c_3 s^2 \vec \nabla_{23}\cdot \vec 
\nabla_{31} \fet \tau_2\cdot \fet \tau_3\nn
&-&2 c_4 \vec \nabla_{12}\cdot 
\vec \nabla_{23} \vec \nabla_{31}\cdot \vec \sigma_1 \vec \nabla_{31}
\cdot \vec \sigma_3 \fet \tau_1\cdot \fet \tau_2-2 c_4 \vec 
\nabla_{12}\cdot \vec \nabla_{23} \vec \nabla_{31}\cdot \vec \sigma_1 
\vec \nabla_{31}\cdot \vec \sigma_3 \fet \tau_2\cdot \fet \tau_3\nn
&-&2 
c_4 \vec \nabla_{12}\cdot \vec \nabla_{23} \vec \nabla_{23}\cdot \vec 
\nabla_{31} \vec \sigma_2\cdot \vec \sigma_3 \fet \tau_1\cdot \fet 
\tau_2-2 c_4 \vec \nabla_{12}\cdot \vec \nabla_{23} \vec 
\nabla_{23}\cdot \vec \nabla_{31} \vec \sigma_2\cdot \vec \sigma_3 \fet 
\tau_1\cdot \fet \tau_3\nn
&+&c_4 \vec \nabla_{23}\cdot \vec \nabla_{31} 
\vec \nabla_{12}\cdot \vec \nabla_{23}\times \vec \sigma_2 \fet 
\tau_1\cdot \fet \tau_2\times \fet \tau_3-2 c_4 \left(s^2+1\right) \vec 
\nabla_{12}\cdot \vec \sigma_3 \vec \nabla_{23}\cdot \vec \sigma_1 
(\fet \tau_1\cdot \fet \tau_2+\fet \tau_2\cdot \fet \tau_3) \Big]\nn
&\times& U_1^s(x_{12}) U_1^s(x_{23}) U_1^s(x_{31}), 
\eeqa
\beqa
V_{\rm ring}^{(5), g_A^0}&=&-\frac{M_\pi^8 s^2 }{1024 \pi^4 F_\pi^6}\int_{-\infty}^{\infty}ds \Big[4 
\fet \tau_2\cdot \fet \tau_3 \left(2 c_1+s^2 (c_2+c_3)-c_3 \vec 
\nabla_{12}\cdot \vec \nabla_{31}\right)+c_4 \vec \nabla_{12}\cdot \vec 
\nabla_{23}\times \vec \sigma_2 \fet \tau_1\cdot \fet \tau_2\times 
\fet \tau_3\Big] \nn
&\times& U_1^s(x_{12}) U_1^s(x_{23}) U_1^s(x_{31}),
\eeqa
where
\beqa
U_1^s(x)&=&\frac{4\pi}{M_\pi}\int\frac{d^3q}{(2\pi)^3}\frac{1}{q^2+\tilde
  M_\pi^2}e^{i \vec q\cdot \vec x/M_\pi}=\frac{e^{-x\sqrt{1+s^2}}}{x}, \quad
\tilde M_\pi^2=M_\pi^2+l_0^2, \quad s=\frac{l_0}{M_\pi}, \nn
  U_2^s(x)&=&8\pi M_\pi\int\frac{d^3q}{(2\pi)^3}\frac{1}{(q^2+\tilde
  M_\pi^2)^2}e^{i \vec q\cdot \vec x/M_\pi}=\frac{1}{\sqrt{1+s^2}}e^{-x\sqrt{1+s^2}}.
\eeqa
The expressions in momentum space are rather lengthy and can be found
in appendix \ref{app1}.

%%%%%%%%%%%%%%%%%%%%%%%%%%%%%%%%%%%%%%%%%%%%%%%%%%%%%%%%%%%%%%%%%%%%%%%%%%%%%%%%%
\section{General operator structure of a local three-nucleon force}
\def\theequation{\arabic{section}.\arabic{equation}}
\label{sec:structure}

As already emphazized in the introduction, having derived explicit
expressions for the long-range part of the 3NF at the three first orders
in the chiral expansion, it is interesting to test convergence in
coordinate space. One generally expects for the chiral expansion of
nuclear potentials to converge at distances of the order of or larger
than $r \sim M_\pi^{-1}$. In order to analyze the convergence of the
chiral expansion for three-nucleon potentials in a meaningful way, we first need
to define a basis in the space of isospin-spin-position or, equivalently,
isospin-spin-momentum three-nucleon operators. Thus, we need to work
out the most general structure of the three-nucleon force. To the best
of our knowledge, this task has not been accomplished yet, see however
Ref.~\cite{Epelbaum:2004xf}, where the most general isospin structure of the 3NF is
given. 

Given that a general 3NF depends, in the center of mass system, 
on four independent momenta in addition to the spin and isospin Pauli
matrices,  its structure is obviously rather rich. Fortunately, 
even at such a high order in the chiral expansion as N$^4$LO,  the most
complicated part of the three-nucleon force (before
antisymmetrization) is still local. For the long-range part, the only
non-localities in the power counting scheme we adopt arise from
the leading relativistic corrections to the $2\pi$ exchange diagrams
discussed in Ref.~\cite{Bernard:2011zr}.   We, therefore, restrict
ourselves here to the most general structure of a local 3NF. 
We, furthermore,  require in the following that the 3NF is
invariant under parity, time-reversal and isospin transformations.

Every operator appearing in the 3NF can be written as a linear combination of
spin-momentum terms multiplied with isospin structures.
We remind the reader that according to the standard convention, the
expressions for nuclear forces are to be understood
as matrix elements with respect to momenta and operators in the spin
and isospin spaces. The building blocks for the spin-momentum structures are
\beq
\vec \sigma_1, \; \vec \sigma_2, \; \vec \sigma_3, \; \vec q_1,\;  \vec q_3,
\eeq
where $\vec \sigma_i$ are the Pauli spin matrices of the 
nucleons while $\vec q_1$, $\vec q_3$ denote the two independent
relative momenta.\footnote{The momentum transfer $\vec q_2$ can be
  expressed in terms of $\vec q_{1,3}$ via $\vec q_2 = - \vec q_1 -
  \vec q_3$.} These building blocks have to be
contracted with the tensors $\delta_ {ab}$ and $\epsilon_{abc}$ to
build scalar operators. We have the following symmetry constraints:
\begin{itemize}
\item 
Parity invariance of the force allows only for spin-momentum structures which are invariant under
$$
\vec{q}_1\rightarrow -\vec{q}_1\quad\mbox{and}\quad \vec{q}_3\rightarrow -\vec{q}_3\,.
$$
\item 
Time-reversal invariance implies that only those structures contribute
which are invariant under
$$
\vec{\sigma}_i\rightarrow -\vec{\sigma}_i \,, \quad \vec{ q}_i\rightarrow \vec{ q}_i \quad \mbox{and}
\quad \tau_i^y\rightarrow -\tau_i^y, \quad i=1,2,3, 
$$
see Eq.~(2.47) of Ref.~\cite{GloeckleBook}\footnote{The invariance under $\tau_i^y\rightarrow -\tau_i^y$ follows directly from the invariance of the matrix element under 
 $\langle t^\prime| \fet\tau_i| t\rangle\rightarrow \langle t| \fet\tau_i |t^\prime\rangle$}.
 \item 
Isospin conservation requires any structure to be a product of a spin-momentum operator with one of the following isospin-structures:
 $$
 1, \quad\fet \tau_1\cdot\fet \tau_2, \quad\fet \tau_1\cdot\fet
 \tau_3,\quad \fet \tau_2\cdot\fet \tau_3\quad \mbox{and} \quad
 \fet \tau_1\cdot(\fet \tau_2\times\fet \tau_3).
 $$
\end{itemize}
In addition to the symmetry constraints, we will also employ the Schouten identity
 $$
 \delta_{i,j}\epsilon_{k,l,m}-\delta_{i,k}\epsilon_{l,m,j}+\delta_{i,l}\epsilon_{m,j,k}-\delta_{i,m}\epsilon_{j,k,l}=0
 $$
 to eliminate redundant structures. A general local three-nucleon force can be written in a form
 $$
 \sum_i {O}_i(\vec\sigma_1,\vec \sigma_2, \vec\sigma_3,\fet
 \tau_1,\fet \tau_2, \fet \tau_3,\vec q_1, \vec q_3) \, F_i(q_1, q_3, \vec q_1\cdot \vec q_3)\,,
 $$
 where ${O}_i$ are spin-momentum-isospin operators and the scalar
 structure functions $F_i$ depend only on absolute values $|\vec q_1|,
 |\vec q_3|$ and on the scalar product $\vec{q}_1\cdot\vec{q}_3$. The
 three-nucleon force $V_{3N}^{\rm full}$ in Eq.~(\ref{def_perm}) is obviously invariant under any permutation $P\in S_3$,
 with $S_3$ denoting
 the permutation group:
 \beqa
 \sum_i P{O}_i PF_i=\sum_i {O}_i F_i,
\label{symmetry}
 \eeqa
where 
\beqa
P{O}_i(\vec\sigma_1,\vec \sigma_2, \vec\sigma_3,\fet
 \tau_1,\fet \tau_2, \fet \tau_3,\vec q_1, \vec q_3)&=&{O}_i(\vec\sigma_{P[1]},\vec \sigma_{P[2]}, \vec\sigma_{P[3]},\fet
 \tau_{P[1]},\fet \tau_{P[2]}, \fet \tau_{P[3]},\vec q_{P[1]}, \vec
 q_{P[3]}),\nonumber\\
P{F}_i(q_1, q_3, \vec q_1\cdot \vec q_3)&=&{F}_i(q_{P[1]}, q_{P[3]}, \vec q_{P[1]}\cdot \vec q_{P[3]}).
\eeqa
To understand the behavior of the structure functions under
permutations of momenta it is necessary to analyze the behavior of the
operators ${O}_i$ under permutations. Since the operator set we
consider here  is complete, the permuted operator $P{O}_i$ is just a linear combination of ${O}_j$'s:
$$
P{O}_i=\sum_j {O}_j D_{ji}(P),
$$ 
where $D(P)$ are some invertible matrices. It is easy to see that the
set of matrices $D$ builds a representation of $S_3$. Indeed
$$
P'P{O}_i=P'\sum_k {O}_k D_{ki}(P)=\sum_{j,k} {O}_j \, D_{jk}(P')\,
D_{ki}(P)=\sum_j {O}_j D_{ji} (P'P),
$$
from which immediately follows
$$
D(P' P)=D(P')D(P).
$$
Transformations of the structure functions $F_i$ with respect to
permutations $P$ of the nucleon labels can now be read off from
 $$
 \sum_i {O}_i F_i=\sum_i P{O}_i PF_i=\sum_{i,j} {O}_j D_{ji}(P)
 PF_i=\sum_i{O}_i\left(\sum_j D_{ij}(P) PF_j \right),
 $$
from which we obtain the identity
 $$
 F_i=\sum_j D_{ij}(P) PF_j.
 $$

It is advantageous to choose the basis in the space of operators $O_i$
such that the representation matrices $D$ are block-diagonal
corresponding to irreducible representations of the group $S_3$. 
There are three inequivalent irreducible representations of $S_3$:
\begin{itemize}
\item The trivial (identity) and antisymmetric $(-1)^{w(P)}$
  representations with $w(P)= \pm 1$ for even/odd
  permutations are one
  dimensional. 
\item The third irreducible representation is two dimensional. The
  representation matrices can e.g.~be chosen as 
\beq
\begin{array}{lll}
{\cal D}(())=\left(
\begin{array}{cc}
1&0\\
0&1
\end{array}
\right),
&{\cal D}((12))=\frac{1}{2}\left(
\begin{array}{cc}
1&\sqrt{3}\\
\sqrt{3}&-1
\end{array}
\right),
&{\cal D}((13))=\left(
\begin{array}{cc}
-1& 0\\
0 & 1
\end{array}
\right),\\
{\cal D}((23))=-\frac{1}{2}\left(
\begin{array}{cc}
-1&\sqrt{3}\\
\sqrt{3}&1
\end{array}
\right),
&{\cal D}((123))=-\frac{1}{2}\left(
\begin{array}{cc}
1&\sqrt{3}\\
-\sqrt{3}&1
\end{array}
\right),
&
{\cal D}((132))=-\frac{1}{2}\left(
\begin{array}{cc}
1&-\sqrt{3}\\
\sqrt{3}&1
\end{array}
\right), 
\end{array}
\eeq
where we used the cyclic notation for
permutations:
\beqa
()[1]&=&1, \quad ()[2]=2, \quad ()[3]=3,\nn
(12)[1]&=&2, \quad  (12)[2]=1, \quad  (12)[3]=3,\nn
(13)[1]&=&3, \quad  (13)[2]=2, \quad  (13)[3]=1,\nn
(23)[1]&=&1, \quad  (23)[2]=3, \quad  (23)[3]=2,\nn
(123)[1]&=&2, \quad  (123)[2]=3, \quad  (123)[3]=1,\nn
(132)[1]&=&3, \quad  (132)[2]=1, \quad  (132)[3]=2.
\eeqa
\end{itemize}
%Notice that for all permutations $P\in
%S_3$, the matrices ${\cal D}(P)$ defined above are orthogonal.
%Consequently, if a given function $F_i$
%transforms under permutations via irreducible representation ${\cal D}$
%we get
%\beq
%P F_i = \sum_j F_j {\cal D}(P^{-1})_{j, i} = \sum_j{\cal D}(P)_{i, j} F_j.
%\eeq

To construct the operators
${O}_i$ for which $D(P)$ is block-diagonal,  we introduce the
symmetrizing and antisymmetrizing functions:
$$
S({O})=\frac{1}{6}\sum_{P\in S_3}P{O}, \quad A({O})=\frac{1}{6}\sum_{P\in S_3}(-1)^{w(P)}P{O}.
$$ 
It is obvious that 
$$
P S({O})=S({O})\quad {\rm and}\quad P A({O})=(-1)^{w(P)} A({O})
$$
for all $P\in S_3$ such that $S({O})$ and $A({O})$ transform under one-dimensional representations. To construct operators which transform
under two-dimensional irreducible representation we introduce the
functions 
$$
G_{ij}({O})=\frac{1}{3}\sum_{P\in S_3}{\cal D}_{ij}(P) PO, \quad i,j=1,2.
$$
It is easy to verify that the resulting operators $G_{ij}(O)$ indeed transform under two-dimensional irreducible
representation:
$$
P G_{ij}(O)=\frac{1}{3}\sum_{Q\in S_3}{\cal
  D}_{ij}(Q)PQO=\frac{1}{3}\sum_{Q\in S_3}{\cal
  D}_{ij}(P^{-1}Q)QO=
%\frac{1}{3}\sum_{Q\in S_3}\sum_k{\cal
%  D}(P^{-1})_{ik}{\cal D}(Q)_{kj}QO=
\sum_k G_{kj}(O){\cal D}_{ki}(P).
$$

With all the symmetry constraints introduced above, we found that the
most general structure of a local 3NF can be written in terms of 89
operators ${O}_1, \dots, {O}_{89}$, which transform with respect to
permutations according to irreducible representations of $S_3$. These
89 operators can be generated from a set of 22 independent operators 
 ${\cal G}_1,\dots,{\cal G}_{22}$ using the functions $S$, $A$
 and $G_{ij}$ defined above. The explicit form of generating operators
 ${\cal G}_1,\dots,{\cal G}_{22}$ we adopt in this work and their relation to
 the generated operators ${O}_1, \dots, {O}_{89}$ are given in  Table~\ref{generalstr}.
\begin{table}[t]
\begin{tabular}{|c|c|c|c|c|c|c|}
\hline
Generators ${\cal G}$ of 89 independent operators & $S$ & $A$ & $G_{12}$ & $G_{22}$ & $G_{11}$ & $G_{21}$\\
\hline
${\cal G}_{1} = 1$ & $O_1$ & 0 & 0 & 0 & 0& 0\\
\hline
${\cal G}_{2} =\fet \tau_1\cdot\fet \tau_3$ & $O_2$ & 0 & $O_3$ & $O_4$& $0$ & $0$\\
\hline 
${\cal G}_{3} =\vec{\sigma}_1\cdot\vec{\sigma}_3$ & $O_5$ & 0 & $O_6$ &  $O_7$& $0$ & $0$\\
\hline 
${\cal G}_{4} =\fet \tau_1\cdot\fet \tau_3\vec{\sigma}_1\cdot\vec{\sigma}_3$ &  $O_8$ & 0 &  $O_9$ &  $O_{10}$& $0$ & $0$\\
\hline 
${\cal G}_{5} =\fet \tau_2\cdot\fet \tau_3\vec{\sigma}_1\cdot\vec{\sigma}_2$ &  $O_{11}$ &   $O_{12}$ &  $O_{13}$&  $O_{14}$&   $O_{15}$ &   $O_{16}$\\
\hline 
${\cal G}_{6} =\fet \tau_1\cdot(\fet \tau_2\times\fet
\tau_3)\vec{\sigma}_1\cdot(\vec{\sigma}_2\times\vec{\sigma}_3)$ &
$O_{17}$ &  0 & 0 & 0 &  0 &  0\\
\hline 
${\cal G}_{7} =\fet \tau_1\cdot(\fet \tau_2\times\fet
\tau_3)\vec{\sigma}_2\cdot(\vec{q}_1\times\vec{q}_3)$ &
$O_{18}$ &  0 &  ${O}_{19}$&  $O_{20}$&  $0$ &  $0$\\
\hline 
${\cal G}_{8} =\vec{q}_1\cdot\vec{\sigma}_1\vec{q}_1\cdot\vec{\sigma}_3$ &  $O_{21}$ &   $O_{22}$ &   $O_{23}$ &  $O_{24}$&   $O_{25}$ &   $O_{26}$\\
\hline 
${\cal G}_{9} =\vec{q}_1\cdot\vec{\sigma}_3\vec{q}_3\cdot\vec{\sigma}_1$ &  $O_{27}$ &  0 &   $O_{28}$ &  $O_{29}$&  $0$&  $0$\\
\hline 
${\cal G}_{10} =\vec{q}_1\cdot\vec{\sigma}_1\vec{q}_3\cdot\vec{\sigma}_3$ &  $O_{30}$ &  0 &   $O_{31}$ &  $O_{32}$&  $0$&  $0$\\
\hline 
${\cal G}_{11} =\fet \tau_2\cdot\fet \tau_3\vec{q}_1\cdot\vec{\sigma}_1\vec{q}_1\cdot\vec{\sigma}_2$ &  $O_{33}$ &    $O_{34}$  &   $O_{35}$ &  $O_{36}$&   $O_{37}$ &    $O_{38}$ \\
\hline 
${\cal G}_{12} =\fet \tau_2\cdot\fet \tau_3\vec{q}_1\cdot\vec{\sigma}_1\vec{q}_3\cdot\vec{\sigma}_2$ &  $O_{39}$ &    $O_{40}$  &   $O_{41}$ &  $O_{42}$&   $O_{43}$ &    $O_{44}$ \\
\hline 
${\cal G}_{13} =\fet \tau_2\cdot\fet \tau_3\vec{q}_3\cdot\vec{\sigma}_1\vec{q}_1\cdot\vec{\sigma}_2$ &  $O_{45}$ &   $O_{46}$  &   $O_{47}$ &  $O_{48}$&   $O_{49}$ &    $O_{50}$ \\
\hline 
${\cal G}_{14} =\fet \tau_2\cdot\fet \tau_3\vec{q}_3\cdot\vec{\sigma}_1\vec{q}_3\cdot\vec{\sigma}_2$ &  $O_{51}$ &    $O_{52}$  &   $O_{53}$ &  $O_{54}$&   $O_{55}$ &    $O_{56}$ \\
\hline 
${\cal G}_{15} =\fet \tau_1\cdot\fet
\tau_3\vec{q}_2\cdot\vec{\sigma}_1\vec{q}_2\cdot\vec{\sigma}_3$ &
$O_{57}$ &   0  &   ${O}_{58}$ &  $O_{59}$&$0$&   $0$\\
\hline 
${\cal G}_{16} =\fet \tau_2\cdot\fet \tau_3\vec{q}_3\cdot\vec{\sigma}_2\vec{q}_3\cdot\vec{\sigma}_3$ &  $O_{60}$ &   $O_{61}$  &   $O_{62}$ &  $O_{63}$&  $O_{64}$ &    $O_{65}$\\
\hline 
${\cal G}_{17} =\fet \tau_1\cdot\fet \tau_3\vec{q}_1\cdot\vec{\sigma}_1\vec{q}_3\cdot\vec{\sigma}_3$ &  $O_{66}$ & $0$  &   $O_{67}$ &  $O_{68}$& $0$ &   $0$\\
\hline 
${\cal G}_{18} =\fet \tau_1\cdot(\fet \tau_2\times\fet \tau_3)\vec{\sigma}_1\cdot\vec{\sigma}_3\vec{\sigma}_2\cdot(\vec{q}_1\times\vec{q}_3)$ &  $O_{69}$ & 0  &   $O_{70}$ &  $O_{71}$& $0$ &   $0$\\
\hline 
${\cal G}_{19} =\fet \tau_1\cdot(\fet \tau_2\times\fet
\tau_3)\vec{\sigma}_3\cdot\vec{q}_1\vec{q}_1\cdot(\vec{\sigma}_1\times\vec{\sigma}_2)$
& $\; O_{72} \; $ &  $ \;  O_{73} \; $ &   $ \;  O_{74} \; $ &  $ \;  O_{75} \; $&  $ \;  O_{76} \; $ &   $ \;  O_{77} \;$\\
\hline 
${\cal G}_{20} =\fet \tau_1\cdot(\fet \tau_2\times\fet \tau_3)\vec{\sigma}_1\cdot\vec{q}_1\vec{\sigma}_2\cdot\vec{q}_1\vec{\sigma}_3\cdot(\vec{q}_1\times\vec{q}_3)$ &  $O_{78}$ &  $O_{79}$ &   $O_{80}$ &  $O_{81}$&  $O_{82}$ &   $O_{83}$\\
\hline 
${\cal G}_{21} =\fet \tau_1\cdot(\fet \tau_2\times\fet \tau_3)\vec{\sigma}_1\cdot\vec{q}_2\vec{\sigma}_3\cdot\vec{q}_2\vec{\sigma}_2\cdot(\vec{q}_1\times\vec{q}_3)$ &  $O_{84}$ & 0 &   $O_{85}$ &  $O_{86}$& $0$ &  $0$\\
\hline 
$\quad {\cal G}_{22} =\fet \tau_1\cdot(\fet \tau_2\times\fet
\tau_3)\vec{\sigma}_1\cdot\vec{q}_1\vec{\sigma}_3\cdot\vec{q}_3\vec{\sigma}_2\cdot(\vec{q}_1\times\vec{q}_3)
\quad$
& $O_{87}$ & $0$ &   ${O}_{88}$ &  ${O}_{89}$& $0$ &  $0$\\
\hline 
\end{tabular}
\caption{The set of 22 generating operators ${\cal G}_i$ and their
  relation to 89 independent operators ${O}_1,\dots,{O}_{89}$ which parametrize the most
  general structure of a local 3NF.  The operators ${O}_i$ are
  generated by application of one of the 6 functions $S, A, G_{11},
  G_{12}, G_{21}, G_{22}$ on the corresponding operator ${\cal
    G}_j$. The 22 operators are constructed to be either
    totally symmetric, symmetric under $1\leftrightarrow 3$ or
    unsymmetric.
\label{generalstr}}
\end{table}
The complete expression for a  local three-nucleon force in our
notation can be written in the symmetric form
\beq
V_{\rm 3N}^{\rm full} = \sum_{i=1}^{89}{O}_i(\vec\sigma_1,\vec \sigma_2,
\vec\sigma_3,\fet \tau_1,\fet \tau_2, \fet \tau_3,\vec q_1, \vec q_3)F_i(q_1, q_3, \vec q_1\cdot \vec q_3).\label{strf89}
\eeq
In this representation, the structure functions $F_i$ have simple
transformation properties with respect to permutations\footnote{The
  only exception is $F_{17}$ which mixes different contributions from other
  structure functions. This is due to the use of the  Schouten
  identities.}. An alternative  way to express the three-nucleon
force is given by 
\beq
V_{\rm 3N}^{\rm full} = \sum_{i=1}^{22}{\cal G}_i(\vec\sigma_1,\vec \sigma_2,
\vec\sigma_3,\fet \tau_1,\fet \tau_2, \fet \tau_3,\vec q_1, \vec q_3){\cal F}_i(q_1, q_3, \vec q_1\cdot \vec q_3)+5\,{\rm permutations}.
\label{strf22}
\eeq
%The price to pay for using $22$ operators instead of $89$ is the
%more complicated transformation properties of the functions ${\cal F}$ with respect to
%permutations. 
%where we can not say anything about the transformation properties of the structure functions
%${\cal F}_i$ under permutation of the three nucleons.
It is easy to see that the two representations (\ref{strf22}) and
(\ref{strf89}) are equivalent. While Eq.~(\ref{strf22}) can obviously
be brought into the form of Eq.~(\ref{strf89}) we now show that the
converse is also true.
%(i.e. Eq.~(\ref{strf89}) can be recast in the form 
%of Eq.~(\ref{strf22}), the inverse transformation is trivial) can be easily seen from the following consideration.
Eq.~(\ref{strf89}) can be rewritten in the form
\beqa
V_{\rm 3N}^{\rm full} &=& \sum_{i=1}^{22} \left\{ S ( {\cal G}_i)M_i + A ( {\cal G}_i)N_i + \sum_{j,k=1}^2 G_{j k} ( {\cal
  G}_i) L_{j k}^i \right\}\nonumber\\
&=& \sum_{P\in S_3}\sum_{i=1}^{22}P ( {\cal G}_i) \left\{ \frac{1}{6}M_i +\frac{1}{6}(-1)^{w(P)}N_i + \sum_{j,k=1}^2  \frac{1}{3}{\cal D}_{jk}(P) L_{j k}^i \right\},
\label{strf89a}
\eeqa
where $M_i, N_i, L_{j k}^i$ are some of the structure functions $F_l (l=1,\dots,
89)$ from Eq.~(\ref{strf89}).
From the symmetry property (\ref{symmetry}) we get
\beqa
V_{\rm 3N}^{\rm full} &=& \frac{1}{6}\sum_{P'\in S_3}P'(V_{\rm 3N}^{\rm full})\nonumber\\
&=&\frac{1}{6}\sum_{P',P\in S_3}\sum_{i=1}^{22} P'P ( {\cal G}_i)\left\{ \frac{1}{6}P'(M_i) +
\frac{1}{6}(-1)^{w(P)}P'(N_i) + \sum_{j,k=1}^2  \frac{1}{3}{\cal D}_{jk}(P) P'(L_{j k}^i) \right\}\nonumber\\
  &=&\sum_{P''\in S_3}P''\sum_{i=1}^{22}{\cal G}_i \sum_{P\in S_3}\left\{ \frac{1}{36}P^{-1}(M_i) +
\frac{1}{36}(-1)^{w(P)}P^{-1}(N_i) + \sum_{j,k=1}^2  \frac{1}{18}{\cal D}_{jk}(P) P^{-1}(L_{j k}^i) \right\}\,,
\label{equivalence}
\eeqa
where we made a change of variable
$P''=P'P$ in the last line.
% of Eq.~(\ref{equivalence}).
This equation has the form of Eq.~(\ref{strf22}) with
\beq
{\cal F}_i\;:=\; \sum_{P\in S_3}\left\{ \frac{1}{36}P^{-1}(M_i) +
\frac{1}{36}(-1)^{w(P)}P^{-1}(N_i) + \sum_{j,k=1}^2  \frac{1}{18}{\cal D}_{jk}(P) P^{-1}(L_{j k}^i) \right\}\,.
\eeq

\section{Chiral expansion of the long-range tail of the 3NF}
\def\theequation{\arabic{section}.\arabic{equation}}
\label{sec:numerics}

\begin{figure}[tb]
\vskip 1 true cm
%  \begin{center} 
%    \epsfxsize=4.3cm
%    \epsffile{fig2.eps}
\includegraphics[width=\textwidth,keepaspectratio,angle=0,clip]{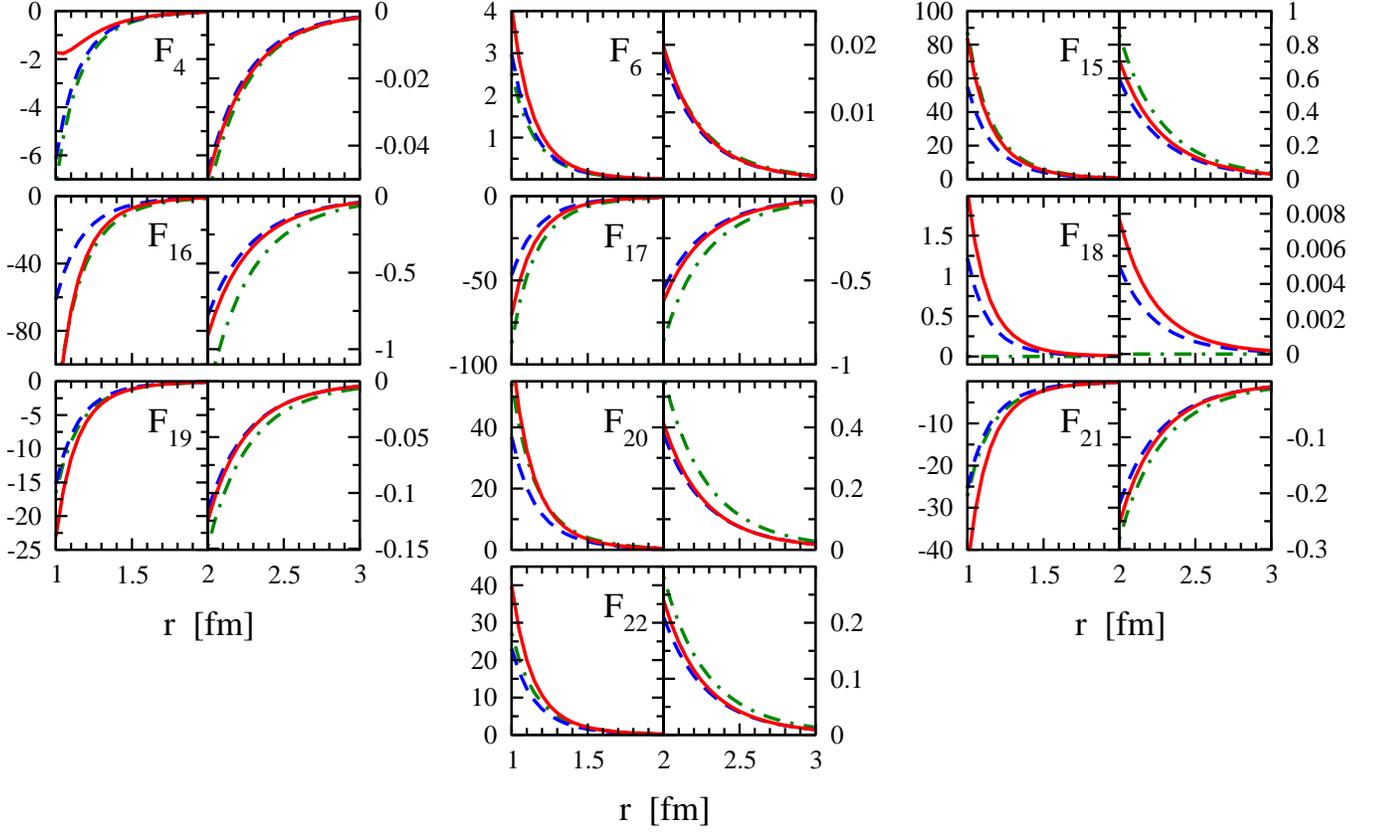}
    \caption{Chiral expansion of the profile functions ${\cal F}_i
      (r)$ in MeV generated by the two-pion exchange 3NF topology up to
      N$^4$LO (in the equilateral triangle configuration). Dashed-dotted, dashed and solid lines correspond to
      ${\cal F}_i^{(3)}$,   ${\cal F}_i^{(3)} + {\cal F}_i^{(4)}$ and
      ${\cal F}_i^{(3)} + {\cal F}_i^{(4)} + {\cal F}_i^{(5)}$,
      respectively. 
\label{fig:TPEr} 
 }
%  \end{center}
\end{figure}

With  these preparations we are now in the position to address the
convergence of the chiral expansion for the long-range tail of the
3NF. It is clear that all arguments of the previous section can also be
applied to operators in coordinate space. Here and in what follows, we
use the following basis of $22$ operators:
\beqa
\label{operatorsR}
\tilde  {\cal G}_{1} &=& 1 \,, \nn
\tilde  {\cal G}_{2} &=& \fet \tau_1\cdot\fet \tau_3 \,, \nn
\tilde  {\cal G}_{3} &=&  \vec{\sigma}_1\cdot\vec{\sigma}_3   \,,  \nn
\tilde  {\cal G}_{4} &=&  \fet \tau_1\cdot\fet \tau_3\, \vec{\sigma}_1\cdot\vec{\sigma}_3   \,,  \nn
\tilde  {\cal G}_{5} &=&  \fet \tau_2\cdot\fet \tau_3\, \vec{\sigma}_1\cdot\vec{\sigma}_2    \,, \nn
\tilde  {\cal G}_{6} &=&   \fet \tau_1\cdot(\fet \tau_2\times\fet
\tau_3)\, \vec{\sigma}_1\cdot(\vec{\sigma}_2\times\vec{\sigma}_3)   \,, \nn
\tilde  {\cal G}_{7} &=&   \fet \tau_1\cdot(\fet \tau_2\times\fet
\tau_3)\, \vec{\sigma}_2\cdot(\hat{r}_{12}\times\hat{r}_{23})   \,, \nn
\tilde  {\cal G}_{8} &=&   \hat{r}_{23}\cdot\vec{\sigma}_1\, \hat{r}_{23}\cdot\vec{\sigma}_3   \,, \nn
\tilde  {\cal G}_{9} &=&    \hat{r}_{23}\cdot\vec{\sigma}_3\, \hat{r}_{12}\cdot\vec{\sigma}_1   \,,\nn
\tilde  {\cal G}_{10} &=&  \hat{r}_{23}\cdot\vec{\sigma}_1\, \hat{r}_{12}\cdot\vec{\sigma}_3    \,, \nn
\tilde  {\cal G}_{11} &=&   \fet \tau_2\cdot\fet \tau_3\,
\hat{r}_{23}\cdot\vec{\sigma}_1\, \hat{r}_{23}\cdot\vec{\sigma}_2   \,, \nn
\tilde  {\cal G}_{12} &=&  \fet \tau_2\cdot\fet \tau_3\,
\hat{r}_{23}\cdot\vec{\sigma}_1\, \hat{r}_{12}\cdot\vec{\sigma}_2    \,, \nn
\tilde  {\cal G}_{13} &=&   \fet \tau_2\cdot\fet \tau_3\,
\hat{r}_{12}\cdot\vec{\sigma}_1\, \hat{r}_{23}\cdot\vec{\sigma}_2   \,, \nn
\tilde  {\cal G}_{14} &=&   \fet \tau_2\cdot\fet \tau_3\,
\hat{r}_{12}\cdot\vec{\sigma}_1\, \hat{r}_{12}\cdot\vec{\sigma}_2   \,, \nn
\tilde  {\cal G}_{15} &=&    \fet \tau_1\cdot\fet
\tau_3\, \hat{r}_{13}\cdot\vec{\sigma}_1\, \hat{r}_{13}\cdot\vec{\sigma}_3   \,,\nn
\tilde  {\cal G}_{16} &=&    \fet \tau_2\cdot\fet \tau_3\,
\hat{r}_{12}\cdot\vec{\sigma}_2\, \hat{r}_{12}\cdot\vec{\sigma}_3  \,, \nn
\tilde  {\cal G}_{17} &=&   \fet \tau_1\cdot\fet \tau_3\,
\hat{r}_{23}\cdot\vec{\sigma}_1\, \hat{r}_{12}\cdot\vec{\sigma}_3   \,, \nn
\tilde  {\cal G}_{18} &=&   \fet \tau_1\cdot(\fet \tau_2\times\fet
\tau_3)\, \vec{\sigma}_1\cdot\vec{\sigma}_3\, \vec{\sigma}_2\cdot(\hat{r}_{12}\times\hat{r}_{23})   \,, \nn
\tilde  {\cal G}_{19} &=&   \fet \tau_1\cdot(\fet \tau_2\times\fet
\tau_3)\, \vec{\sigma}_3\cdot\hat{r}_{23}\, \hat{r}_{23}\cdot(\vec{\sigma}_1\times\vec{\sigma}_2)  \,,  \nn
\tilde  {\cal G}_{20} &=&  \fet \tau_1\cdot(\fet \tau_2\times\fet
\tau_3)\, \vec{\sigma}_1\cdot\hat{r}_{23}\,
\vec{\sigma}_2\cdot\hat{r}_{23}\, \vec{\sigma}_3\cdot(\hat{r}_{12}\times\hat{r}_{23})    \,, \nn
\tilde  {\cal G}_{21} &=&  \fet \tau_1\cdot(\fet \tau_2\times\fet
\tau_3)\, \vec{\sigma}_1\cdot\hat{r}_{13}\,
\vec{\sigma}_3\cdot\hat{r}_{13}\, \vec{\sigma}_2\cdot(\hat{r}_{12}\times\hat{r}_{23})    \,, \nn
\tilde  {\cal G}_{22} &=&      \fet \tau_1\cdot(\fet \tau_2\times\fet
\tau_3)\, \vec{\sigma}_1\cdot\hat{r}_{23}\,
\vec{\sigma}_3\cdot\hat{r}_{12}\, \vec{\sigma}_2\cdot(\hat{r}_{12}\times\hat{r}_{23})\,,
\eeqa  
where $\hat r_{ij} \equiv \vec r_{ij}/| \vec r_{ij |}$ and $\vec
r_{ij}= \vec r_i - \vec r_j$ denotes the position of nucleon $i$
with respect to nucleon $j$. The 3NF is a linear combination of the
operators $\tilde  {\cal G}_{i}$ with the coefficients given by  scalar functions ${\cal F}_i (r_{12},  r_{23},
r_{31})$. These functions have the dimension of energy 
and can be interpreted as the potential energy between three static
nucleons projected onto the corresponding operator. The profile functions ${\cal F}_i $ receive
contributions from the long-range and the intermediate-range 3NF topologies and are
predicted (at long distances) in terms of the chiral expansion. In
order to explore the convergence, we plot these
functions for the equilateral triangle configuration of the nucleons
given by the condition 
\beq
r_{12}= r_{23}= r_{31} =r\,. 
\eeq
Restricting ourselves to this particular configuration allows us to
stay with
simple one-dimensional plots. We emphasize, however, that the
conclusions about the convergence of the chiral expansion for the 3NF
drawn in this section apply to this particular configuration. 
We begin with the longest-range $2\pi$ exchange topology. 
Projecting the
coordinate-space expressions given in section \ref{sec:TPE} onto the
operators in Eq.~(\ref{operatorsR}) and evaluating the
three-dimensional integrals in Eqs.~(\ref{AB4}) and (\ref{AB5})
numerically we compute the corresponding contributions to the profile functions ${\cal F}^{(3)} (r)$,  ${\cal F}^{(4)} (r)$ and ${\cal
  F}^{(5)} (r)$ at N$^2$LO, N$^3$LO and N$^4$LO, respectively. Our
results for the 3NF profile functions generated by the $2\pi$ exchange
diagrams are visualized in Fig.~\ref{fig:TPEr}.   
Here and in what follows, we use the values of the low-energy
constants corresponding to the order-$Q^4$ KH fit to the pion-nucleon
partial wave analysis of our work \cite{Krebs:2012yv}. In particular, we employ the following values
of $c_i$ (all in units of GeV$^{-1}$): 
\beq
c_1 = -0.75, \quad 
c_2 = 3.49, \quad 
c_3 = -4.77, \quad 
c_4 = 3.34\,.  
\eeq 
The results for the functions ${\cal F}_i (r)$ plotted in
Fig.~\ref{fig:TPEr}  resemble the findings of our work
\cite{Krebs:2012yv}, where a good convergence of the chiral
expansion in momentum space was observed by looking at the functions ${\cal A} (q_2)$ and ${\cal B}
(q_2)$ for low values of the momentum transfer. While there are large
corrections at N$^4$LO to some of the profile functions and,
especially, to ${\cal F}_4 (r)$ at short distances of the order of  $r \sim 1$ fm, 
we observe a very good convergence at long distances of the order of
$r \gtrsim 2$ fm.  At such large distances, the N$^4$LO
results appear to be very close to N$^3$LO ones. As already pointed out in the
introduction, fast convergence 
of the longest-range 3NF is not surprising given that effects of the
$\Delta$-isobar are, to a large extent, accounted for already in the
leading contributions ${\cal F}_i^{(3)} (r)$  to this topology  through resonance
saturation of the LECs $c_{3,4}$.  We further emphasize that the
operator structure of the $2\pi$ exchange topology is fairly
restricted: only $10$ out  of $22$ functions ${\cal F}_i (r)$ get
non-vanishing contributions. Notice that the larger number of nonvanishing 
functions  ${\cal F}_i$ in coordinate space compared to momentum space
has to be expected due to the appearance of gradients when performing the
Fourier transform. In contrast to the momentum space representation,
the number of nonvanishing structures in the coordinate space
representation of a 3NF is not representative for estimating the number
of affected nucleon-deuteron polarization observables at a fixed
kinematics.   

It is instructive to compare the strength of the three- and
two-nucleon potentials. While the long-range
three-nucleon potentials are considerably weaker than the two-nucleon
potentials, they are still not negligible. For example, the isovector-tensor and
isoscalar central nucleon-nucleon potentials governed by one-pion
exchange and (subleading) two-pion exchange, respectively, have the
strength of the order of $3\ldots 4$ MeV at distances $r \sim 2$ fm
\cite{Epelbaum:2012vx}.  These are the strongest 
two-nucleon forces at large distances.  The strongest three-nucleon potentials 
${\cal F}_{15} (r), {\cal F}_{16} (r)$ and ${\cal F}_{17} (r)$ reach at such distances the
strength of $\sim 0.7 \ldots 1$ MeV.  We remind the reader that 
nuclear potentials become  scheme dependent at short
distances below $r \sim 1 \ldots 1.5$ fm, where the contributions of
short-range topologies start playing important role.  An estimation of
this intrinsic scheme dependence for nucleon-nucleon potentials is
provided in Fig.~3 of Ref.~\cite{Epelbaum:2012vx}. 

\begin{figure}[tb]
\vskip 1 true cm
%  \begin{center} 
%    \epsfxsize=4.3cm
%    \epsffile{fig2.eps}
\includegraphics[width=\textwidth,keepaspectratio,angle=0,clip]{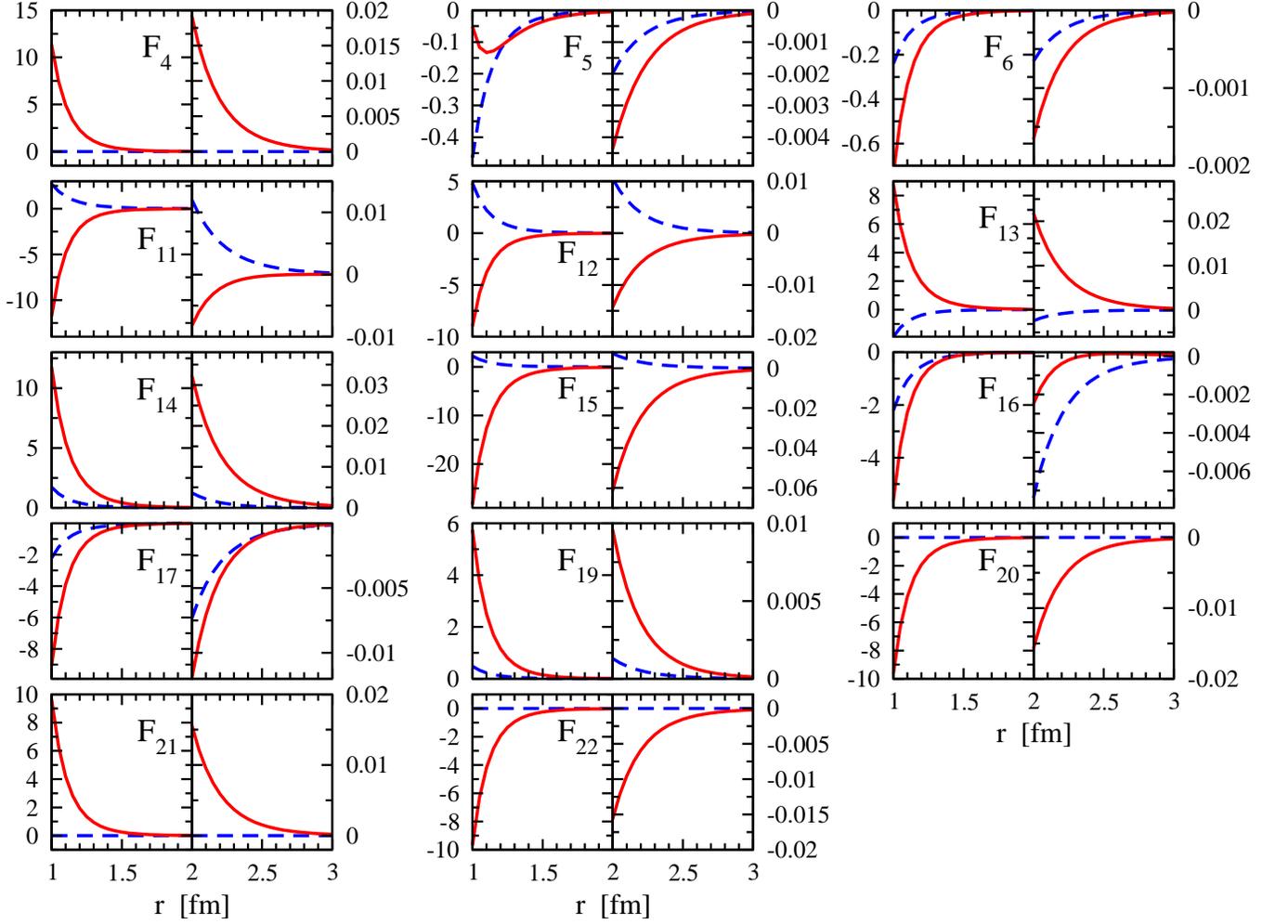}
    \caption{Chiral expansion of the profile functions ${\cal F}_i
      (r)$ in MeV generated by the two-pion-one-pion exchange  3NF topology up to
      N$^4$LO (in the equilateral triangle configuration). Dashed and solid lines correspond to
      ${\cal F}_i^{(4)}$ and  ${\cal F}_i^{(4)} + {\cal F}_i^{(5)}$,
      respectively. 
\label{fig:TPEOPEr} 
 }
%  \end{center}
\end{figure}

\begin{figure}[tb]
\vskip 1 true cm
%  \begin{center} 
%    \epsfxsize=4.3cm
%    \epsffile{fig2.eps}
\includegraphics[width=\textwidth,keepaspectratio,angle=0,clip]{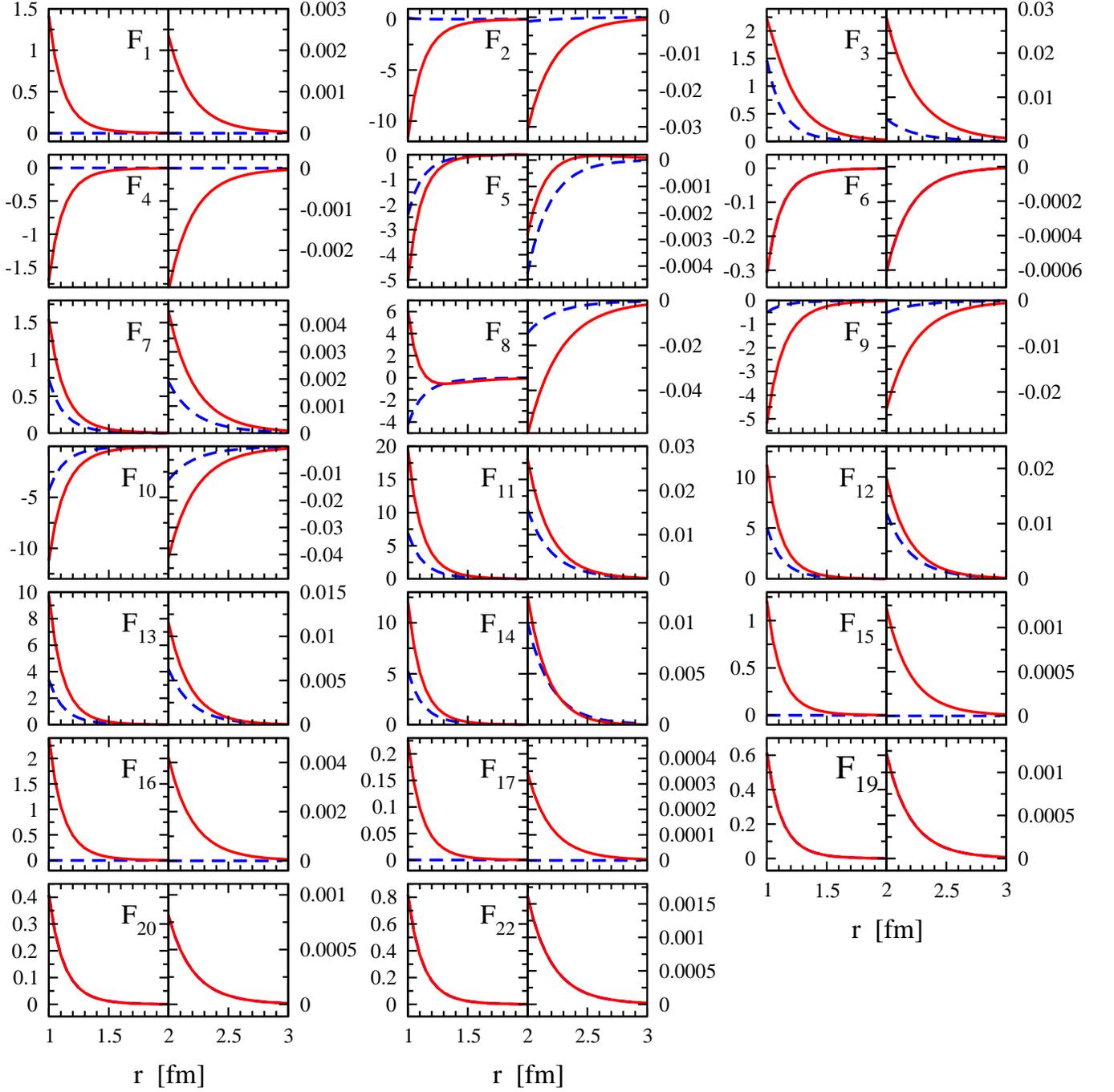}
    \caption{Chiral expansion of the profile functions ${\cal F}_i
      (r)$ in MeV generated by the ring  3NF topology up to
      N$^4$LO (in the equilateral triangle configuration). Dashed and solid lines correspond to
      ${\cal F}_i^{(4)}$ and  ${\cal F}_i^{(4)} + {\cal F}_i^{(5)}$,
      respectively. 
\label{fig:ringr} 
 }
%  \end{center}
\end{figure}

The convergence of the chiral expansion for the two-pion-one-pion exchange and ring
topologies is, as
expected, much worse, see Figs.~\ref{fig:TPEOPEr} and \ref{fig:ringr}.  
In nearly all cases, the subleading contributions at N$^4$LO dominate
over the nominally leading ones at N$^3$LO even at large
distances. This is analogous to the pattern
observed for the two-pion exchange two-nucleon potential. In that
case, the strong dominance of the subleading terms appears because of
several reasons including the large numerical coefficients, an
enhancement by one power of $\pi$ as compared to the standard chiral
power counting which is characteristic to the triangle diagrams, see
also Ref.~\cite{Baru:2012iv}, as well as the large numerical values of
the LECs $c_{3,4}$ from the subleading pion-nucleon effective
Lagrangian which are governed by the $\Delta$ isobar.  In the case of
the 3NF  $2\pi$-$1\pi$ exchange and ring potentials the situation is
less dramatic. In particular, the enhancement by a power of $\pi$
actually affects the leading contributions at N$^3$LO which involve 
the loop function $A(q_2)$. Still, the corrections at N$^4$LO are
large which can presumably be attributed to the large numerical values of the
LECs $c_i$.   One should, however, emphasize that the potentials
generated by the  $2\pi$-$1\pi$ exchange and ring diagrams have a
considerably shorter range as compared to the $2\pi$ exchange
ones and only reach at most $\sim 50$ keV at distances of the order
of $r\sim 2$ fm. It is, therefore, not clear whether the lack of convergence
will have any significant phenomenological impact. Clearly, to answer
this question one needs to carry out complete calculations of few- and
many-nucleon observables. This work is in progress.   
Last but not least, we emphasize that especially the ring topology
generates a very rich structure in the 3NF and  gives rise to $20$ out of
$22$ profile functions. 

\begin{figure}[tb]
\vskip 1 true cm
%  \begin{center} 
%    \epsfxsize=4.3cm
%    \epsffile{fig2.eps}
\includegraphics[width=0.95\textwidth,keepaspectratio,angle=0,clip]{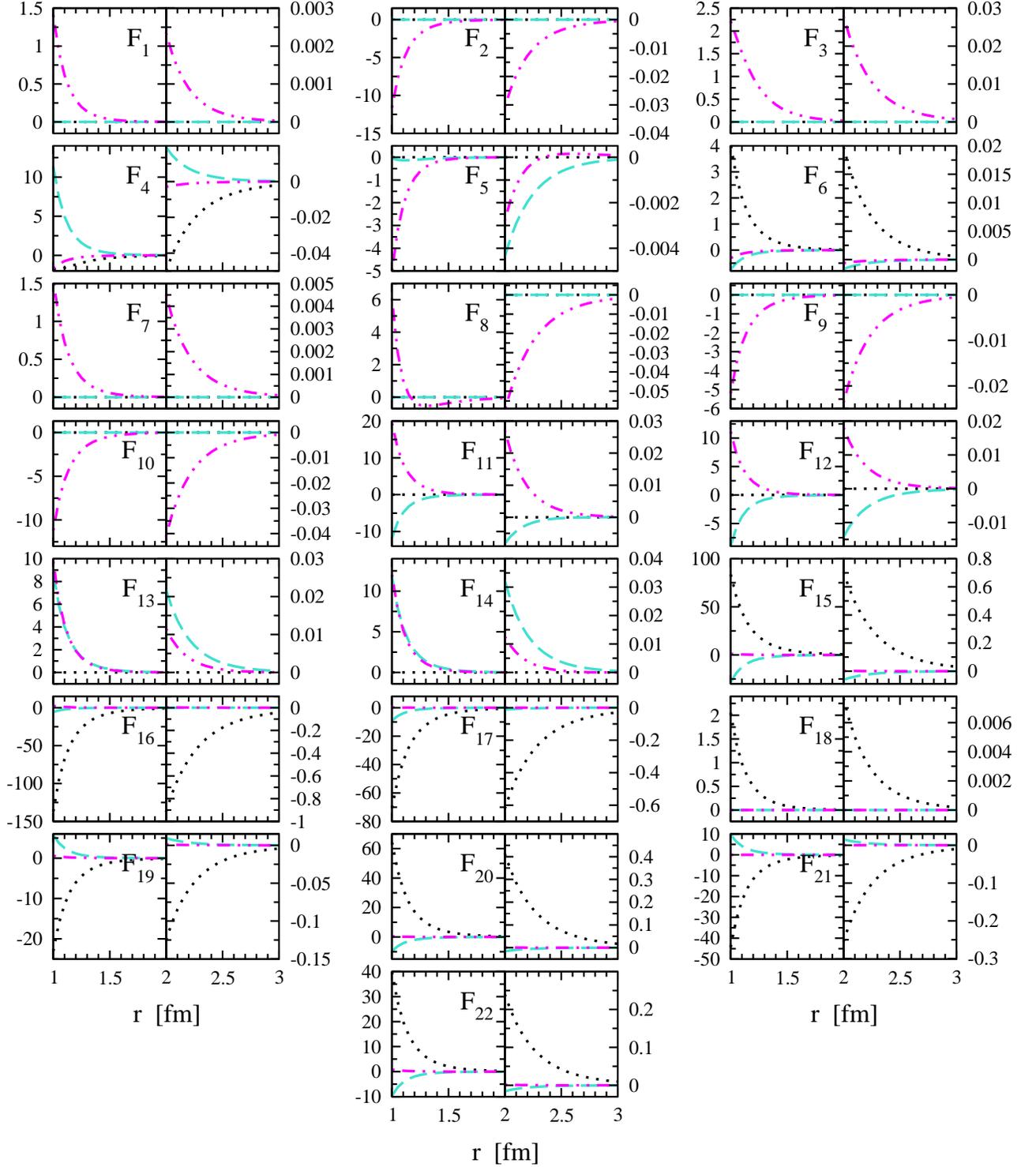}
    \caption{Individual contributions of the two-pion exchange (dotted
      lines), two-pion-one-pion exchange (long-dashed lines) and ring
      (dashed-double-dotted lines) topologies to the profile functions  ${\cal F}_i
      (r)$ in MeV at N$^4$LO in the equilateral triangle configuration. 
\label{fig:topologiesr} 
 }
%  \end{center}
\end{figure}

It is also instructive to compare the 3NF potentials corresponding to
the individual topologies with each other. 
Such a comparison is given in Fig.~\ref{fig:topologiesr},
where we restrict ourselves to N$^4$LO, i.e.~we only show ${\cal F}_i^{(3)} + {\cal F}_i^{(4)} + {\cal F}_i^{(5)}$ . We observe that the
$2\pi$-$1\pi$ exchange and ring potentials are  of a
comparable size. However, in all cases where the
longest-range $2\pi$ exchange topology
contribute, it clearly dominates  at $r\gtrsim 2 $ fm over the two
other topologies. 
At shorter distances of the order of $r \sim 1$ fm the impact of
the $2\pi$-$1\pi$  exchange and ring terms becomes more significant
with, e.g.~$|{\cal F}_{11, 15} (1\mbox{ fm}) | \sim 20$ MeV  to be
compared with the strongest $2\pi$ exchange potentials  
$|{\cal F}_{15, 16, 17} (1\mbox{ fm}) | \sim 100$ MeV.   As pointed
out before, it is difficult to draw conclusions on the
phenomenological importance of the new structures based on this
comparison alone since one generally expects that (scheme-dependent) short-range
contributions to the 3NF not considered in the present work would
become significant at $r \lesssim 1$ fm.   

\begin{figure}[tb]
\vskip 1 true cm
%  \begin{center} 
%    \epsfxsize=4.3cm
%    \epsffile{fig2.eps}
\includegraphics[width=0.95\textwidth,keepaspectratio,angle=0,clip]{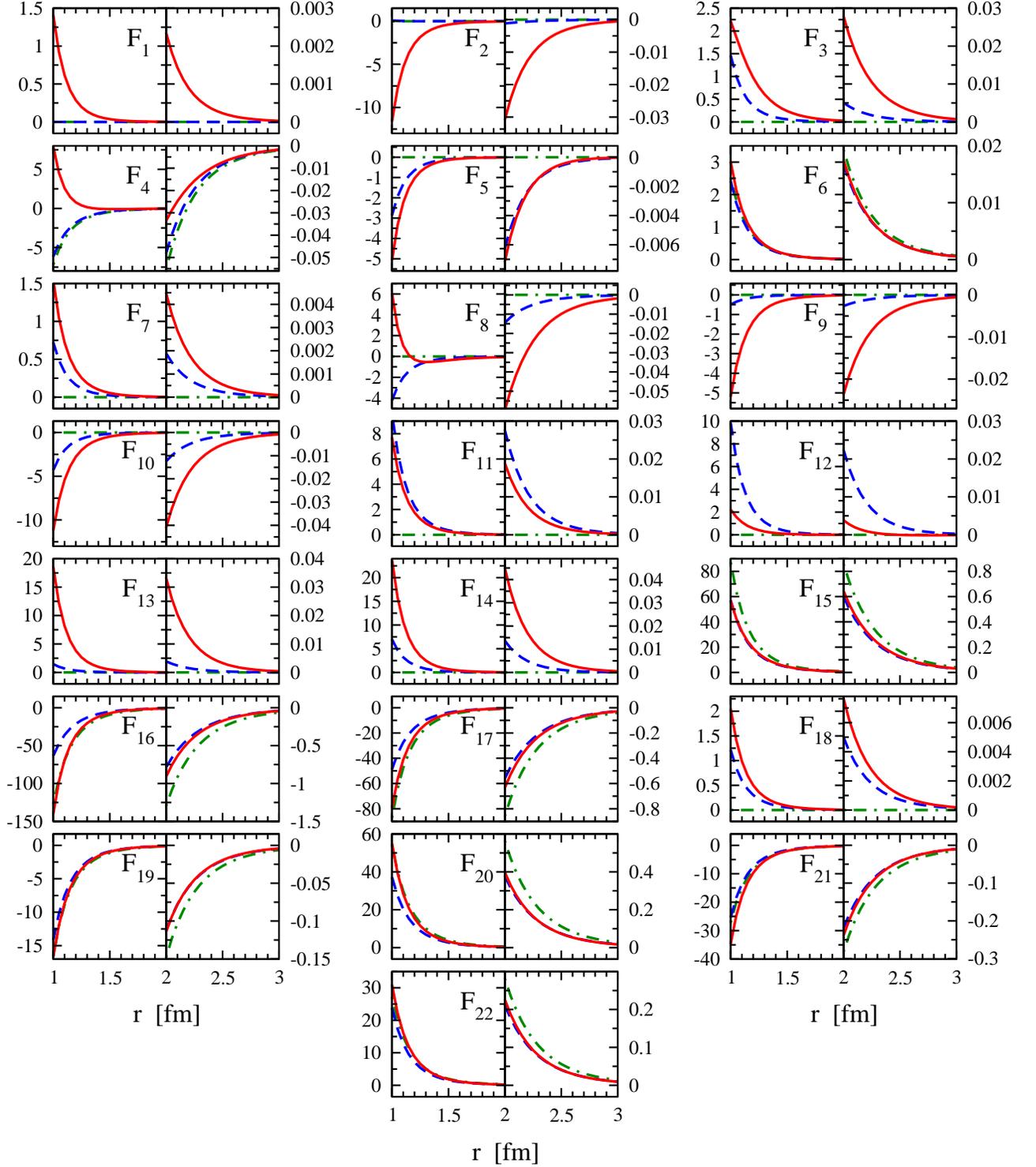}
    \caption{Chiral expansion of the profile functions ${\cal F}_i
      (r)$ in MeV emerging from all long-range 3NF topologies up to
      N$^4$LO (in the equilateral triangle configuration). Dashed-dotted, dashed and solid lines correspond to
      ${\cal F}_i^{(3)}$,   ${\cal F}_i^{(3)} + {\cal F}_i^{(4)}$ and
      ${\cal F}_i^{(3)} + {\cal F}_i^{(4)} + {\cal F}_i^{(5)}$,
      respectively. 
\label{fig:completer} 
 }
%  \end{center}
\end{figure}

Last but not least, Fig.~\ref{fig:completer} shows the chiral
expansion of the complete results for the functions ${\cal F}_{i} (r)$
emerging from adding the contributions from all three topologies together.  
The interpretation follows directly from the above discussion. At
long distances of the order of $r \gtrsim 2$ fm dominated by
the $2\pi$ exchange  one observes  a good convergence for all cases
where the potentials are numerically sizable. On the other hand, those profile
functions which are not affected by the $2\pi$ exchange are typically 
dominated by the N$^4$LO contributions which might be still not converged
at this order in the chiral expansion. The corresponding potentials
are, however, rather weak. At shorter distances $r \sim 1 \ldots 2$
fm, the  $2\pi$-$1\pi$ exchange and ring contributions start becoming more
important relative to the $2\pi$ exchange terms.  One again 
observes the dominance of the N$^4$LO contributions which supports the
assumption about the important role played by $\Delta$ excitations,
whose effects are partially taken into account at N$^4$LO through
resonance saturation of the LECs $c_2$, $c_3$ and $c_4$.

%%%%%%%%%%%%%%%%%%%%%%%%%%%%%%%%%%%%%%%%%%%%%%%%%%%%%%%%%%%%%%%%%%%%%%%%%%%%%%%%%
\section{Summary and outlook}
\def\theequation{\arabic{section}.\arabic{equation}}
\label{sec:summary}

In this paper, we have worked out and analyzed in detail the
intermediate-range contributions to the three-nucleon force  
at N$^4$LO, which emerge from the $2\pi$-$1\pi$ exchange and ring
topologies. We used here the heavy-baryon formulation of
chiral EFT with pions and nucleons being the only explicit degrees of
freedom. The pertinent results of our study can be summarized  
as follows. 
\begin{itemize}
\item
We worked out the coordinate-space representation of the N$^4$LO
corrections to the $2\pi$ exchange 3NF calculated in momentum space in
Ref.~\cite{Krebs:2012yv}. 
\item
We derived the N$^4$LO contributions to the intermediate-range
$2\pi$-$1\pi$ exchange and ring topologies. These represent the first
corrections to the leading results which appear at N$^3$LO and have
been worked out in Ref.~\cite{Bernard:2007sp}. We provide explicit
analytical expressions in both momentum and coordinate spaces.   
The obtained corrections do not involve any unknown low-energy
constants. 
\item 
We have demonstrated that the most general  structure of an
isospin-invariant local three-nucleon force involves 89 independent
isospin-spin-momentum operators. We proposed a set of 22
linearly-independent operators 
which can serve as a basis and gives rise to all 89 structures in the
3NF upon making permutations of  nucleon labels. We also discussed
the properties of the corresponding scalar structure functions ${\cal F}_{1
  \ldots 22}$ parametrizing the 3NF with respect to permutations. 
\item
Finally, using the above mentioned operator basis, we addressed the convergence of the chiral
expansion for the long-range tail of the 3NF in the equilateral
triangle configuration with $r_{12}=r_{23}=r_{31}=r$ by comparing our
predictions for the potentials ${\cal F}_{i}$ at different
orders in the chiral expansion.  Consistently with the 
momentum-space results of Ref.~\cite{Krebs:2012yv}, we observe a good
convergence for the longest-range $2\pi$ exchange topology which
clearly dominates the 3NF at distances of the order $r\gtrsim 2$ fm.    
The intermediate-range $2\pi$-$1\pi$ exchange and ring diagrams
provide sizable corrections to ${\cal F}_{i}$ at $r\sim 1$ fm and also
contribute to those $12$ profile functions which vanish for the $2\pi$
exchange. 
As expected, we found that N$^4$LO corrections to the
intermediate-range topologies are numerically large and in most cases
dominate over the nominally leading N$^3$LO terms. This can be traced
back to the role played by the $\Delta$(1232) isobar whose excitations
provide an important 3NF mechanism. In the standard, delta-less formulation
of chiral EFT we employ here, effects of the $\Delta$ isobar are not
incorporated in N$^3$LO contributions to the 3NF. For the
intermediate-range topologies we are primarily interested in here, 
first effects of the $\Delta$ appear at N$^4$LO through resonance
saturation of the LECs $c_2$,  $c_3$ and $c_4$ which accompany the 
subleading pion-nucleon vertices in the effective Lagrangian. The
importance of the $\Delta$ isobar is reflected in the numerically
large values of these LECs which are responsible for large N$^4$LO
corrections we observe.  
\end{itemize}

The results of our work provide important step towards precise,
quantitative theoretical description of the 3NF in the framework of
chiral EFT. The long-range part of the 3NF is governed by exchange of
pions, the Goldstone bosons of the spontaneously broken chiral
symmetry of QCD, and can be rigorously calculated in the framework of
chiral EFT. It is expected to affect the
energy dependence of the nucleon-deuteron scattering amplitude at low
energies and might be responsible for the long-standing puzzles such
as e.g.~the $A_y$ puzzle in elastic three- and four-nucleon scattering \cite{Viviani:2010mf}.
Although the resulting intermediate-range potentials are
significantly weaker than the $2\pi$ exchange terms, the appearance of
new structures might lead to large effects in certain 
nucleon-deuteron scattering observables.  It would be interesting in
the future to explore this possibility in a systematic way.     
Clearly, the N$^4$LO corrections to the short-range part of the 3NF
should also be worked out. This work is in progress. Notice that subleading
contributions to the three-nucleon contact interactions at N$^4$LO
are discussed in Ref.~\cite{Girlanda:2011fh}. Finally, the
large N$^4$LO corrections for the intermediate-range terms 
raise an obvious question in regard to whether the chiral expansion for these
quantities can be truncated at this order. One should especially keep
in mind that while the obtained N$^4$LO corrections do include some of the
$1/(m_\Delta - m_N)$ contributions through values of the LECs
$c_{2,3,4}$ and, in this sense, take into account physics associated
with intermediate excitation of  a single $\Delta$ isobar,  double and
triple $\Delta$ excitations start contributing only at orders N$^5$LO
and N$^6$LO, respectively. Phenomenological studies of Ref.~\cite{Pieper:2001ap}
indicate that at least double $\Delta$ excitations might induce 
sizable 3NFs. This issue must be investigated in the future. Rather
than calculating N$^5$LO and N$^6$LO corrections to the 3NF in the
delta-less formulation of chiral EFT, which correspond to the two-loop
level, it is more feasible and probably also more efficient to include
the $\Delta$ isobar as an explicit degree of freedom in the effective
Lagrangian, see Refs.~\cite{Krebs:2007rh,Ordonez:1995rz,Kaiser:1998wa,Epelbaum:2007sq,Epelbaum:2008td}  for some promising steps in this
directions. In such a delta-full formulation, the leading types of  
$1/(m_\Delta - m_N)$,  $1/(m_\Delta - m_N)^2$, $\ldots$ contributions
are resumed and the 3NF mechanisms associated with single, double and
triple intermediate $\Delta$-excitations are taken into account
already at N$^3$LO.  Work along these lines is underway.

%%%%%%%%%%%%%%%%%%%%%%%%%%%%%%%%%%%%%%%%%%%%%%%%%%%%%%%%%%%%%%%%%%%%%%%%%%%%%%%%%
\section*{Acknowledgments}

We are grateful to Ulf-G.~Mei{\ss}ner for many valuable comments on the manuscript. 
This work is supported by the EU HadronPhysics3 project ``Study of strongly interacting matter'', 
by the European Research Council (ERC-2010-StG 259218 NuclearEFT) and 
by the DFG (TR 16, ``Subnuclear Structure of Matter'').

%%%%%%%%%%%%%%%%%%%%%%%%%%%%%%%%%%%%%%%%%%%%%%%%%%%%%%%%%%%%%%%%%%%%%%%%%%%%%%%%%
\appendix

%%%%%%%%%%%%%%%%%%%%%%%%%%%%%%%%%%%%%%%%%%%%%%%%%%%%%%%%%%%%%%%%%%%%%%%%%%%%%%%%%
\def\theequation{\Alph{section}.\arabic{equation}}
\setcounter{equation}{0}
\section{Ring contributions in momentum space}
\label{app1}

In this appendix we give the lengthy expressions for subleading
contributions to ring diagrams  in momentum space. We employ here 
 the general parametrization of local three-body-force: 
\beqa
V_{{\rm ring}}&=&\sum_{i=1}^{22}{\cal G}_i \; {\cal R}_i (q_1, \, q_3,\, z),
\eeqa
where ${\cal G}_1, \; \dots\; , {\cal G}_{22}$ are the
spin-isospin-momentum operators which we defined in Table~\ref{generalstr}.
The N$^4$LO contributions to the structure functions ${\cal R}_i (q_1, \, q_3,\, z)$ with $z=\hat{q}_1\cdot\hat{q}_3$
proportional to $g_A^4$ are given by
\beqa
{\cal R}_1^{(5), g_A^4}&=&\frac{g_A^4 }{512 
F_\pi^6 \left(z^2-1\right)^2} I(4:0,-q_1,q_3;0) (c_2+c_3) \left(144 M_\pi^4 \left(z^2-1
\right)^2+8 M_\pi^2 \left(z^2-1\right) \left(16 z^2 \left(q_1^2+q_3^2
\right)\nnrlrl
-13 \left(q_1^2+q_3^2\right)+16 q_1 q_3 z^3-10 q_1 q_3 z
\right)+q_1^4 \left(32 z^4-32 z^2+9\right)+4 q_1^3 q_3 z \left(1-4 z^2
\right)^2\nnrl
+2 q_1^2 q_3^2 \left(80 z^4-78 z^2+25\right)+4 q_1 q_3^3 z 
\left(1-4 z^2\right)^2+q_3^4 \left(32 z^4-32
  z^2+9\right)\right)\nn
&&-\frac{g_A^4 }{12288 \pi^2 F_\pi^6 q_1 \left(z^2-1
\right)^2} L(q_3) (c_2+c_3) 
\left(4 M_\pi^2 \left(z^2-1\right) \left(q_1 \left(128
      z^2-101\right)+27 q_3 z
\right)\nnrl
+3 q_1^3 \left(32 z^4-32 z^2+9\right)+3 q_1^2 q_3 z \left(32 
z^4-5\right)+q_1 q_3^2 \left(320 z^4-412 z^2+173\right)+3 q_3^3 z 
\left(26 z^2-17\right)\right)\nn
&&-\frac{g_A^4 }{12288 \pi^2 F_\pi^6 q_3 \left(z^2-1
\right)^2} L(q_1) (c_2+c_3) \left(4 M_\pi^2 \left(z^2-1
\right) \left(27 q_1 z+q_3 \left(128 z^2-101\right)\right)\nnrl+q_1^3 
\left(78 z^3-51 z\right)
+q_1^2 q_3 \left(320 z^4-412 z^2+173\right)+3 
q_1 q_3^2 z \left(32 z^4-5\right)+3 q_3^3 \left(32 z^4-32 z^2+9
\right)\right)\nn
&&-\frac{g_A^4 }{12288 \pi^2 F_\pi^6 q_1 q_3 \left(z^2-1\right)^2} L(q_2) (c_2+c_3) \left(-4 M_\pi^2 \left(z^2-1
\right) \left(27 z \left(q_1^2+q_3^2\right)-74 q_1 q_3 z^2+128 q_1 q_3
\right)\nnrl
+q_1^4 \left(51 z-78 z^3\right)+4 q_1^3 q_3 \left(2 z^4-85 
z^2+56\right)+2 q_1^2 q_3^2 z \left(116 z^4-424 z^2+227\right)\nnrl
+4 q_1 
q_3^3 \left(2 z^4-85 z^2+56\right)+3 q_3^4 z \left(17-26 z^2\right)
\right)-\frac{3 
g_A^4 (c_2+c_3) \left(q_1^2+2 q_1 q_3 z+q_3^2\right)}{4096 \pi^2 
F_\pi^6 \left(z^2-1\right)},\nn
{\cal R}_2^{(5), g_A^4}&=&\frac{g_A^4}{256 F_\pi^6 \left(z^2-1
\right)^2} I(4:0,-q_1,q_3;0) \left(32 c_1 \left(z^2-1\right) \left(4 
\left(3 M_\pi^2+q_1^2+q_3^2\right) z^2+2 q_1 q_3 z\nnrlrl
-3 \left(4 
M_\pi^2+q_1^2+q_3^2\right)\right) M_\pi^2+c_3 \left(-240 \left(z^2-1
\right)^2 M_\pi^4-24 \left(z^2-1\right) \left(8 q_1 q_3 z^3+8 
\left(q_1^2+q_3^2\right) z^2\nnrlrlrl
-6 q_1 q_3 z-7 \left(q_1^2+q_3^2\right)
\right) M_\pi^2+q_1^4 \left(-32 z^4+48 z^2-19\right)+q_3^4 \left(-32 
z^4+48 z^2-19\right)\nnrlrl
-4 q_1 q_3^3 z \left(16 z^4-20 z^2+7\right)-4 
q_1^3 q_3 z \left(16 z^4-20 z^2+7\right)-2 q_1^2 q_3^2 \left(64 
z^4-90 z^2+35\right)\right)\right)\nn
&& +\frac{ g_A^4}{6144 
F_\pi^6 \pi^2 \left(4 M_\pi^2+q_1^2\right) q_3 \left(z^2-1\right)^2 
\left(-4 \left(z^2-1\right) M_\pi^2+q_1^2+q_3^2+2 q_1 q_3 z
\right)}L(q_1)\left.
%(
\nnrl
\times \big(
96 c_1
\left(z^2-1\right) \left(16 \left(z^2-1
\right) \left(3 q_1 z+q_3 \left(8 z^2-5\right)\right) M_\pi^4+4 \left(z 
\left(5 z^2-6\right) q_1^3+q_3 \left(12 z^4-29 z^2+14\right)
q_1^2\nnrlrlrl
+3 
q_3^2 z \left(3-4 z^2\right) q_1+q_3^3 \left(3-4 z^2\right)\right) 
M_\pi^2-q_1^2 \left(z q_1^3+q_3 \left(14 z^2-11\right) q_1^2+q_3^2 z 
\left(16 z^2-13\right) q_1\nnrlrlrl
+q_3^3 \left(4 z^2-3\right)\right)\right) 
M_\pi^2+c_3 \left(-64 \left(z^2-1\right)^2 \left(45 q_1 z+q_3 
\left(256 z^2-211\right)\right) M_\pi^6\nnrlrl
-16 \left(z^2-1\right) \left(3 
z \left(53 z^2-57\right) q_1^3+q_3 \left(848 z^4-1673 z^2+789\right) 
q_1^2\nnrlrlrl
+4 q_3^2 z \left(48 z^4-152 z^2+95\right) q_1+4 q_3^3 \left(48 
z^4-124 z^2+73\right)\right) M_\pi^4-4 \left(3 z \left(46 z^4-110 
z^2+61\right) q_1^5\nnrlrlrl
+q_3 \left(688 z^6-2546 z^4+2782 z^2-969\right) 
q_1^4+2 q_3^2 z \left(144 z^6-992 z^4+1357 z^2-554\right) q_1^3\nnrlrlrl
+2 
q_3^3 \left(48 z^6-640 z^4+929 z^2-382\right) q_1^2-3 q_3^4 z 
\left(96 z^4-128 z^2+47\right) q_1\nnrlrlrl
-3 q_3^5 \left(32 z^4-48 z^2+19
\right)\right) M_\pi^2+q_1^2 \left(q_1^2+2 q_3 z q_1+q_3^2\right) 
\left(\left(42 z^3-33 z\right) q_1^3\nnrlrlrl
+q_3 \left(496 z^4-884 z^2+415
\right) q_1^2+3 q_3^2 z \left(32 z^4-32 z^2+9\right) q_1+3 q_3^3 
\left(32 z^4-48 z^2+19\right)\right)\right)\right)\nn
&& -\frac{ 
g_A^4}{6144 F_\pi^6 \pi^2 q_1 q_3 \left(4 M_\pi^2+q_1^2+q_3^2+2 q_1 
q_3 z\right) \left(z^2-1\right)^2 \left(-4 \left(z^2-1\right) 
M_\pi^2+q_1^2+q_3^2+2 q_1 q_3 z\right)}L(q_2)
\left.
\nnrl
%\left
\times \big(
c_3 \left(-64 \left(z^2-1\right)^2 \left(-166 q_1 
q_3 z^2+45 \left(q_1^2+q_3^2\right) z+256 q_1 q_3\right) M_\pi^6+16 
\left(z^2-1\right) \left(3 z \left(57-53 z^2\right) q_1^4\nnrlrlrl
+4 q_3 
\left(53 z^4-281 z^2+240\right) q_1^3+2 q_3^2 z \left(434 z^4-1433 
z^2+1035\right) q_1^2+4 q_3^3 \left(53 z^4-281 z^2+240\right) q_1\nnrlrlrl
+3 
q_3^4 z \left(57-53 z^2\right)\right) M_\pi^4-4 \left(q_1^2+2 q_3 z 
q_1+q_3^2\right) \left(3 z \left(46 z^4-110 z^2+61\right) q_1^4\nnrlrlrl
-4 q_3 
\left(34 z^6-314 z^4+577 z^2-288\right) q_1^3+2 q_3^2 z \left(-268 
z^6+1574 z^4-2476 z^2+1143\right) q_1^2\nnrlrlrl
-4 q_3^3 \left(34 z^6-314 
z^4+577 z^2-288\right) q_1+3 q_3^4 z \left(46 z^4-110 z^2+61\right)
\right) M_\pi^2\nnrlrl
+\left(q_1^2+2 q_3 z q_1+q_3^2\right)^2 \left(\left(42 
z^3-33 z\right) q_1^4-4 q_3 \left(82 z^4-203 z^2+112\right) q_1^3\nnrlrlrl
-2 
q_3^2 z \left(364 z^4-824 z^2+433\right) q_1^2
-4 q_3^3 \left(82 
z^4-203 z^2+112\right) q_1+3 q_3^4 z \left(14 z^2-11\right)\right)
\right)\nnrl
-96 c_1 M_\pi^2 \left(z^2-1\right) \left(-16 \left(z^2-1\right) 
\left(3 z q_1^2+2 q_3 \left(8-5 z^2\right) q_1+3 q_3^2 z\right) 
M_\pi^4+4 \left(z \left(6-5 z^2\right) q_1^4\nnrlrlrl
+4 q_3 \left(z^4-8 z^2+8
\right) q_1^3+2 q_3^2 z \left(10 z^4-41 z^2+34\right) q_1^2+4 q_3^3 
\left(z^4-8 z^2+8\right) q_1+q_3^4 z \left(6-5 z^2\right)\right) 
M_\pi^2\nnrlrl
+\left(q_1^2+2 q_3 z q_1+q_3^2\right)^2 \left(z q_1^2+2 q_3 
\left(8-7 z^2\right) q_1+q_3^2 z\right)\right)\right) \nn
&&+\frac{ 
g_A^4}{6144 F_\pi^6 \pi^2 q_1 \left(4 M_\pi^2+q_3^2\right) 
\left(z^2-1\right)^2 \left(-4 \left(z^2-1\right) 
M_\pi^2+q_1^2+q_3^2+2 q_1 q_3 z\right)}L(q_3) \left.
%(
\nnrl
\times \big(
96 c_1 
\left(z^2-1\right) \left(16 \left(z^2-1\right) \left(3 q_3 z+q_1 
\left(8 z^2-5\right)\right) M_\pi^4-4 \left(\left(4 z^2-3\right) 
q_1^3+3 q_3 z \left(4 z^2-3\right) q_1^2\nnrlrlrl
+q_3^2 \left(-12 z^4+29 
z^2-14\right) q_1+q_3^3 z \left(6-5 z^2\right)\right) M_\pi^2-q_3^2 
\left(\left(4 z^2-3\right) q_1^3+q_3 z \left(16 z^2-13\right) 
q_1^2\nnrlrlrl
+q_3^2 \left(14 z^2-11\right) q_1+q_3^3 z\right)\right) 
M_\pi^2+c_3 \left(-64 \left(z^2-1\right)^2 \left(45 q_3 z+q_1 
\left(256 z^2-211\right)\right) M_\pi^6\nnrlrl
-16 \left(z^2-1\right) \left(4 
\left(48 z^4-124 z^2+73\right) q_1^3+4 q_3 z \left(48 z^4-152 z^2+95
\right) q_1^2\nnrlrlrl
+q_3^2 \left(848 z^4-1673 z^2+789\right) q_1+3 q_3^3 z 
\left(53 z^2-57\right)\right) M_\pi^4+4 \left(3 \left(32 z^4-48 
z^2+19\right) q_1^5\nnrlrlrl
+3 q_3 z \left(96 z^4-128 z^2+47\right) q_1^4+2 
q_3^2 \left(-48 z^6+640 z^4-929 z^2+382\right) q_1^3\nnrlrlrl
+2 q_3^3 z 
\left(-144 z^6+992 z^4-1357 z^2+554\right) q_1^2+q_3^4 \left(-688 
z^6+2546 z^4-2782 z^2+969\right) q_1\nnrlrlrl
-3 q_3^5 z \left(46 z^4-110 
z^2+61\right)\right) M_\pi^2+q_3^2 \left(q_1^2+2 q_3 z q_1+q_3^2
\right) \left(3 \left(32 z^4-48 z^2+19\right) q_1^3\nnrlrlrl
+3 q_3 z \left(32 
z^4-32 z^2+9\right) q_1^2+q_3^2 \left(496 z^4-884 z^2+415\right) 
q_1+3 q_3^3 z \left(14 z^2-11\right)\right)\right)\right)\nn
&&+\frac{c_3 \left(q_1^2+2 q_3 z 
q_1+q_3^2\right) g_A^4}{2048 F_\pi^6 \pi^2 \left(z^2-1\right)},\nn
{\cal R}_3^{(5), g_A^4}&=&-\frac{3 g_A^4 }{16 F_\pi^6 \left(z^2-1\right)} I(4:0,-q_1,q_3;0) c_3 q_1 q_3 (q_1 z+q_3) (q_1+q_3 
z)\nn
&&+\frac{3 g_A^4 }{128 
\pi^2 F_\pi^6 \left(z^2-1\right) \left(-4 M_\pi^2 \left(z^2-1
\right)+q_1^2+2 q_1 q_3 z+q_3^2\right)} q_1 L(q_1) \left(8 
c_1 M_\pi^2 \left(z^2-1\right) (q_1+q_3 z)\nnrl
+c_3 \left(-8 M_\pi^2 
\left(z^2-1\right) \left(q_1+q_3 z^3\right)-\left(q_1^2+2 q_1 q_3 
z+q_3^2\right) \left(2 q_1 z^2-3 q_1-q_3 z\right)\right)\right)\nn
&&+\frac{3 g_A^4 }{128 \pi^2 F_\pi^6 \left(z^2-1\right) \left(-4 M_\pi^2 
\left(z^2-1\right)+q_1^2+2 q_1 q_3 z+q_3^2\right)} q_3 L(q_3) 
\left(8 c_1 M_\pi^2 \left(z^2-1\right) (q_1 z+q_3)\nnrl
+c_3 
\left(\left(q_1^2+2 q_1 q_3 z+q_3^2\right) \left(q_1 z-2 q_3 z^2+3 
q_3\right)-8 M_\pi^2 \left(z^2-1\right) \left(q_1 z^3+q_3\right)
\right)\right)\nn
&&-\frac{3 g_A^4 }{128 \pi^2 F_\pi^6 \left(z^2-1\right) 
\left(4 M_\pi^2+q_1^2+2 q_1 q_3 z+q_3^2\right) \left(-4 M_\pi^2 
\left(z^2-1\right)+q_1^2+2 q_1 q_3 z+q_3^2\right)}
L(q_2) \left.
%(
\nnrl
\times \big(
8 c_1 M_\pi^2 \left(z^2-1\right) \left(4 M_\pi^2 
\left(q_1 \left(q_1-q_3 z^3+3 q_3 z\right)+q_3^2\right)+\left(q_1^2+2 
q_1 q_3 z+q_3^2\right)^2\right)\nnrl
+c_3 \left(-32 M_\pi^4 \left(z^2-1
\right) \left(q_1^2-q_1 q_3 z \left(z^2-3\right)+q_3^2\right)-4 M_\pi^2 
\left(q_1^2+2 q_1 q_3 z+q_3^2\right) \left(q_1^2 \left(4 z^2-5
\right)\nnrlrlrl
+q_1 q_3 z \left(-3 z^4+15 z^2-14\right)+q_3^2 \left(4 z^2-5
\right)\right)-\left(q_1^2+2 q_1 q_3 z+q_3^2\right)^2 \left(2 z^2 
\left(q_1^2+q_3^2\right)-3 \left(q_1^2+q_3^2\right)\nnrlrlrl
+5 q_1 q_3 z^3-7 
q_1 q_3 z\right)\right)\right),\nn
{\cal R}_4^{(5), g_A^4}&=&\frac{g_A^4 }{8 F_\pi^6 \left(z^2-1\right)} I(4:0,-q_1,q_3;0) q_1 q_3 (c_2+c_3) (q_1 z+q_3) (q_1+q_3 
z)\nn
&&+\frac{g_A^4 }{64 \pi^2 
F_\pi^6 \left(z^2-1\right)} L(q_2) (c_2+c_3) 
\left(q_1^2-q_1 q_3 z \left(z^2-3\right)+q_3^2\right)-\frac{g_A^4 q_1 L(q_1) (c_2+c_3) (q_1+q_3 
z)}{64 \pi^2 F_\pi^6 \left(z^2-1\right)}\nn
&&-\frac{g_A^4 q_3 L(q_3) 
(c_2+c_3) (q_1 z+q_3)}{64 \pi^2 F_\pi^6 \left(z^2-1\right)},\nn
{\cal R}_5^{(5), g_A^4}&=&-\frac{g_A^4 }{8 F_\pi^6 \left(z^2-1\right)} I(4:0,-q_1,q_3;0) c_4 q_1 z (q_1 z+q_3) \left(q_1^2+2 
q_1 q_3 z+q_3^2\right)\nn
&&+\frac{g_A^4 }{64 \pi^2 F_\pi^6 q_3 \left(z^2-1\right) \left(-4 
M_\pi^2 \left(z^2-1\right)+q_1^2+2 q_1 q_3 z+q_3^2
\right)} c_4 
q_1 L(q_1) \left(\left(q_1^2+2 q_1 q_3 z+q_3^2\right) \left(z 
\left(q_1^2+q_3^2\right)\nnrlrl
+4 q_1 q_3 z^2-2 q_1 q_3\right)-8 M_\pi^2 
\left(z^2-1\right) \left(q_1^2 z+q_1 q_3 \left(3 z^2-1\right)+q_3^2 
z^3\right)\right)\nn
&&+\frac{g_A^4 }{64 \pi^2 F_\pi^6 q_3 
\left(z^2-1\right) \left(-4 M_\pi^2 \left(z^2-1\right)+q_1^2+2 q_1 
q_3 z+q_3^2\right)} c_4 L(q_2) \left(8 M_\pi^2 \left(z^2-1\right) 
\left(q_1^3 z+q_1^2 q_3 \left(z^2+2\right)\nnrlrl
-q_1 q_3^2 z \left(z^2-4
\right)+q_3^3\right)-\left(q_1^2+2 q_1 q_3 z+q_3^2\right)^2 \left(q_1 
z+q_3 \left(3-2 z^2\right)\right)\right)\nn
&&+\frac{g_A^4 }{64 \pi^2 
F_\pi^6 \left(z^2-1\right) \left(4 M_\pi^2+q_3^2\right) \left(-4 
M_\pi^2 \left(z^2-1\right)+q_1^2+2 q_1 q_3 z+q_3^2\right)} c_4 L(q_3) \left(-32 M_\pi^4 
\left(z^2-1\right)\nnrl
\times \left(q_1^2 z^2+q_1 q_3 z \left(z^2+1\right)+q_3^2
\right)+4 M_\pi^2 \left(q_1^4 z^2+q_1^3 q_3 \left(3 z^3+z
\right)+q_1^2 q_3^2 \left(-z^4+5 z^2+2\right)\nnrlrl
-q_1 q_3^3 z \left(3 
z^4+z^2-8\right)+q_3^4 \left(5-4 z^2\right)\right)+q_3^2 
\left(q_1^2+2 q_1 q_3 z+q_3^2\right) \left(q_1^2 z^2+q_1 q_3 z 
\left(z^2+1\right)\nnrlrl
+q_3^2 \left(3-2 z^2\right)\right)\right),\nn
{\cal R}_7^{(5), g_A^4}&=&\frac{g_A^4}{128 F_\pi^6 q_1 q_3 
\left(z^2-1\right)^2} I(4:0,-q_1,q_3;0) c_4 z \left(q_1^2+2 q_3 z q_1+q_3^2\right) 
\left(4 \left(3 M_\pi^2+q_1^2+q_3^2\right) z^2+2 q_1 q_3 z\nnrl
-3 \left(4 
M_\pi^2+q_1^2+q_3^2\right)\right)\nn
&& -\frac{g_A^4}{3072 F_\pi^6 \pi^2 q_1 \left(4 M_\pi^2+q_1^2\right)
q_3^2 \left(z^2-1\right)^2 \left(-4 \left(z^2-1\right) 
M_\pi^2+q_1^2+q_3^2+2 q_1 q_3 z\right)} c_4 L(q_1)\left.
%(
\nnrl
\times \big(
-1024 (q_1+q_3 z) \left(z^2-1
\right)^2 M_\pi^6-16 \left(z^2-1\right) \left(\left(41 z^2-48\right) 
q_1^3+q_3 z \left(50 z^2-71\right) q_1^2\nnrlrl
+q_3^2 \left(24 z^4-29 z^2-16
\right) q_1+q_3^3 z \left(24 z^2-31\right)\right) M_\pi^4-4 
\left(\left(25 z^4-70 z^2+42\right) q_1^5\nnrlrl
+q_3 z \left(40 z^4-143 
z^2+88\right) q_1^4+q_3^2 \left(36 z^6-115 z^4+17 z^2+32\right) 
q_1^3+q_3^3 z \left(12 z^4-101 z^2+59\right) q_1^2\nnrlrl
+3 q_3^4 z^2 
\left(7-12 z^2\right) q_1+3 q_3^5 z \left(3-4 z^2\right)\right) 
M_\pi^2+q_1^2 \left(q_1^2+2 q_3 z q_1+q_3^2\right) \left(\left(13 
z^2-10\right) q_1^3\nnrlrl
+q_3 z \left(16 z^2-7\right) q_1^2+3 q_3^2 z^2 
\left(4 z^2-1\right) q_1+3 q_3^3 z \left(4 z^2-3\right)\right)\right) 
\nn
&& -\frac{ g_A^4}{3072 F_\pi^6 
\pi^2 q_1^2 q_3^2 \left(z^2-1\right)^2 \left(-4 \left(z^2-1\right) 
M_\pi^2+q_1^2+q_3^2+2 q_1 q_3 z\right)} c_4  L(q_2)\left(256 
\left(q_1^2+2 q_3 z q_1+q_3^2\right)\nnrl
\times \left(z^2-1\right)^2 M_\pi^4+4 
\left(z^2-1\right) \left(\left(25 z^2-32\right) q_1^4+4 q_3 z 
\left(13 z^2-20\right) q_1^3+2 q_3^2 \left(14 z^4-15 z^2-20\right) 
q_1^2\nnrlrl
+4 q_3^3 z \left(13 z^2-20\right) q_1+q_3^4 \left(25 z^2-32
\right)\right) M_\pi^2-\left(q_1^2+2 q_3 z q_1+q_3^2\right)^2 \left(2 
q_1 q_3 z^3+13 \left(q_1^2+q_3^2\right) z^2\nnrlrl
+4 q_1 q_3 z-10 
\left(q_1^2+q_3^2\right)\right)\right)\nn
&&-\frac{ g_A^4}{3072 F_\pi^6 \pi^2 q_1^2 q_3 \left(4 
M_\pi^2+q_3^2\right) \left(z^2-1\right)^2 \left(-4 \left(z^2-1\right) 
M_\pi^2+q_1^2+q_3^2+2 q_1 q_3 z\right)} c_4 L(q_3)\left.
%(
\nnrl
\times \big(
-1024 (q_3+q_1 
z) \left(z^2-1\right)^2 M_\pi^6-16 \left(z^2-1\right) \left(z 
\left(24 z^2-31\right) q_1^3+q_3 \left(24 z^4-29 z^2-16\right) 
q_1^2\nnrlrl
+q_3^2 z \left(50 z^2-71\right) q_1+q_3^3 \left(41 z^2-48\right)
\right) M_\pi^4+4 \left(3 z \left(4 z^2-3\right) q_1^5+3 q_3 z^2 
\left(12 z^2-7\right) q_1^4\nnrlrl
+q_3^2 z \left(-12 z^4+101 z^2-59\right) 
q_1^3-q_3^3 \left(36 z^6-115 z^4+17 z^2+32\right) q_1^2+q_3^4 z 
\left(-40 z^4+143 z^2-88\right) q_1\nnrlrl
+q_3^5 \left(-25 z^4+70 z^2-42
\right)\right) M_\pi^2+q_3^2 \left(q_1^2+2 q_3 z q_1+q_3^2\right) 
\left(3 z \left(4 z^2-3\right) q_1^3+3 q_3 z^2 \left(4 z^2-1\right) 
q_1^2\nnrlrl
+q_3^2 z \left(16 z^2-7\right) q_1+q_3^3 \left(13 z^2-10\right)
\right)\right) -\frac{c_4 \left(2 q_1 
q_3+\left(q_1^2+q_3^2\right) z\right) g_A^4}{3072 F_\pi^6 \pi^2 q_1 
q_3 \left(z^2-1\right)},\nn
{\cal R}_8^{(5), g_A^4}&=&\frac{3 g_A^4 }{16 F_\pi^6 q_1 \left(z^2-1
\right)^2} I(4:0,-q_1,q_3;0) q_3 \left(c_3 \left(4 z^3 \left(3 
M_\pi^2+q_1^2+q_3^2\right)-12 M_\pi^2 z-7 z \left(q_1^2+q_3^2
\right)+8 q_1 q_3 z^4\nnrlrl
-12 q_1 q_3 z^2-2 q_1 q_3\right)-16 c_1 M_\pi^2 z 
\left(z^2-1\right)\right)\nn
&&-\frac{3 g_A^4 }{128 \pi^2 F_\pi^6 q_1 \left(z^2-1\right)^2 \left(-4 M_\pi^2 
\left(z^2-1\right)+q_1^2+2 q_1 q_3 z+q_3^2\right)} L(q_1) \left(c_3 \left(4 M_\pi^2 \left(z^2-1
\right) \left(q_1 \left(z^2+4\right)\nnrlrlrl
+5 q_3 z\right)+\left(q_1^2+2 q_1 
q_3 z+q_3^2\right) \left(3 q_1 \left(z^2-2\right)+q_3 z \left(4 z^2-7
\right)\right)\right)-16 c_1 M_\pi^2 \left(z^2-1\right) (q_1+q_3 z)
\right)\nn
&&-\frac{3 g_A^4 }{128 \pi^2 
F_\pi^6 q_1^2 \left(z^2-1\right)^2 \left(-4 M_\pi^2 \left(z^2-1
\right)+q_1^2+2 q_1 q_3 z+q_3^2\right)} q_3 
L(q_3) \left(c_3 \left(4 M_\pi^2 \left(z^2-1\right) \left(5 q_1 z\nnrlrlrl
+q_3 
\left(z^2+4\right)\right)+\left(q_1^2+2 q_1 q_3 z+q_3^2\right) 
\left(q_1 z \left(4 z^2-7\right)+3 q_3 \left(z^2-2\right)\right)
\right)-16 c_1 M_\pi^2 \left(z^2-1\right) (q_1 z+q_3)\right)\nn
&&-\frac{3 g_A^4 }{128 \pi^2 F_\pi^6 q_1^2 
\left(z^2-1\right)^2 \left(4 M_\pi^2+q_1^2+2 q_1 q_3 z+q_3^2\right) 
\left(-4 M_\pi^2 \left(z^2-1\right)+q_1^2+2 q_1 q_3 z+q_3^2
\right)} L(q_2) \left.
%(
\nnrl
\times \big(
16 
c_1 M_\pi^2 \left(z^2-1\right) \left(4 M_\pi^2 \left(q_1 
\left(q_1-q_3 z^3+3 q_3 z\right)+q_3^2\right)+\left(q_1^2+2 q_1 q_3 
z+q_3^2\right)^2\right)\nnrl
+c_3 \left(-16 M_\pi^4 \left(z^2-1\right) 
\left(q_1^2 \left(z^2+4\right)-2 q_1 q_3 z \left(z^2-6\right)+q_3^2 
\left(z^2+4\right)\right)-4 M_\pi^2 \left(q_1^2+2 q_1 q_3 z+q_3^2
\right) \nnrlrl
\times\left(q_1^2 \left(z^4+6 z^2-10\right)-2 q_1 q_3 z \left(2 
z^4-13 z^2+14\right)+q_3^2 \left(z^4+6 z^2-10\right)
\right)\nnrlrl
-\left(q_1^2+2 q_1 q_3 z+q_3^2\right)^2 \left(3 q_1^2 
\left(z^2-2\right)+2 q_1 q_3 z \left(4 z^2-7\right)+3 q_3^2 
\left(z^2-2\right)\right)\right)\right)+\frac{3 g_A^4 c_3 q_3 z}{128 \pi^2 F_\pi^6 q_1 \left(z^2-1
\right)},\nn
{\cal R}_9^{(5), g_A^4}&=&\frac{3 g_A^4 }{32 F_\pi^6 \left(z^2-1\right)^2} I(4:0,-q_1,q_3;0) \left(16 c_1
    M_\pi^2 \left(z^2-1
\right)+c_3 \left(-12 M_\pi^2 \left(z^2-1\right)-2 z^2 \left(q_1^2+3 
q_1 q_3 z+q_3^2\right)\nnrlrl
+5 \left(q_1^2+q_3^2\right)+12 q_1 q_3 z\right)
\right)\nn
&&-\frac{3 g_A^4 }{256 \pi^2 F_\pi^6 q_3 
\left(z^2-1\right)^2 \left(-4 M_\pi^2 \left(z^2-1\right)+q_1^2+2 q_1 
q_3 z+q_3^2\right)} L(q_1) 
\left(16 c_1 M_\pi^2 z \left(z^2-1\right) (q_1+q_3 z)\nnrl
+c_3 \left(-4 
M_\pi^2 \left(z^2-1\right) \left(5 q_1 z+q_3 \left(4 z^4-4 z^2+5\right)
\right)-\left(q_1^2+2 q_1 q_3 z+q_3^2\right)\nnrlrl
\times \left(4 q_1 z^3-7 q_1 
z+2 q_3 z^2-5 q_3\right)\right)\right)\nn
&&-\frac{3 g_A^4 }{256 \pi^2 F_\pi^6 q_1 
\left(z^2-1\right)^2 \left(-4 M_\pi^2 \left(z^2-1\right)+q_1^2+2 q_1 
q_3 z+q_3^2\right)} L(q_3) \left(16 c_1 M_\pi^2 z 
\left(z^2-1\right) (q_1 z+q_3)\nnrl
+c_3 \left(-4 M_\pi^2 \left(z^2-1
\right) \left(q_1 \left(4 z^4-4 z^2+5\right)+5 q_3 z
\right)-\left(q_1^2+2 q_1 q_3 z+q_3^2\right)\nnrlrl
\times \left(2 q_1 z^2-5 q_1+4 
q_3 z^3-7 q_3 z\right)\right)\right)\nn
&&+\frac{3 g_A^4 }{256 \pi^2 F_\pi^6 q_1 q_3 \left(z^2-1
\right)^2 \left(4 M_\pi^2+q_1^2+2 q_1 q_3 z+q_3^2\right) \left(-4 
M_\pi^2 \left(z^2-1\right)+q_1^2+2 q_1 q_3 z+q_3^2\right)} z L(q_2)
\left.
%(
\nnrl
\times \big(
16 c_1 M_\pi^2 
\left(z^2-1\right) \left(4 M_\pi^2 \left(q_1 \left(q_1-q_3 z^3+3 q_3 
z\right)+q_3^2\right)+\left(q_1^2+2 q_1 q_3 z+q_3^2\right)^2
\right)\nnrl
+c_3 \left(-16 M_\pi^4 \left(z^2-1\right) \left(5 
\left(q_1^2+q_3^2\right)-4 q_1 q_3 z^3+14 q_1 q_3 z\right)-4 M_\pi^2 
\left(3 z^2-4\right) \left(q_1^2+2 q_1 q_3 z+q_3^2\right)\nnrlrl
\times \left(3 
\left(q_1^2+q_3^2\right)-2 q_1 q_3 z^3+8 q_1 q_3 z
\right)-\left(q_1^2+2 q_1 q_3 z+q_3^2\right)^2 \left(4 z^2 
\left(q_1^2+q_3^2\right)-7 \left(q_1^2+q_3^2\right)\nnrlrlrl
+10 q_1 q_3 z^3-16 
q_1 q_3 z\right)\right)\right)-\frac{3 
g_A^4 c_3}{256 \pi^2 F_\pi^6 \left(z^2-1\right)},\nn
{\cal R}_{10}^{(5), g_A^4}&=&\frac{3 g_A^4 }{32 F_\pi^6 \left(z^2-1\right)^2} I(4:0,-q_1,q_3;0) z
  \left(16 c_1 M_\pi^2 z \left(z^2-1
\right)+c_3 \left(-4 z^3 \left(3 M_\pi^2+q_1^2+q_3^2\right)+12 M_\pi^2 
z\nnrlrl
+7 z \left(q_1^2+q_3^2\right)-8 q_1 q_3 z^4+12 q_1 q_3 z^2+2 q_1 q_3
\right)\right)\nn
&&-\frac{3 g_A^4 }{256 \pi^2 F_\pi^6 q_3 \left(z^2-1\right)^2 \left(-4 M_\pi^2 
\left(z^2-1\right)+q_1^2+2 q_1 q_3 z+q_3^2\right)} z 
L(q_1) \left(16 c_1 M_\pi^2 \left(z^2-1\right) (q_1+q_3 z)\nnrl
+c_3 
\left(-4 M_\pi^2 \left(z^2-1\right) \left(q_1 \left(z^2+4\right)+5 
q_3 z\right)-\left(q_1^2+2 q_1 q_3 z+q_3^2\right) \left(3 q_1 
\left(z^2-2\right)+q_3 z \left(4 z^2-7\right)\right)\right)
\right)\nn
&&-\frac{3 g_A^4 }{256 \pi^2 
F_\pi^6 q_1 \left(z^2-1\right)^2 \left(-4 M_\pi^2 \left(z^2-1
\right)+q_1^2+2 q_1 q_3 z+q_3^2\right)} z 
L(q_3) \left(16 c_1 M_\pi^2 \left(z^2-1\right) (q_1 z+q_3)\nnrl
+c_3 
\left(-4 M_\pi^2 \left(z^2-1\right) \left(5 q_1 z+q_3 \left(z^2+4
\right)\right)-\left(q_1^2+2 q_1 q_3 z+q_3^2\right) \left(q_1 z \left(4 
z^2-7\right)+3 q_3 \left(z^2-2\right)\right)\right)\right)\nn
&&+\frac{3 g_A^4 }{256 \pi^2 F_\pi^6 q_1 q_3 
\left(z^2-1\right)^2 \left(4 M_\pi^2+q_1^2+2 q_1 q_3 z+q_3^2\right) 
\left(-4 M_\pi^2 \left(z^2-1\right)+q_1^2+2 q_1 q_3 z+q_3^2
\right)} z L(q_2) \left.
%(
\nnrl
\times \big(
16 
c_1 M_\pi^2 \left(z^2-1\right) \left(4 M_\pi^2 \left(q_1 
\left(q_1-q_3 z^3+3 q_3 z\right)+q_3^2\right)+\left(q_1^2+2 q_1 q_3 
z+q_3^2\right)^2\right)\nnrl
+c_3 \left(-16 M_\pi^4 \left(z^2-1\right) 
\left(q_1^2 \left(z^2+4\right)-2 q_1 q_3 z \left(z^2-6\right)+q_3^2 
\left(z^2+4\right)\right)-4 M_\pi^2 \left(q_1^2+2 q_1 q_3 z+q_3^2
\right)\nnrlrl
\times \left(q_1^2 \left(z^4+6 z^2-10\right)-2 q_1 q_3 z \left(2 
z^4-13 z^2+14\right)+q_3^2 \left(z^4+6 z^2-10\right)
\right)\nnrlrl
-\left(q_1^2+2 q_1 q_3 z+q_3^2\right)^2 \left(3 q_1^2 
\left(z^2-2\right)+2 q_1 q_3 z \left(4 z^2-7\right)+3 q_3^2 
\left(z^2-2\right)\right)\right)\right)-\frac{3 g_A^4 c_3}{256 \pi^2
F_\pi^6 \left(z^2-1\right)},\nn
{\cal R}_{11}^{(5), g_A^4}&=&-\frac{g_A^4 }{16 F_\pi^6 q_1 \left(z^2-1\right)^2} I(4:0,-q_1,q_3;0) c_4 z (q_1 z+q_3) \left(-12 M_\pi^2 
\left(z^2-1\right)+3 q_1^2+6 q_1 q_3 z+q_3^2 \left(7-4 z^2\right)
\right)\nn
&&-\frac{g_A^4 }{128 
\pi^2 F_\pi^6 q_1 q_3 \left(z^2-1\right)^2 \left(-4 M_\pi^2 
\left(z^2-1\right)+q_1^2+2 q_1 q_3 z+q_3^2\right)} c_4 L(q_1) 
\left(4 M_\pi^2 \left(z^2-1\right) \left(q_1^2 z
    \left(z^2+4\right)\nnrlrl
-2 
q_1 q_3 \left(2 z^4-9 z^2+2\right)+5 q_3^2 z\right)-\left(q_1^2+2 q_1 
q_3 z+q_3^2\right) \left(q_1^2 z \left(z^2+2\right)-2 q_1 q_3 \left(2 
z^4-7 z^2+2\right)\nnrlrl
+q_3^2 z \left(7-4 z^2\right)\right)\right)\nn
&&-\frac{g_A^4 }{128 \pi^2 F_\pi^6 q_1^2 q_3 \left(z^2-1\right)^2 
\left(-4 M_\pi^2 \left(z^2-1\right)+q_1^2+2 q_1 q_3 z+q_3^2
\right)} c_4 
L(q_2) \left(\left(q_1^2+2 q_1 q_3 z+q_3^2\right) \nnrl
\times\left(q_1^3 z 
\left(z^2+2\right)+3 q_1^2 q_3 \left(2 z^4-z^2+2\right)-q_1 q_3^2 z 
\left(z^2-10\right)-3 q_3^3 \left(z^2-2\right)\right)\nnrl
-4 M_\pi^2 
\left(z^2-1\right) \left(q_1^3 z \left(z^2+4\right)+q_1^2 q_3 \left(6 
z^4+z^2+8\right)+3 q_1 q_3^2 z \left(z^2+4\right)+q_3^3 \left(z^2+4
\right)\right)\right)\nn
&&-\frac{g_A^4 }{128 \pi^2 F_\pi^6 q_1^2 \left(z^2-1\right)^2 \left(4 
M_\pi^2+q_3^2\right) \left(-4 M_\pi^2 \left(z^2-1\right)+q_1^2+2 q_1
q_3 z+q_3^2\right)} c_4 L(q_3) \left(16 M_\pi^4 \left(z^2-1\right) \nnrl
\times\left(5 q_1^2 z^2+5 q_1 q_3 z \left(z^2+1\right)+q_3^2 \left(z^2+4
\right)\right)-4 M_\pi^2 \left(3 q_1^4 z^2+3 q_1^3 q_3 \left(3 z^3+z
\right)+q_1^2 q_3^2 \left(-z^4+15 z^2+4\right)\nnrlrl
+q_1 q_3^3 z \left(-7 
z^4+z^2+18\right)-q_3^4 \left(z^4+6 z^2-10\right)\right)-3 q_3^2 
\left(q_1^2+2 q_1 q_3 z+q_3^2\right) \left(q_1^2 z^2+q_1 q_3 z 
\left(z^2+1\right)\nnrlrl
-q_3^2 \left(z^2-2\right)\right)
\right)+\frac{g_A^4 c_4 (q_1+q_3 z)}{128 \pi^2 F_\pi^6 
q_1 \left(z^2-1\right)},\nn
{\cal R}_{12}^{(5), g_A^4}&=&-\frac{g_A^4 }{16 F_\pi^6 q_3 \left(z^2-1\right)^2} I(4:0,-q_1,q_3;0) c_4 z
  \left(12 M_\pi^2 \left(z^2-1
\right) (q_1+q_3 z)+q_1^3 \left(-\left(2 z^2+1\right)\right)\nnrl
-q_1^2 q_3 
z \left(4 z^2+5\right)-q_1 q_3^2 \left(4 z^2+5\right)+q_3^3 z \left(4 
z^2-7\right)\right)\nn
&&-\frac{g_A^4 }{128 \pi^2 F_\pi^6 q_3^2 
\left(z^2-1\right)^2 \left(-4 M_\pi^2 \left(z^2-1\right)+q_1^2+2 q_1 
q_3 z+q_3^2\right)}
c_4 z L(q_1) \left(\left(q_1^2+2 q_1 q_3 z+q_3^2\right) \nnrl
\times\left(3 q_1^2 
z+q_1 \left(5 q_3 z^2+q_3\right)+q_3^2 z \left(7-4 z^2\right)
\right)-4 M_\pi^2 \left(z^2-1\right) \left(5 z \left(q_1^2+q_3^2
\right)+9 q_1 q_3 z^2+q_1 q_3\right)\right)\nn
&&-\frac{g_A^4 }{128 \pi^2 F_\pi^6 q_1 q_3^2 \left(z^2-1\right)^2 \left(-4 
M_\pi^2 \left(z^2-1\right)+q_1^2+2 q_1 q_3 z+q_3^2
\right)} c_4 z L(q_2) \left(q_1^2+2 q_1 q_3 
z+q_3^2\right) \nn
&&\times\left(4 M_\pi^2 \left(z^2-1\right) \left(5 q_1 z+q_3 
\left(z^2+4\right)\right)-3 q_1^3 z-q_1^2 q_3 \left(7 z^2+2
\right)+q_1 q_3^2 z \left(2 z^2-11\right)+3 q_3^3 \left(z^2-2\right)
\right)\nn
&&-\frac{g_A^4 }{128 
\pi^2 F_\pi^6 q_1 q_3 \left(z^2-1\right)^2 \left(4 M_\pi^2+q_3^2
\right) \left(-4 M_\pi^2 \left(z^2-1\right)+q_1^2+2 q_1 q_3 z+q_3^2
\right)} c_4 z L(q_3) \left.
%(
\nnrl
\times \big(
-16 M_\pi^4 \left(z^2-1\right) 
\left(q_1^2 \left(4 z^2+1\right)+2 q_1 q_3 z \left(2 z^2+3
\right)+q_3^2 \left(z^2+4\right)\right)+4 M_\pi^2 \left(q_1^4 \left(2 
z^2+1\right)\nnrlrl
+6 q_1^3 q_3 z \left(z^2+1\right)-2 q_1^2 q_3^2 
\left(z^4-7 z^2-3\right)-2 q_1 q_3^3 z \left(3 z^4+z^2-10
\right)-q_3^4 \left(z^4+6 z^2-10\right)\right)\nnrl
+q_3^2 \left(q_1^2+2 q_1 
q_3 z+q_3^2\right) \left(q_1^2 \left(2 z^2+1\right)+2 q_1 q_3 z 
\left(z^2+2\right)-3 q_3^2 \left(z^2-2\right)\right)\right)-\frac{g_A^4 c_4 (q_1 z+q_3)}{128 \pi^2 F_\pi^6 q_3 \left(z^2-1
\right)},\nn
{\cal R}_{13}^{(5), g_A^4}&=&-\frac{g_A^4 }{16 F_\pi^6 q_3 \left(z^2-1
\right)^2} I(4:0,-q_1,q_3;0) c_4 (q_1 z+q_3) \left(12 M_\pi^2 
\left(z^2-1\right)-q_1^2 \left(2 z^2+1\right)\nnrl
-2 q_1 q_3 \left(2 z^3+z
\right)+q_3^2 \left(2 z^2-5\right)\right)\nn
&&-\frac{g_A^4 }{128 \pi^2 F_\pi^6 q_3^2 \left(z^2-1\right)^2 
\left(-4 M_\pi^2 \left(z^2-1\right)+q_1^2+2 q_1 q_3 z+q_3^2
\right)} c_4 L(q_1) \left(\left(q_1^2+2 q_1 q_3 z+q_3^2
\right) \nnrl
\times\left(3 q_1^2 z^2+2 q_1 q_3 \left(2 z^3+z\right)+q_3^2 
\left(5-2 z^2\right)\right)-4 M_\pi^2 \left(z^2-1\right) \left(5 
q_1^2 z^2+2 q_1 q_3 \left(4 z^3+z\right)\nnrlrl
+q_3^2 \left(4 z^4-4 z^2+5
\right)\right)\right)\nn
&&-\frac{g_A^4 }{128 \pi^2 F_\pi^6 q_1 q_3^2 \left(z^2-1\right)^2 \left(-4 
M_\pi^2 \left(z^2-1\right)+q_1^2+2 q_1 q_3 z+q_3^2
\right)} c_4 L(q_2) \left(4 M_\pi^2 \left(z^2-1\right) 
\left(5 q_1^3 z^2\nnrlrl
+5 q_1^2 q_3 \left(2 z^3+z\right)+q_1 q_3^2 \left(-4 
z^4+23 z^2-4\right)+5 q_3^3 z\right)-\left(q_1^2+2 q_1 q_3 z+q_3^2
\right) \left(3 q_1^3 z^2+3 q_1^2 q_3 \left(2 z^3+z\right)\nnrlrl
+q_1 q_3^2 
\left(-8 z^4+21 z^2-4\right)+q_3^3 z \left(7-4 z^2\right)\right)
\right)\nn
&&-\frac{g_A^4 }{128 \pi^2 F_\pi^6 
q_1 q_3 \left(z^2-1\right)^2 \left(4 M_\pi^2+q_3^2\right) \left(-4 
M_\pi^2 \left(z^2-1\right)+q_1^2+2 q_1 q_3 z+q_3^2\right)} c_4 L(q_3) \left(-16 M_\pi^4 \left(z^2-1\right) 
\nnrl
\times\left(q_1^2 \left(4 z^3+z\right)+q_1 q_3 \left(4 z^4+5 z^2+1\right)+5 
q_3^2 z\right)+4 M_\pi^2 \left(q_1^4 \left(2 z^3+z\right)+q_1^3 q_3 
\left(6 z^4+5 z^2+1\right)\nnrlrl
+q_1^2 q_3^2 z \left(-2 z^4+11 z^2+9
\right)+q_1 q_3^3 \left(-6 z^6-3 z^4+19 z^2+2\right)+3 q_3^4 z 
\left(4-3 z^2\right)\right)\nnrl
+q_3^2 \left(q_1^2+2 q_1 q_3 z+q_3^2
\right) \left(q_1^2 \left(2 z^3+z\right)+q_1 q_3 \left(2 z^4+3 z^2+1
\right)+q_3^2 z \left(7-4 z^2\right)\right)\right)\nn
&&-\frac{g_A^4 
c_4 (q_1 z+q_3)}{128 \pi^2 F_\pi^6 q_3 \left(z^2-1\right)},\nn
{\cal R}_{14}^{(5), g_A^4}&=&-\frac{g_A^4 }{16 F_\pi^6 q_3^2 \left(z^2-1\right)^2} I(4:0,-q_1,q_3;0) c_4 q_1
  \left(-12 M_\pi^2 \left(z^2-1
\right) (q_1+q_3 z)+q_1^3 \left(2 z^2+1\right)\nnrl
+q_1^2 q_3 z \left(4 
z^2+5\right)+q_1 q_3^2 \left(4 z^2+5\right)+q_3^3 z \left(7-4 z^2
\right)\right)\nn
&&+\frac{g_A^4 }{128 \pi^2 F_\pi^6 q_3^3 
\left(z^2-1\right)^2 \left(-4 M_\pi^2 \left(z^2-1\right)+q_1^2+2 q_1 
q_3 z+q_3^2\right)} c_4 
q_1 L(q_1) \left(\left(q_1^2+2 q_1 q_3 z+q_3^2\right) \nnrl
\times\left(3 q_1^2 
z+q_1 \left(5 q_3 z^2+q_3\right)+q_3^2 z \left(7-4 z^2\right)
\right)-4 M_\pi^2 \left(z^2-1\right) \left(5 z \left(q_1^2+q_3^2
\right)+9 q_1 q_3 z^2+q_1 q_3\right)\right)\nn
&&-\frac{g_A^4 }{128 \pi^2 F_\pi^6 q_3^3 \left(z^2-1\right)^2 \left(-4 M_\pi^2 
\left(z^2-1\right)+q_1^2+2 q_1 q_3 z+q_3^2\right)} c_4 L(q_2) \left(q_1^2+2 q_1 q_3 
z+q_3^2\right) \nn
&&\times\left(-4 M_\pi^2 \left(z^2-1\right) \left(5 q_1 z+q_3 
\left(z^2+4\right)\right)+3 q_1^3 z+q_1^2 q_3 \left(7 z^2+2
\right)+q_1 q_3^2 z \left(11-2 z^2\right)-3 q_3^3 \left(z^2-2\right)
\right)\nn
&&-\frac{g_A^4 }{128 \pi^2 F_\pi^6 q_3^2 \left(z^2-1\right)^2 
\left(4 M_\pi^2+q_3^2\right) \left(-4 M_\pi^2 \left(z^2-1
\right)+q_1^2+2 q_1 q_3 z+q_3^2\right)} c_4 
L(q_3) \left(16 M_\pi^4 \left(z^2-1\right) \nnrl
\times\left(q_1^2 \left(4 z^2+1
\right)+2 q_1 q_3 z \left(2 z^2+3\right)+q_3^2 \left(z^2+4\right)
\right)-4 M_\pi^2 \left(q_1^4 \left(2 z^2+1\right)+6 q_1^3 q_3 z 
\left(z^2+1\right)\nnrlrl
-2 q_1^2 q_3^2 \left(z^4-7 z^2-3\right)-2 q_1 q_3^3 
z \left(3 z^4+z^2-10\right)-q_3^4 \left(z^4+6 z^2-10\right)
\right)-q_3^2 \left(q_1^2+2 q_1 q_3 z+q_3^2\right) \nnrl
\times\left(q_1^2 \left(2 
z^2+1\right)+2 q_1 q_3 z \left(z^2+2\right)-3 q_3^2 \left(z^2-2
\right)\right)\right)+\frac{g_A^4 c_4 q_1 (q_1+q_3 
z)}{128 \pi^2 F_\pi^6 q_3^2 \left(z^2-1\right)},\nn
{\cal R}_{15}^{(5), g_A^4}&=&-\frac{g_A^4 }{16 F_\pi^6 \left(z^2-1\right)^2} I(4:0,-q_1,q_3;0) (c_2+c_3) \left(2 z^2 \left(-2 
M_\pi^2+q_1^2+q_3^2\right)+4 M_\pi^2+q_1^2\nnrl
+2 q_1 q_3 z^3+4 q_1 q_3 
z+q_3^2\right)\nn
&&-\frac{g_A^4 z L(q_2) 
(c_2+c_3) \left(3 q_1^2-2 q_1 q_3 z \left(z^2-4\right)+3 q_3^2
\right)}{128 \pi^2 F_\pi^6 q_1 q_3 \left(z^2-1\right)^2}+\frac{g_A^4 
L(q_1) (c_2+c_3) \left(3 q_1 z+2 q_3 z^2+q_3\right)}{128 \pi^2 
F_\pi^6 q_3 \left(z^2-1\right)^2}\nn
&&+\frac{g_A^4 L(q_3) (c_2+c_3) 
\left(2 q_1 z^2+q_1+3 q_3 z\right)}{128 \pi^2 F_\pi^6 q_1 \left(z^2-1
\right)^2}+\frac{g_A^4 (c_2+c_3)}{128 \pi^2 F_\pi^6 \left(z^2-1
\right)},\nn
{\cal R}_{16}^{(5), g_A^4}&=&-\frac{g_A^4 }{8 F_\pi^6 q_1 q_3 \left(z^2-1\right)^2} I(4:0,-q_1,q_3;0) z (c_2+c_3) \left(q_1^2+2 q_1 q_3 
z+q_3^2\right)\nn
&&\times \left(-4 M_\pi^2 \left(z^2-1\right)+3 q_1 (q_1+2 q_3 
z)+2 q_3^2 z^2+q_3^2\right)\nn
&&-
\frac{g_A^4 L(q_2) (c_2+c_3) \left(q_1^2+2 q_1 q_3 z+q_3^2\right) 
\left(q_1^2 \left(z^2+2\right)+2 q_1 q_3 z \left(z^2+2\right)+3 q_3^2 
z^2\right)}{64 \pi^2 F_\pi^6 q_1^2 q_3^2 \left(z^2-1
\right)^2}\nn
&&+\frac{g_A^4 L(q_1) (c_2+c_3) \left(q_1^3 \left(z^2+2
\right)+q_1^2 q_3 z \left(4 z^2+5\right)+q_1 q_3^2 z^2 \left(2 z^2+7
\right)+q_3^3 z \left(2 z^2+1\right)\right)}{64 \pi^2 F_\pi^6 q_1 q_3^2 
\left(z^2-1\right)^2}\nn
&&+\frac{3 g_A^4 z L(q_3) (c_2+c_3) (q_1+q_3 z) \left(q_1^2+2 
q_1 q_3 z+q_3^2\right)}{64 \pi^2 F_\pi^6 q_1^2 q_3 \left(z^2-1
\right)^2}+\frac{g_A^4 (c_2+c_3) \left(z \left(q_1^2+q_3^2
\right)+2 q_1 q_3\right)}{64 \pi^2 F_\pi^6 q_1 q_3
\left(z^2-1\right)},\nn
{\cal R}_{17}^{(5), g_A^4}&=&-\frac{g_A^4 }{16 F_\pi^6 \left(z^2-1
\right)} I(4:0,-q_1,q_3;0) (c_2+c_3) \left(-4 M_\pi^2 \left(z^2-1
\right)+q_1^2+2 q_1 q_3 z+q_3^2\right)\nn
&&-\frac{g_A^4 z L(q_2) (c_2+c_3) \left(q_1^2+2 q_1 q_3 z+q_3^2
\right)}{128 \pi^2 F_\pi^6 q_1 q_3 \left(z^2-1\right)}+\frac{g_A^4 
L(q_1) (c_2+c_3) (q_1 z+q_3)}{128 \pi^2 F_\pi^6 q_3 \left(z^2-1
\right)}\nn
&&+\frac{g_A^4 L(q_3) (c_2+c_3) (q_1+q_3 z)}{128 \pi^2 F_\pi^6 
q_1 \left(z^2-1\right)},\nn
{\cal R}_{6}^{(5), g_A^4}&=&{\cal R}_{18}^{(5), g_A^4}\;=\;{\cal
  R}_{19}^{(5), g_A^4}\;=\;
{\cal R}_{20}^{(5), g_A^4}\;=\;{\cal R}_{21}^{(5), g_A^4}\;=\;{\cal R}_{22}^{(5), g_A^4}\;=\;0.
\eeqa
In the above expressions, $q_1$ and $q_3$ are always to be understood as the
magnitudes of the corresponding three-momenta (except in the arguments of the
function $I$), $q_1 \equiv | \vec q_1 \, |$, 
$q_3 \equiv | \vec q_3 \, |$. Further, the function $I(d:p_1,p_2,p_3;p_4)$ 
refers to the scalar loop integral 
\beqa
I(d:p_1,p_2,p_3; p_4)&=&\frac{1}{i}\int
\frac{d^d l}{(2\pi)^d}\frac{1}{(l+p_1)^2-M_\pi^2+i \epsilon}
\frac{1}{(l+p_2)^2-M_\pi^2+i \epsilon}
\frac{1}{(l+p_3)^2-M_\pi^2+i \epsilon}\frac{1}{v\cdot(l+p_4)+i \epsilon}\,.
\quad\quad
\eeqa
This expression involves four-momenta $p_i$. For the
case $p_i^0 =0$ which we are interested in, it can be expressed in terms
of the three-point function in  Euclidean space 
$J\left(d:\vec{p}_1,\vec{p}_2,\vec{p}_3\right)$
\beq
J\left(d:\vec{p}_1,\vec{p}_2,\vec{p}_3\right)=
\int\frac{d^d
  l}{(2\pi)^d}\frac{1}{(\vec{l}+\vec{p}_1)^2+M_\pi^2}\frac{1}{(\vec{l}+\vec{p}_2)^2+M_\pi^2}\frac{1}{(\vec{l}+\vec{p}_3)^2+M_\pi^2}.
\eeq
In particular, the function $I\left(4:0,-q_1,q_3;0\right)$ which
enters the expressions for $R_i$ can be written as
\beq
I\left(4:0,-q_1,q_3;0\right) = \frac{1}{2}J\left(3:\vec{0},-\vec{q}_1,\vec{q}_3\right).
\eeq

The N$^4$LO contributions to the structure functions proportional to 
$g_A^2$ vanish for ${\cal R}_{1,3,4,6,8,9,10,15,\ldots ,22}$.  The
nonvanishing contributions have the form
\beqa
{\cal R}_{2}^{(5), g_A^2}&=&-\frac{g_A^2}{128 F_\pi^6 \left(z^2-1
\right)^2} I(4:0,-q_1,q_3;0) \left(32 c_1 \left(z^2-1\right) \left(4 
\left(z^2-1\right) M_\pi^2-q_1^2-q_3^2+2 \left(q_1^2+q_3^2\right) 
z^2\nnrlrl
+2 q_1 q_3 z\right) M_\pi^2-8 c_3 \left(2 M_\pi^2+q_1^2+q_3^2+2 
q_1 q_3 z\right) \left(z^2-1\right) \left(-4 M_\pi^2-q_1^2-q_3^2+2 
\left(2 M_\pi^2+q_1^2+q_3^2\right) z^2\nnrlrl
+2 q_1 q_3 z\right)-c_2 
\left(-4 M_\pi^2-q_1^2-q_3^2+4 \left(M_\pi^2+q_1^2+q_3^2\right) z^2+6 
q_1 q_3 z\right)\nnrl
\times \left(4 \left(z^2-1\right) M_\pi^2-q_3^2-q_1 (q_1+2 
q_3 z)\right)\right) \nn
&&-\frac{ g_A^2}{3072 F_\pi^6 \pi^2 q_3 \left(z^2-1
\right)^2}L(q_1)\left(-96 c_1 \left(z^2-1\right) \left(q_1 z+q_3 
\left(2 z^2-1\right)\right) M_\pi^2+c_2 \left(-3 \left(2 z^3+z\right) 
q_1^3\nnrlrl
+q_3 \left(8 \left(z^2-5\right) z^2+5\right) q_1^2-3 q_3^2 z 
\left(4 z^2+5\right) q_1+3 q_3^3 \left(1-4 z^2\right)+4 M_\pi^2 
\left(z^2-1\right) \left(3 q_1 z+q_3 \left(8 z^2-5\right)\right)
\right)\nnrl
+8 c_3 \left(z^2-1\right) \left(3 z q_1^3+q_3 \left(16 z^2-7
\right) q_1^2+3 q_3^2 z \left(2 z^2+1\right) q_1+3 q_3^3 \left(2 z^2-1
\right)\nnrlrl
+M_\pi^2 \left(6 q_1 z+q_3 \left(28 z^2-22\right)\right)\right)
\right)\nn
&&-\frac{ g_A^2}{3072 F_\pi^6 \pi^2 q_1 q_3 \left(z^2-1\right)^2}L(q_2)\left(96 c_1 \left(z^2-1\right) \left(z q_1^2-2 q_3 
\left(z^2-2\right) q_1+q_3^2 z\right) M_\pi^2\nnrl
-24 c_3 \left(2 
M_\pi^2+q_1^2+q_3^2+2 q_1 q_3 z\right) \left(z^2-1\right) \left(z 
q_1^2-2 q_3 \left(z^2-2\right) q_1+q_3^2 z\right)\nnrl
+c_2 
\left(\left(q_1^2+2 q_3 z q_1+q_3^2\right) \left(4 q_1 q_3 z^4+6 
\left(q_1^2+q_3^2\right) z^3-2 q_1 q_3 z^2+3 \left(q_1^2+q_3^2\right) 
z+16 q_1 q_3\right)\nnrlrl
-4 M_\pi^2 \left(z^2-1\right) \left(-10 q_1 q_3 
z^2+3 \left(q_1^2+q_3^2\right) z+16 q_1 q_3\right)\right)\right) \nn
&&-\frac{
g_A^2}{3072 F_\pi^6 \pi^2 q_1 \left(z^2-1\right)^2}L(q_3) 
\left(-96 c_1 \left(z^2-1\right) \left(q_3 z+q_1 \left(2 z^2-1\right)
\right) M_\pi^2+c_2 \left(\left(3-12 z^2\right) q_1^3\nnrlrl
-3 q_3 z \left(4 
z^2+5\right) q_1^2+q_3^2 \left(8 \left(z^2-5\right) z^2+5\right) 
q_1-3 q_3^3 z \left(2 z^2+1\right)+4 M_\pi^2 \left(z^2-1\right) \nnrlrl
\times
\left(3 q_3 z+q_1 \left(8 z^2-5\right)\right)\right)+8 c_3 
\left(z^2-1\right) \left(\left(6 z^2-3\right) q_1^3+3 q_3 \left(2 
z^3+z\right) q_1^2+q_3^2 \left(16 z^2-7\right) q_1\nnrlrl
+3 q_3^3 z+M_\pi^2 
\left(6 q_3 z+q_1 \left(28 z^2-22\right)\right)\right)\right) +\frac{c_2 
\left(q_1^2+2 q_3 z q_1+q_3^2\right) g_A^2}{1024 F_\pi^6 \pi^2 
\left(z^2-1\right)},\nn
{\cal R}_{5}^{(5), g_A^2}&=&\frac{g_A^2 }{8 F_\pi^6 \left(z^2-1\right)} I(4:0,-q_1,q_3;0) c_4 q_1 z (q_1 z+q_3) \left(q_1^2+2 q_1 
q_3 z+q_3^2\right)\nn
&&+\frac{g_A^2 c_4 
L(q_2) (q_1 z+q_3) \left(q_1^2+2 q_1 q_3 z+q_3^2\right)}{64 \pi^2 
F_\pi^6 q_3 \left(z^2-1\right)}-\frac{g_A^2 c_4 q_1 z L(q_1) 
\left(q_1^2+2 q_1 q_3 z+q_3^2\right)}{64 \pi^2 F_\pi^6 q_3 
\left(z^2-1\right)}\nn
&&-\frac{g_A^2 c_4 L(q_3) \left(q_1^2 z^2+q_1 q_3 z 
\left(z^2+1\right)+q_3^2\right)}{64 \pi^2 F_\pi^6
\left(z^2-1\right)},\nn
{\cal R}_{7}^{(5), g_A^2}&=&-\frac{g_A^2 }{64 F_\pi^6 q_1 q_3 
\left(z^2-1\right)^2} I(4:0,-q_1,q_3;0) c_4 z
  \left(q_1^2+2 q_1 q_3 z+q_3^2
\right) \left(4 M_\pi^2 \left(z^2-1\right)+q_1^2 \left(2
    z^2-1\right)\nnrl
+2 
q_1 q_3 z+q_3^2 \left(2 z^2-1\right)\right)\nn
&&-\frac{g_A^2 }{1536 \pi^2 F_\pi^6 q_1^2 q_3^2 \left(z^2-1
\right)^2} c_4 L(q_2) \left(q_1^2+2 q_1 q_3 
z+q_3^2\right) \left(16 M_\pi^2 \left(z^2-1\right)+q_1^2 \left(7 
z^2-4\right)\nnrl
+2 q_1 q_3 z \left(z^2+2\right)+q_3^2 \left(7 z^2-4
\right)\right)\nn
&&+\frac{g_A^2 }{1536 \pi^2 F_\pi^6 q_1 q_3^2 \left(z^2-1
\right)^2} c_4 L(q_1) \left(16 M_\pi^2 \left(z^2-1\right) 
(q_1+q_3 z)+q_1^3 \left(7 z^2-4\right)+q_1^2 q_3 z \left(10 z^2-1
\right)\nnrl
+3 q_1 q_3^2 z^2 \left(2 z^2+1\right)+3 q_3^3 z \left(2 z^2-1
\right)\right)\nn
&&+\frac{g_A^2 }{1536 \pi^2 F_\pi^6 q_1^2 q_3 \left(z^2-1
\right)^2} c_4 L(q_3) \left(16 M_\pi^2 \left(z^2-1\right) 
(q_1 z+q_3)+3 q_1^3 z \left(2 z^2-1\right)+3 q_1^2 q_3 z^2 \left(2 
z^2+1\right)\nnrl
+q_1 q_3^2 z \left(10 z^2-1\right)+q_3^3 \left(7 z^2-4
\right)\right)+\frac{g_A^2 c_4 \left(z \left(q_1^2+q_3^2\right)+2 q_1 q_3
\right)}{1536 \pi^2 F_\pi^6 q_1 q_3 \left(z^2-1\right)},\nn
{\cal R}_{11}^{(5), g_A^2}&=&\frac{g_A^2 }{16 
F_\pi^6 q_1 \left(z^2-1\right)^2} I(4:0,-q_1,q_3;0) c_4 z (q_1 z+q_3) \left(3 \left(q_1^2+2 
q_1 q_3 z+q_3^2\right)-4 M_\pi^2 \left(z^2-1\right)\right)\nn
&&+\frac{g_A^2 c_4 \left(z^2+2\right) 
L(q_2) (q_1 z+q_3) \left(q_1^2+2 q_1 q_3 z+q_3^2\right)}{128 \pi^2 
F_\pi^6 q_1^2 q_3 \left(z^2-1\right)^2}-\frac{g_A^2 c_4 z L(q_1) 
\left(q_1^2 \left(z^2+2\right)+6 q_1 q_3 z+3 q_3^2\right)}{128 \pi^2 
F_\pi^6 q_1 q_3 \left(z^2-1\right)^2}\nn
&&-\frac{g_A^2 c_4 L(q_3) \left(3 
q_1^2 z^2+3 q_1 q_3 z \left(z^2+1\right)+q_3^2 \left(z^2+2\right)
\right)}{128 \pi^2 F_\pi^6 q_1^2 \left(z^2-1\right)^2}-\frac{g_A^2 c_4 
(q_1+q_3 z)}{128 \pi^2 F_\pi^6 q_1 \left(z^2-1\right)},\nn
{\cal R}_{12}^{(5), g_A^2}&=&-\frac{g_A^2 }{16 F_\pi^6 q_3 \left(z^2-1
\right)^2} I(4:0,-q_1,q_3;0) c_4 z \left(\left(q_1^2+2 q_1 q_3 
z+q_3^2\right) \left(2 q_1 z^2+q_1+3 q_3 z\right)\nnrl
-4 M_\pi^2 
\left(z^2-1\right) (q_1+q_3 z)\right)-\frac{g_A^2 c_4 z L(q_2) \left(q_1^2+2 q_1 q_3 z+q_3^2
\right) \left(3 q_1 z+q_3 \left(z^2+2\right)\right)}{128 \pi^2 F_\pi^6 
q_1 q_3^2 \left(z^2-1\right)^2}\nn
&&+\frac{g_A^2 c_4 z L(q_1) \left(3 z 
\left(q_1^2+q_3^2\right)+5 q_1 q_3 z^2+q_1 q_3\right)}{128 \pi^2 
F_\pi^6 q_3^2 \left(z^2-1\right)^2}\nn
&&+\frac{g_A^2 c_4 z L(q_3) 
\left(q_1^2 \left(2 z^2+1\right)+2 q_1 q_3 z \left(z^2+2\right)+q_3^2 
\left(z^2+2\right)\right)}{128 \pi^2 F_\pi^6 q_1 q_3 \left(z^2-1
\right)^2}+\frac{g_A^2 c_4 (q_1 z+q_3)}{128 \pi^2 F_\pi^6 q_3 
\left(z^2-1\right)},\nn
{\cal R}_{13}^{(5), g_A^2}&=&-\frac{g_A^2 }{16 F_\pi^6 q_3 \left(z^2-1\right)^2} I(4:0,-q_1,q_3;0) c_4 (q_1
  z+q_3) \left(\left(2 z^2+1
\right) \left(q_1^2+2 q_1 q_3 z+q_3^2\right)-4 M_\pi^2 \left(z^2-1
\right)\right)\nn
&&-\frac{3 g_A^2 c_4 
z L(q_2) (q_1 z+q_3) \left(q_1^2+2 q_1 q_3 z+q_3^2\right)}{128 \pi^2 
F_\pi^6 q_1 q_3^2 \left(z^2-1\right)^2}+\frac{g_A^2 c_4 L(q_1) 
\left(3 q_1^2 z^2+2 q_1 q_3 \left(2 z^3+z\right)+q_3^2 \left(2 z^2+1
\right)\right)}{128 \pi^2 F_\pi^6 q_3^2 \left(z^2-1
\right)^2}\nn
&&+\frac{g_A^2 c_4 L(q_3) \left(q_1^2 \left(2 z^3+z\right)+q_1 
q_3 \left(2 z^4+3 z^2+1\right)+3 q_3^2 z\right)}{128 \pi^2 F_\pi^6 
q_1 q_3 \left(z^2-1\right)^2}+\frac{g_A^2 c_4 (q_1 z+q_3)}{128 \pi^2 
F_\pi^6 q_3 \left(z^2-1\right)},\nn
{\cal R}_{14}^{(5), g_A^2}&=&\frac{g_A^2 }{16 F_\pi^6 q_3^2 \left(z^2-1
\right)^2} I(4:0,-q_1,q_3;0) c_4 q_1 \left(\left(q_1^2+2 q_1 q_3 
z+q_3^2\right) \left(2 q_1 z^2+q_1+3 q_3 z\right)\nnrl
-4 M_\pi^2 
\left(z^2-1\right) (q_1+q_3 z)\right)+\frac{g_A^2 c_4 L(q_2) \left(q_1^2+2 q_1 q_3 z+q_3^2\right) 
\left(3 q_1 z+q_3 \left(z^2+2\right)\right)}{128 \pi^2 F_\pi^6 q_3^3 
\left(z^2-1\right)^2}\nn
&&-\frac{g_A^2 c_4 L(q_3) \left(q_1^2 \left(2 
z^2+1\right)+2 q_1 q_3 z \left(z^2+2\right)+q_3^2 \left(z^2+2\right)
\right)}{128 \pi^2 F_\pi^6 q_3^2 \left(z^2-1\right)^2}\nn
&&-\frac{g_A^2 c_4 
q_1 L(q_1) \left(3 z \left(q_1^2+q_3^2\right)+5 q_1 q_3 z^2+q_1 q_3
\right)}{128 \pi^2 F_\pi^6 q_3^3 \left(z^2-1\right)^2}-\frac{g_A^2 c_4 
q_1 (q_1+q_3 z)}{128 \pi^2 F_\pi^6 q_3^2 \left(z^2-1\right)}.
\eeqa

Finally, the N$^4$LO contributions to the structure functions proportional to 
$g_A^0$ vanish for ${\cal R}_{1,3,\ldots ,6,8,\ldots ,22}$.  The nonvanishing contributions have the form 
\beqa
{\cal R}_{2}^{(5), g_A^0}&=&-\frac{1}{256 F_\pi^6 \left(z^2-1\right)^2} I(4:0,-q_1,q_3;0) \left(4 M_\pi^2 \left(z^2-1\right)-q_1 
(q_1+2 q_3 z)-q_3^2\right) \left(32 c_1 M_\pi^2
\left(z^2-1\right)\nnrl
+3 
c_2 \left(-4 M_\pi^2 \left(z^2-1\right)+q_1^2+2 q_1 q_3 z+q_3^2
\right)-8 c_3 \left(z^2-1\right) \left(2 M_\pi^2+q_1^2+2 q_1 q_3 
z+q_3^2\right)\right)\nn
&&-\frac{1 }{2048 \pi^2 F_\pi^6 q_1 q_3 \left(z^2-1\right)^2} z 
L(q_2) \left(q_1^2+2 q_1 q_3 z+q_3^2\right) \left(-32 c_1 M_\pi^2 
\left(z^2-1\right)+c_2 \left(20 M_\pi^2 \left(z^2-1\right)\nnrlrl
+\left(2 
z^2-5\right) \left(q_1^2+2 q_1 q_3 z+q_3^2\right)\right)+8 c_3 
\left(z^2-1\right) \left(2 M_\pi^2+q_1^2+2 q_1 q_3 z+q_3^2\right)
\right)\nn
&&-\frac{1}{6144 \pi^2 F_\pi^6 
q_3 \left(z^2-1\right)^2} L(q_1) 
\left(96 c_1 M_\pi^2 \left(z^2-1\right) (q_1 z+q_3)+3 c_2 (q_1 z+q_3) 
\left(-20 M_\pi^2 \left(z^2-1\right)\nnrlrl
+q_1^2 \left(5-2 z^2\right)+6 q_1 
q_3 z+3 q_3^2\right)-8 c_3 \left(z^2-1\right) \left(2 M_\pi^2 \left(3 
q_1 z+q_3 \left(7-4 z^2\right)\right)+3 q_1^3 z\nnrlrl
+q_1^2 q_3 \left(4 
z^2+5\right)
+9 q_1 q_3^2 z+3 q_3^3\right)\right)\nn
&&-\frac{1 }{6144 \pi^2 F_\pi^6 q_1 
\left(z^2-1\right)^2} L(q_3) \left(96 c_1 M_\pi^2 
\left(z^2-1\right) (q_1+q_3 z)+3 c_2 (q_1+q_3 z) \left(-20 M_\pi^2 
\left(z^2-1\right)\nnrlrl
+3 q_1^2+6 q_1 q_3 z+q_3^2 \left(5-2 z^2\right)
\right)-8 c_3 \left(z^2-1\right) \left(2 M_\pi^2 \left(q_1 \left(7-4 
z^2\right)+3 q_3 z\right)+3 q_1^3+9 q_1^2 q_3 z\nnrlrl
+q_1 q_3^2 \left(4 
z^2+5\right)+3 q_3^3 z\right)\right)-\frac{c_2 \left(q_1^2+2 q_1 q_3 z+q_3^2
\right)}{2048 \pi^2 F_\pi^6 \left(z^2-1\right)},\nn
{\cal R}_{7}^{(5), g_A^0}&=&\frac{1 }{128 F_\pi^6 q_1 q_3 \left(z^2-1\right)^2} I(4:0,-q_1,q_3;0) c_4 z \left(q_1^2+2 q_1 q_3 z+q_3^2\right) 
\left(-4 M_\pi^2 \left(z^2-1\right)+q_1^2+2 q_1 q_3 z+q_3^2
\right)\nn
&&+\frac{c_4 L(q_2) 
\left(q_1^2+2 q_1 q_3 z+q_3^2\right) \left(\left(z^2+2\right) 
\left(q_1^2+2 q_1 q_3 z+q_3^2\right)-8 M_\pi^2 \left(z^2-1\right)
\right)}{3072 \pi^2 F_\pi^6 q_1^2 q_3^2 \left(z^2-1\right)^2}\nn
&&-\frac{c_4 
L(q_1) \left(-8 M_\pi^2 \left(z^2-1\right) (q_1+q_3 z)+q_1^3 
\left(z^2+2\right)+q_1^2 q_3 z \left(4 z^2+5\right)+9 q_1 q_3^2 z^2+3 
q_3^3 z\right)}{3072 \pi^2 F_\pi^6 q_1 q_3^2 \left(z^2-1
\right)^2}\nn
&&-\frac{c_4 L(q_3) \left(-8 M_\pi^2 \left(z^2-1\right) (q_1 
z+q_3)+3 q_1^3 z+9 q_1^2 q_3 z^2+q_1 q_3^2 z \left(4 z^2+5
\right)+q_3^3 \left(z^2+2\right)\right)}{3072 \pi^2 F_\pi^6 q_1^2 q_3 
\left(z^2-1\right)^2}\nn
&&-\frac{c_4 \left(z \left(q_1^2+q_3^2\right)+2 
q_1 q_3\right)}{3072 \pi^2 F_\pi^6 q_1 q_3 \left(z^2-1\right)}.
\eeqa
A Mathematica notebook which contains the above expressions for the
structure functions in momentum space is available from the authors
upon request.


\begin{thebibliography}{99}

%\cite{Stephan:2010zz}
\bibitem{Stephan:2010zz} 
  E.~Stephan, S.~Kistryn, R.~Sworst, A.~Biegun, K.~Bodek, I.~Ciepal, A.~Deltuva and E.~Epelbaum {\it et al.},
  %``Vector and tensor analyzing powers in deuteron-proton breakup at 130 MeV,''
  Phys.\ Rev.\ C {\bf 82}, 014003 (2010).
  %%CITATION = PHRVA,C82,014003;%%


%\cite{Ciepal:2012zz}
\bibitem{Ciepal:2012zz} 
  I.~Ciepal, S.~.Kistryn, E.~Stephan, A.~Biegun, K.~Bodek, A.~Deltuva, E.~Epelbaum and M.~Eslami-Kalantari {\it et al.},
  %``Vector analyzing powers of deuteron-proton elastic scattering and breakup at 130 MeV,''
  Phys.\ Rev.\ C {\bf 85}, 017001 (2012).
  %%CITATION = PHRVA,C85,017001;%%

%\cite{Viviani:2010mf}
\bibitem{Viviani:2010mf} 
  M.~Viviani, L.~Girlanda, A.~Kievsky, L.~E.~Marcucci and S.~Rosati,
  %``Proton-^3He elastic scattering at low energies and the 'A_y Puzzle',''
  arXiv:1004.1306 [nucl-th].
  %%CITATION = ARXIV:1004.1306;%%

%\cite{Navratil:2007we}
\bibitem{Navratil:2007we} 
  P.~Navratil, V.~G.~Gueorguiev, J.~P.~Vary, W.~E.~Ormand and A.~Nogga,
  %``Structure of A=10-13 nuclei with two plus three-nucleon interactions from chiral effective field theory,''
  Phys.\ Rev.\ Lett.\  {\bf 99}, 042501 (2007)
  [nucl-th/0701038].
  %%CITATION = NUCL-TH/0701038;%%

%\cite{Gazit:2008ma}
\bibitem{Gazit:2008ma} 
  D.~Gazit, S.~Quaglioni and P.~Navratil,
  %``Three-Nucleon Low-Energy Constants from the Consistency of Interactions and Currents in Chiral Effective Field Theory,''
  Phys.\ Rev.\ Lett.\  {\bf 103}, 102502 (2009)
  [arXiv:0812.4444 [nucl-th]].
  %%CITATION = ARXIV:0812.4444;%%

%\cite{Roth:2011ar}
\bibitem{Roth:2011ar} 
  R.~Roth, J.~Langhammer, A.~Calci, S.~Binder and P.~Navratil,
  %``Similarity-Transformed Chiral NN+3N Interactions for the Ab Initio Description of 12-C and 16-O,''
  Phys.\ Rev.\ Lett.\  {\bf 107}, 072501 (2011)
  [arXiv:1105.3173 [nucl-th]].
  %%CITATION = ARXIV:1105.3173;%%

%\cite{Hebeler:2010jx}
\bibitem{Hebeler:2010jx} 
  K.~Hebeler, J.~M.~Lattimer, C.~J.~Pethick and A.~Schwenk,
  %``Constraints on neutron star radii based on chiral effective field theory interactions,''
  Phys.\ Rev.\ Lett.\  {\bf 105}, 161102 (2010)
  [arXiv:1007.1746 [nucl-th]].
  %%CITATION = ARXIV:1007.1746;%%

%\cite{Hagen:2012sh}
\bibitem{Hagen:2012sh} 
  G.~Hagen, M.~Hjorth-Jensen, G.~R.~Jansen, R.~Machleidt and T.~Papenbrock,
  %``Continuum effects and three-nucleon forces in neutron-rich oxygen isotopes,''
  arXiv:1202.2839 [nucl-th].
  %%CITATION = ARXIV:1202.2839;%%

%\cite{Holt:2012fr}
\bibitem{Holt:2012fr} 
  J.~D.~Holt, J.~Menendez and A.~Schwenk,
  %``Three-body forces and proton-rich nuclei,''
  Phys.\ Rev.\ Lett.\  {\bf 110}, 022502 (2013)
  [arXiv:1207.1509 [nucl-th]].
  %%CITATION = ARXIV:1207.1509;%%

%\cite{Tews:2012fj}
\bibitem{Tews:2012fj} 
  I.~Tews, T.~Krueger, K.~Hebeler and A.~Schwenk,
  %``Neutron matter at next-to-next-to-next-to-leading order in chiral effective field theory,''
  arXiv:1206.0025 [nucl-th].
  %%CITATION = ARXIV:1206.0025;%%


%\cite{KalantarNayestanaki:2011wz}
\bibitem{KalantarNayestanaki:2011wz} 
  N.~Kalantar-Nayestanaki, E.~Epelbaum, J.~G.~Messchendorp and A.~Nogga,
  %``Signatures of three-nucleon interactions in few-nucleon systems,''
  Rept.\ Prog.\ Phys.\  {\bf 75}, 016301 (2012)
  [arXiv:1108.1227 [nucl-th]].
  %%CITATION = ARXIV:1108.1227;%%


%\cite{Hammer:2012id}
\bibitem{Hammer:2012id} 
  H.-W.~Hammer, A.~Nogga and A.~Schwenk,
  %``Three-body forces: From cold atoms to nuclei,''
  arXiv:1210.4273 [nucl-th].
  %%CITATION = ARXIV:1210.4273;%%


%\cite{Epelbaum:2008ga}
\bibitem{Epelbaum:2008ga} 
  E.~Epelbaum, H.~-W.~Hammer and U.-G.~Mei{\ss}ner,
  %``Modern Theory of Nuclear Forces,''
  Rev.\ Mod.\ Phys.\  {\bf 81}, 1773 (2009)
  [arXiv:0811.1338 [nucl-th]].
  %%CITATION = ARXIV:0811.1338;%%

%\cite{Machleidt:2011zz}
\bibitem{Machleidt:2011zz} 
  R.~Machleidt and D.~R.~Entem,
  %``Chiral effective field theory and nuclear forces,''
  Phys.\ Rept.\  {\bf 503}, 1 (2011)
  [arXiv:1105.2919 [nucl-th]].
  %%CITATION = ARXIV:1105.2919;%%

%\cite{Epelbaum:2012vx}
\bibitem{Epelbaum:2012vx} 
  E.~Epelbaum and U.-G.~Mei{\ss}ner,
  %``Chiral dynamics of few- and many-nucleon systems,''
  Ann.\ Rev.\ Nucl.\ Part.\ Sci.\  {\bf 62}, 159 (2012)
  [arXiv:1201.2136 [nucl-th]].
  %%CITATION = ARXIV:1201.2136;%%

%\cite{Entem:2003ft}
\bibitem{Entem:2003ft} 
  D.~R.~Entem and R.~Machleidt,
  %``Accurate charge dependent nucleon nucleon potential at fourth order of chiral perturbation theory,''
  Phys.\ Rev.\ C {\bf 68}, 041001 (2003)
  [nucl-th/0304018].
  %%CITATION = NUCL-TH/0304018;%%

%\cite{Epelbaum:2004fk}
\bibitem{Epelbaum:2004fk} 
  E.~Epelbaum, W.~Gl\"ockle and U.-G.~Mei{\ss}ner,
  %``The Two-nucleon system at next-to-next-to-next-to-leading order,''
  Nucl.\ Phys.\ A {\bf 747}, 362 (2005)
  [nucl-th/0405048].
  %%CITATION = NUCL-TH/0405048;%%


%\cite{vanKolck:1994yi}
\bibitem{vanKolck:1994yi} 
  U.~van Kolck,
  %``Few nucleon forces from chiral Lagrangians,''
  Phys.\ Rev.\ C {\bf 49}, 2932 (1994).
  %%CITATION = PHRVA,C49,2932;%%

%\cite{Epelbaum:2002vt}
\bibitem{Epelbaum:2002vt} 
  E.~Epelbaum, A.~Nogga, W.~Gl\"ockle, H.~Kamada, U.-G.~Mei{\ss}ner and H.~Wita{\l}a,
  %``Three nucleon forces from chiral effective field theory,''
  Phys.\ Rev.\ C {\bf 66}, 064001 (2002)
  [nucl-th/0208023].
  %%CITATION = NUCL-TH/0208023;%%

%\cite{Epelbaum:2009zsa}
\bibitem{Epelbaum:2009zsa} 
  E.~Epelbaum, H.~Krebs, D.~Lee and U.-G.~Mei{\ss}ner,
  %``Lattice chiral effective field theory with three-body interactions at next-to-next-to-leading order,''
  Eur.\ Phys.\ J.\ A {\bf 41}, 125 (2009)
  [arXiv:0903.1666 [nucl-th]].
  %%CITATION = ARXIV:0903.1666;%%

%\cite{Bernard:1996gq}
\bibitem{Bernard:1996gq} 
  V.~Bernard, N.~Kaiser and U.-G.~Mei{\ss}ner,
  %``Aspects of chiral pion - nucleon physics,''
  Nucl.\ Phys.\ A {\bf 615}, 483 (1997)
  [hep-ph/9611253].
  %%CITATION = HEP-PH/9611253;%%

%\cite{Fettes:1998ud}
\bibitem{Fettes:1998ud} 
  N.~Fettes, U.-G.~Mei{\ss}ner and S.~Steininger,
  %``Pion - nucleon scattering in chiral perturbation theory. 1. Isospin symmetric case,''
  Nucl.\ Phys.\ A {\bf 640}, 199 (1998)
  [hep-ph/9803266].
  %%CITATION = HEP-PH/9803266;%%

%\cite{Fettes:2000xg}
\bibitem{Fettes:2000xg} 
  N.~Fettes and U.-G.~Mei{\ss}ner,
  %``Pion nucleon scattering in chiral perturbation theory. 2.: Fourth order calculation,''
  Nucl.\ Phys.\ A {\bf 676}, 311 (2000)
  [hep-ph/0002162].
  %%CITATION = HEP-PH/0002162;%%

%\cite{Buettiker:1999ap}
\bibitem{Buettiker:1999ap} 
  P.~Buettiker and U.-G.~Mei{\ss}ner,
  %``Pion nucleon scattering inside the Mandelstam triangle,''
  Nucl.\ Phys.\ A {\bf 668}, 97 (2000)
  [hep-ph/9908247].
  %%CITATION = HEP-PH/9908247;%%

%\cite{Alarcon:2012kn}
\bibitem{Alarcon:2012kn} 
  J.~M.~Alarcon, J.~M.~Camalich and J.~A.~Oller,
  %``Improved description of the $\pi N$-scattering phenomenology in covariant baryon chiral perturbation theory,''
  arXiv:1210.4450 [hep-ph].
  %%CITATION = ARXIV:1210.4450;%%

%\cite{Chen:2012nx}
\bibitem{Chen:2012nx} 
  Y.-H.~Chen, D.-L.~Yao and H.~Q.~Zheng,
  %``Analyses of pion-nucleon elastic scattering amplitudes up to $O(p^4)$ in extended-on-mass-shell subtraction scheme,''
  arXiv:1212.1893 [hep-ph].
  %%CITATION = ARXIV:1212.1893;%%

%\cite{Rentmeester:2003mf}
\bibitem{Rentmeester:2003mf} 
  M.~C.~M.~Rentmeester, R.~G.~E.~Timmermans and J.~J.~de Swart,
  %``Determination of the chiral coupling constants c(3) and c(4) in new pp and np partial wave analyses,''
  Phys.\ Rev.\ C {\bf 67}, 044001 (2003)
  [nucl-th/0302080].
  %%CITATION = NUCL-TH/0302080;%%



%\cite{Bernard:2007sp}
\bibitem{Bernard:2007sp} 
  V.~Bernard, E.~Epelbaum, H.~Krebs and U.-G.~Mei{\ss}ner,
  %``Subleading contributions to the chiral three-nucleon force. I. Long-range terms,''
  Phys.\ Rev.\ C {\bf 77}, 064004 (2008)
  [arXiv:0712.1967 [nucl-th]].
  %%CITATION = ARXIV:0712.1967;%%

%\cite{Ishikawa:2007zz}
\bibitem{Ishikawa:2007zz} 
  S.~Ishikawa and M.~R.~Robilotta,
  %``Two-pion exchange three-nucleon potential: O(q**4) chiral expansion,''
  Phys.\ Rev.\ C {\bf 76}, 014006 (2007)
  [arXiv:0704.0711 [nucl-th]].
  %%CITATION = ARXIV:0704.0711;%%

%\cite{Bernard:2011zr}
\bibitem{Bernard:2011zr} 
  V.~Bernard, E.~Epelbaum, H.~Krebs and U.-G.~Mei{\ss}ner,
  %``Subleading contributions to the chiral three-nucleon force II: Short-range terms and relativistic corrections,''
  Phys.\ Rev.\ C {\bf 84}, 054001 (2011)
  [arXiv:1108.3816 [nucl-th]].
  %%CITATION = ARXIV:1108.3816;%%

%\cite{Robilotta:2006xq}
\bibitem{Robilotta:2006xq} 
  M.~R.~Robilotta,
  %``Nucleon-Nucleon Potential: Drift Effects,''
  Phys.\ Rev.\ C {\bf 74}, 044002 (2006)
  [Erratum-ibid.\ C {\bf 74}, 059902 (2006)]
  [nucl-th/0610046].
  %%CITATION = NUCL-TH/0610046;%%

%\cite{Golak:2009ri}
\bibitem{Golak:2009ri} 
  J.~Golak, D.~Rozpedzik, R.~Skibinski, K.~Topolnicki, H.~Wita{\l}a, W.~Gl\"ockle, A.~Nogga and E.~Epelbaum {\it et al.},
  %``A new way to perform partial wave decompositions of few-nucleon forces,''
  Eur.\ Phys.\ J.\ A {\bf 43}, 241 (2010)
  [arXiv:0911.4173 [nucl-th]].
  %%CITATION = ARXIV:0911.4173;%%

%\cite{Skibinski:2011vi}
\bibitem{Skibinski:2011vi} 
  R.~Skibinski, J.~Golak, K.~Topolnicki, H.~Wita{\l}a, E.~Epelbaum, W.~Gl\"ockle, H.~Krebs and A.~Nogga {\it et al.},
  %``The triton with long-range chiral N3LO three nucleon forces,''
  Phys.\ Rev.\ C {\bf 84}, 054005 (2011)
  [arXiv:1107.5163 [nucl-th]].
  %%CITATION = ARXIV:1107.5163;%%

%\cite{Ordonez:1993tn}
\bibitem{Ordonez:1993tn} 
  C.~Ordonez, L.~Ray and U.~van Kolck,
  %``Nucleon-nucleon potential from an effective chiral Lagrangian,''
  Phys.\ Rev.\ Lett.\  {\bf 72}, 1982 (1994).
  %%CITATION = PRLTA,72,1982;%%

%\cite{Kaiser:1998wa}
\bibitem{Kaiser:1998wa} 
  N.~Kaiser, S.~Gerstendorfer and W.~Weise,
  %``Peripheral NN scattering: Role of delta excitation, correlated two pion and vector meson exchange,''
  Nucl.\ Phys.\ A {\bf 637}, 395 (1998)
  [nucl-th/9802071].
  %%CITATION = NUCL-TH/9802071;%%

%\cite{Krebs:2007rh}
\bibitem{Krebs:2007rh} 
  H.~Krebs, E.~Epelbaum and U.-G.~Mei{\ss}ner,
  %``Nuclear forces with Delta-excitations up to next-to-next-to-leading order. I. Peripheral nucleon-nucleon waves,''
  Eur.\ Phys.\ J.\ A {\bf 32}, 127 (2007)
  [nucl-th/0703087].
  %%CITATION = NUCL-TH/0703087;%%


%\cite{Kaiser:1997mw}
\bibitem{Kaiser:1997mw} 
  N.~Kaiser, R.~Brockmann and W.~Weise,
  %``Peripheral nucleon-nucleon phase shifts and chiral symmetry,''
  Nucl.\ Phys.\ A {\bf 625}, 758 (1997)
  [nucl-th/9706045].
  %%CITATION = NUCL-TH/9706045;%%

%\cite{Machleidt:2010kb}
\bibitem{Machleidt:2010kb} 
  R.~Machleidt and D.~R.~Entem,
  %``Nuclear forces from chiral EFT: The Unfinished business,''
  J.\ Phys.\ G G {\bf 37}, 064041 (2010)
  [arXiv:1001.0966 [nucl-th]].
  %%CITATION = ARXIV:1001.0966;%%

%\cite{Pieper:2001ap}
\bibitem{Pieper:2001ap} 
  S.~C.~Pieper, V.~R.~Pandharipande, R.~B.~Wiringa and J.~Carlson,
  %``Realistic models of pion exchange three nucleon interactions,''
  Phys.\ Rev.\ C {\bf 64}, 014001 (2001)
  [nucl-th/0102004].
  %%CITATION = NUCL-TH/0102004;%%

%\cite{Krebs:2012yv}
\bibitem{Krebs:2012yv} 
  H.~Krebs, A.~Gasparyan and E.~Epelbaum,
  %``Chiral three-nucleon force at N^4LO I: Longest-range contributions,''
  Phys.\ Rev.\ C {\bf 85}, 054006 (2012)
  [arXiv:1203.0067 [nucl-th]].
  %%CITATION = ARXIV:1203.0067;%%

%\cite{Fettes:2000gb}
\bibitem{Fettes:2000gb} 
  N.~Fettes, U.-G.~Mei{\ss}ner, M.~Mojzis and S.~Steininger,
  %``The Chiral effective pion nucleon Lagrangian of order p**4,''
  Annals Phys.\  {\bf 283}, 273 (2000)
  [Erratum-ibid.\  {\bf 288}, 249 (2001)]
  [hep-ph/0001308].
  %%CITATION = HEP-PH/0001308;%%

%\cite{Epelbaum:2004xf}
\bibitem{Epelbaum:2004xf} 
  E.~Epelbaum, U.-G.~Mei{\ss}ner and J.~E.~Palomar,
  %``Isospin dependence of the three-nucleon force,''
  Phys.\ Rev.\ C {\bf 71}, 024001 (2005)
  [nucl-th/0407037].
  %%CITATION = NUCL-TH/0407037;%%

\bibitem{GloeckleBook} 
W.~Gl\"ockle, {\it The Quantum Mechanical Few-Body Problem},
Springer-Verlag, 1983.

%\cite{Baru:2012iv}
\bibitem{Baru:2012iv} 
  V.~Baru, E.~Epelbaum, C.~Hanhart, M.~Hoferichter, A.~E.~Kudryavtsev and D.~R.~Phillips,
  %``The Multiple-scattering series in pion-deuteron scattering and the nucleon-nucleon potential: Perspectives from effective field theory,''
  Eur.\ Phys.\ J.\ A {\bf 48}, 69 (2012)
  [arXiv:1202.0208 [nucl-th]].
  %%CITATION = ARXIV:1202.0208;%%

%\cite{Girlanda:2011fh}
\bibitem{Girlanda:2011fh} 
  L.~Girlanda, A.~Kievsky and M.~Viviani,
  %``Subleading contributions to the three-nucleon contact interaction,''
  Phys.\ Rev.\ C {\bf 84}, 014001 (2011)
  [arXiv:1102.4799 [nucl-th]].
  %%CITATION = ARXIV:1102.4799;%%

%\cite{Ordonez:1995rz}
\bibitem{Ordonez:1995rz} 
  C.~Ordonez, L.~Ray and U.~van Kolck,
  %``The Two nucleon potential from chiral Lagrangians,''
  Phys.\ Rev.\ C {\bf 53}, 2086 (1996)
  [hep-ph/9511380].
  %%CITATION = HEP-PH/9511380;%%

%\cite{Epelbaum:2007sq}
\bibitem{Epelbaum:2007sq} 
  E.~Epelbaum, H.~Krebs and U.-G.~Mei{\ss}ner,
  %``Delta-excitations and the three-nucleon force,''
  Nucl.\ Phys.\ A {\bf 806}, 65 (2008)
  [arXiv:0712.1969 [nucl-th]].
  %%CITATION = ARXIV:0712.1969;%%

%\cite{Epelbaum:2008td}
\bibitem{Epelbaum:2008td} 
  E.~Epelbaum, H.~Krebs and U.-G.~Mei{\ss}ner,
  %``Isospin-breaking two-nucleon force with explicit Delta-excitations,''
  Phys.\ Rev.\ C {\bf 77}, 034006 (2008)
  [arXiv:0801.1299 [nucl-th]].
  %%CITATION = ARXIV:0801.1299;%%






\end{thebibliography}
\end{document}